\documentclass[showpacs,preprintnumbers,amsmath,amssymb,nofootinbib]{revtex4}

\usepackage{graphicx}% Include figure files
\usepackage{dcolumn}% Align table columns on decimal point
\usepackage{bm}% bold math

\def\simge{\mathrel{%
       \rlap{\raise 0.511ex \hbox{$>$}}{\lower 0.511ex \hbox{$\sim$}}}}
\def\simle{\mathrel{
       \rlap{\raise 0.511ex \hbox{$<$}}{\lower 0.511ex \hbox{$\sim$}}}}

\begin{document}

%%%%%%%%%%%%%%%%%%%%%%%%%%%%%%%%%%%%%%%
%\baselineskip 16pt plus 1pt minus 1pt%
%%%%%%%%%%%%%%%%%%%%%%%%%%%%%%%%%%%%%%%

\preprint{\sf TKYNT-07-02, UTHEP-535, BNL-NT-07/3\ \ 2007/February}

\title{Heavy-Quark Free Energy, Debye Mass,
 and Spatial String Tension \\
 at Finite Temperature 
 in Two Flavor Lattice QCD with Wilson Quark Action}

\author{Y.~Maezawa$^1$, N.~Ukita$^2$,
S.~Aoki$^{3,4}$, S.~Ejiri$^5$, T.~Hatsuda$^1$, 
N.~Ishii$^2$ and K.~Kanaya$^3$ \\
(WHOT-QCD Collaboration)}

\affiliation{$^1$Department of Physics, The University of Tokyo, 
Tokyo 113-0033, Japan \\
$^2$Center for Computational Sciences,
University of Tsukuba, Tsukuba, Ibaraki 305-8577, Japan \\
$^3$Graduate School of Pure and Applied Sciences, 
University of Tsukuba, Tsukuba, Ibaraki 305-8571, Japan \\
$^4$RIKEN BNL Research Center,
Brookhaven National Laboratory, Upton, New York 11973, USA \\
$^5$Physics Department,
Brookhaven National Laboratory, Upton, New York 11973, USA}

\date{\today}

\begin{abstract}
We study Polyakov loop correlations and spatial Wilson loop 
at finite temperature in two-flavor QCD
 simulations with the RG-improved gluon action
and the clover-improved Wilson quark action
 on a $ 16^3 \times 4$ lattice. 
 From the line of constant physics
 at $m_{\rm PS}/m_{\rm V} = 0.65$ and 0.80, we
 extract the heavy-quark free energies, the  effective running coupling 
 $g_{\rm eff}(T)$  and the Debye screening mass $m_D(T)$ 
  for  various color channels of heavy quark--quark
 and quark--anti-quark  pairs
 above the critical temperature.  The free energies are well
  approximated by the screened Coulomb form with 
  the appropriate  Casimir factors at high temperature. 
 The magnitude and the temperature dependence of the 
 Debye mass are compared to those of the  next-to-leading order
  thermal perturbation theory and to a phenomenological
   formula in terms of  $g_{\rm eff}(T)$.
 We make a comparison between  our results with  the Wilson quark action and 
 the previous results with  the staggered quark action.
 The spatial string tension is also studied in the high temperature phase
  and is compared to the next-to-next-leading order 
  prediction in an effective theory with dimensional
  reduction.
\end{abstract}


\maketitle

%%%%%%%%%%%%%%%%%%%%%%%%%%%%%%%%%%%%%%%%%%%%%%%%%%%%%%%%%%%%%%%%%%%%%%
\section{Introduction}
\label{sec:intro}

Recent relativistic heavy-ion experiments
have revealed various remarkable properties of QCD
at finite temperatures and densities, suggesting the realization
of the QCD phase transition from the hadronic matter to the quark-gluon 
plasma (QGP) \cite{YHM}.
In order to extract unambiguous signals for the transition
from the heavy-ion experiments, it is indispensable
to make quantitative calculation of the thermal properties of QGP  from 
first principles.
Currently, the lattice QCD simulation is the only systematic method to 
do so.
By now, most of the lattice QCD studies at finite temperature
and chemical potential have been performed using staggered quark actions
with the fourth-root trick of the quark determinant,
which require less computational costs than others.
However the lattice artifacts of the staggered quark actions
 are not fully understood.
Therefore, it is important to compare the results from other lattice quarks
such as Wilson quark actions to control and estimate the lattice 
discretization errors.

Such a study at finite temperature $(T \neq 0)$ and 
 zero chemical potential $(\mu_q =0)$ 
 has been initiated several years ago
using the Iwasaki (RG) improved gauge action and the $N_f=2$ clover 
improved Wilson quark action by the CP-PACS Collaboration \cite{cp1,cp2}.
The phase structure, the transition temperature and the equation of
state have been investigated in detail, and also the crossover scaling
around the chiral phase transition has been tested.
 In contrast to the case of staggered quark actions,
 the subtracted chiral condensate
 in the standard Wilson quark action \cite{Iwasaki}
 and in the clover-improved Wilson quark action \cite{cp1}
 shows the scaling behavior with the critical exponents and scaling function of
 the three-dimensional O(4) spin model. This suggests that the 
 lattice QCD with Wilson-type quarks
 is  in the same universality class as the O(4) spin model, as expected  
 from the effective sigma model analysis \cite{Pisarski,Rajagopal}.
Moreover, extensive calculations of various physical quantities at $T=0$ such as 
the light hadron masses have been carried out using the same action \cite{cp3,cp4}.

Since a lot of experimental results are obtained by the heavy-ion 
collisions and numbers of technical progresses in treating system at
finite baryon density on the lattice 
have been made after the studies by the CP-PACS Collaboration,
it is worth while to revisit the QCD thermodynamics with
Wilson quark actions. In particular, it is 
 essential to perform simulations along the lines of constant
physics (LCP) to clearly extract the temperature- and density-dependences.
As a first step in this direction, we carry out simulations
of $N_f=2$ QCD on an $N_s^3 \times N_t=16^3 \times 4$ lattice
at $m_{\rm PS}/m_{\rm V}=0.65$ and 0.80 in the range
$T/T_{pc}\sim 0.76$--4.0,
where 
$m_{\rm PS}$ $(m_{\rm V})$ is the pseudo-scalar (vector) meson mass and
$T_{pc}$ is the pseudo-critical point along the LCP.
 Among various topics which can be studied using the above configurations,
 we will focus on two subjects in this paper:
 the free energy between heavy quarks and the spatial string tension.
 They are the fundamental quantities
 to characterize the perturbative and non-perturbative
 properties of the hot QCD medium.

 The heavy-quark free energy
 was recently studied by
  lattice simulations
   in the quenched approximation \cite{Kaczmarek:1999mm,Nakamura1,Nakamura2},
   and in full QCD  with the staggered quark action 
   \cite{Kaczmarek:2005ui,Doring}
   and with the Wilson quark action \cite{Bornyakov:2004ii}.
   An analytic study based on the thermal
   perturbation theory was also reported \cite{Laine:2006ns}.
   It was pointed out that 
   the interaction between heavy quarks is
   intimately related to the fate of charmoniums and bottomoniums
   in QGP created in relativistic heavy-ion collisions \cite{Matsui-Satz}.
   In this paper, we make a systematic study of
    the free energy between a quark ($Q$) and an antiquark ($\overline{Q}$)
   in the color singlet and octet channels,
   and between $Q$ and $Q$ in the color anti-triplet
   and sextet channels.  We adopt the Coulomb gauge fixing for the 
  gauge-non-singlet free energies.
  By fitting the numerical results with a screened Coulomb form,
  we extract an effective running coupling and
  the Debye screening mass in each channel as functions of temperature.
We also study the ``force'' between heavy quarks defined through the 
  derivative of the free energy with respect to the inter-quark distance.
  We show that
(i) the free energies in the different channels at high temperature
($T \simge 2 T_{pc}$)
can be well described by channel-dependent
Casimir factors together with the channel-independent
running coupling $g_{\rm eff}(T)$ and Debye mass $m_D(T)$,
(ii) the next-to-leading order result of the Debye mass in
thermal perturbation theory agrees better 
with the lattice $m_D(T)$ data than that of the leading order,
(iii) $g_{\rm eff}(T)$ and $m_D(T)$ satisfy the
leading order  formula, $m_D(T)= \sqrt{1+N_f/6} \, g_{\rm eff} (T) \, T$ 
for $T> 1.5T_{pc}$,
so that most higher order and non-perturbative
effects on the Debye mass is likely to be absorbed in $g_{\rm eff}(T)$, and
(iv) there is a quantitative discrepancy in $m_D(T)$
between our results using the Wilson quark action and those using
the staggered quark action \cite{Kaczmarek:2005ui,Doring}
 even at $T\sim 4 T_{pc}$, 
and (v) the Casimir scaling law valid above $T_{pc}$ is
 violated below $T_{pc}$, in particular in the color octet channel. 

We also extract the spatial string tension $\sigma_s$ at $T>T_{pc}$ from the spatial Wilson loop. 
We  show that  $\sqrt{\sigma_s}$ has approximate linear increase with $T$, 
as previously reported in quenched studies \cite{bali,karr,karsch,boyd}. 
We find that our result for $\sigma_s$ agree quantitatively well with the analytic result  based on a dimensionally reduced effective theory \cite{Laine:2005ai}.

This paper is organized as follows:
In Section \ref{sec:2},
 we present our lattice action and simulation parameters,
and discuss the line of constant physics.
Results of numerical simulations for the heavy quark free energies
are shown in Section \ref{sec:3}.
The effective running coupling and Debye screening mass
are extracted from the heavy quark free energies and are
compared with the analytic results of thermal perturbation theory 
 and with the numerical results with staggered quark action.
The force between heavy quarks are also discussed in this section.
In section \ref{sec:4}, we show the spatial string tension obtained from
 the spatial Wilson loop and its comparison to analytic results.
The paper is concluded in Section \ref{sec:5}.
We discuss the fit range dependence of the free energies
 in Appendix \ref{sec:apa},
and tabulate numerical data of the free energies in Appendix \ref{sec:apb}.

%%%%%%%%%%%%%%%%%%%%%%%%%%%%%%%%%%%%%%%%%%%%%%%%%%%%%%%%%%%%%%%%%%%%%%%%%%
\section{Simulations with a Wilson-type quark action}
\label{sec:2}

\subsection{Lattice action}
\label{sec:action}

We employ the RG-improved gauge action \cite{rg} and the
$N_f=2$ clover-improved Wilson quark action \cite{cl} defined by
\begin{eqnarray}
  S   &=& S_g + S_q, \\
  S_g &=& 
  -{\beta}\sum_x\left(
   c_0\sum_{\mu<\nu;\mu,\nu=1}^{4}W_{\mu\nu}^{1\times1}(x) 
   +c_1\sum_{\mu\ne\nu;\mu,\nu=1}^{4}W_{\mu\nu}^{1\times2}(x)\right), \\
  S_q &=& \sum_{f=1,2}\sum_{x,y}\bar{q}_x^f D_{x,y}q_y^f,
\end{eqnarray}
where $\beta=6/g^2$, $c_1=-0.331$, $c_0=1-8c_1$ and
\begin{eqnarray}
 D_{x,y} &=& \delta_{xy}
   -{K}\sum_{\mu}\{(1-\gamma_{\mu})U_{x,\mu}\delta_{x+\hat{\mu},y}
    +(1+\gamma_{\mu})U_{x,\mu}^{\dagger}\delta_{x,y+\hat{\mu}}\}
   -\delta_{xy}{c_{SW}}{K}\sum_{\mu<\nu}\sigma_{\mu\nu}F_{\mu\nu}.
\end{eqnarray}
Here $K$ is the hopping parameter and 
$F_{\mu\nu}$ is the lattice field strength,
$F_{\mu\nu} = {1}/{8i}(f_{\mu\nu}-f^{\dagger}_{\mu\nu})$,
 with $f_{\mu\nu}$ the standard clover-shaped combination of gauge links.
For the clover coefficient $c_{SW}$, we adopt a mean field value using
$W^{1\times 1}$ which was calculated in the one-loop perturbation theory
\cite{rg}, 
\begin{eqnarray}
 {c_{SW}}=(W^{1\times 1})^{-3/4}=(1-0.8412\beta^{-1})^{-3/4}.
\end{eqnarray}

\begin{figure}[tbp]
  \begin{center}
    \begin{tabular}{c}
    \includegraphics[width=70mm]{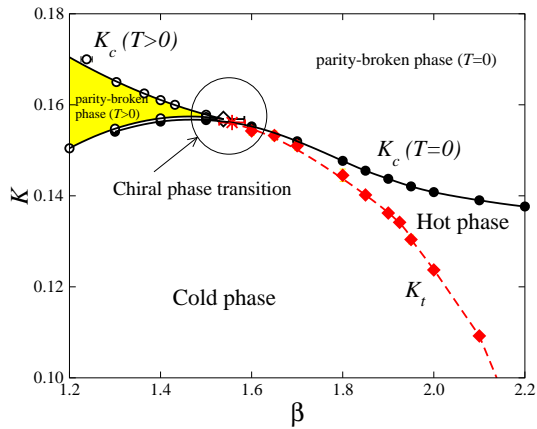}
    \end{tabular}
    \caption{Phase diagram for RG improved gauge action and clover
   improved Wilson quark action for $N_t=4$.}
    \label{fig1}
  \end{center}
\end{figure}

The phase diagram of this action in the $(\beta,K)$ plane has been
obtained by the CP-PACS 
Collaboration \cite{cp1,cp2} as shown in Fig.~\ref{fig1}. 
The solid line $K_c(T=0)$
with filled circles is the chiral limit where pseudo-scalar
mass vanishes at zero temperature. Above the $K_c(T=0)$ line, the parity-flavor
symmetry is spontaneously broken \cite{Aoki:1983qi,Aoki:1986xr,Aoki:1987us}.
 At finite temperatures, the cusp of the
parity-broken phase retracts from the large $\beta$ limit to a finite
$\beta$ \cite{Aoki:1995yf,Aoki:1996pw}. 
The solid line $K_c(T>0)$ connecting open symbols
represents the boundary of the parity-broken phase.
The region below $K_c$ corresponds to the two-flavor QCD with finite quark mass.
We perform simulations in this region after investigating the relation 
between the simulation parameters $(\beta, K)$ and the physical parameters, 
e.g. quark mass and lattice spacing.

 The dashed line $K_t$ with filled diamond
 represents the finite temperature pseudo-critical line determined
from the peak of Polyakov loop susceptibility. This line separates the hot
phase (the quark-gluon plasma phase) and the cold phase (the hadron phase).
The crossing of the $K_t$ and the $K_c(T=0)$ lines is the chiral
phase transition point.

%%%%%%%%%%%%%%%%%%%%%%%%%%%%%%%%%%%%%%%%%%%%%%%%%%%%%%%%%%%%%%%%%%%%%%
\subsection{Determination of lines of constant physics and 
simulation parameters}

\begin{figure}[tbp]
  \begin{center}
    \begin{tabular}{cc}
    \includegraphics[width=70mm]{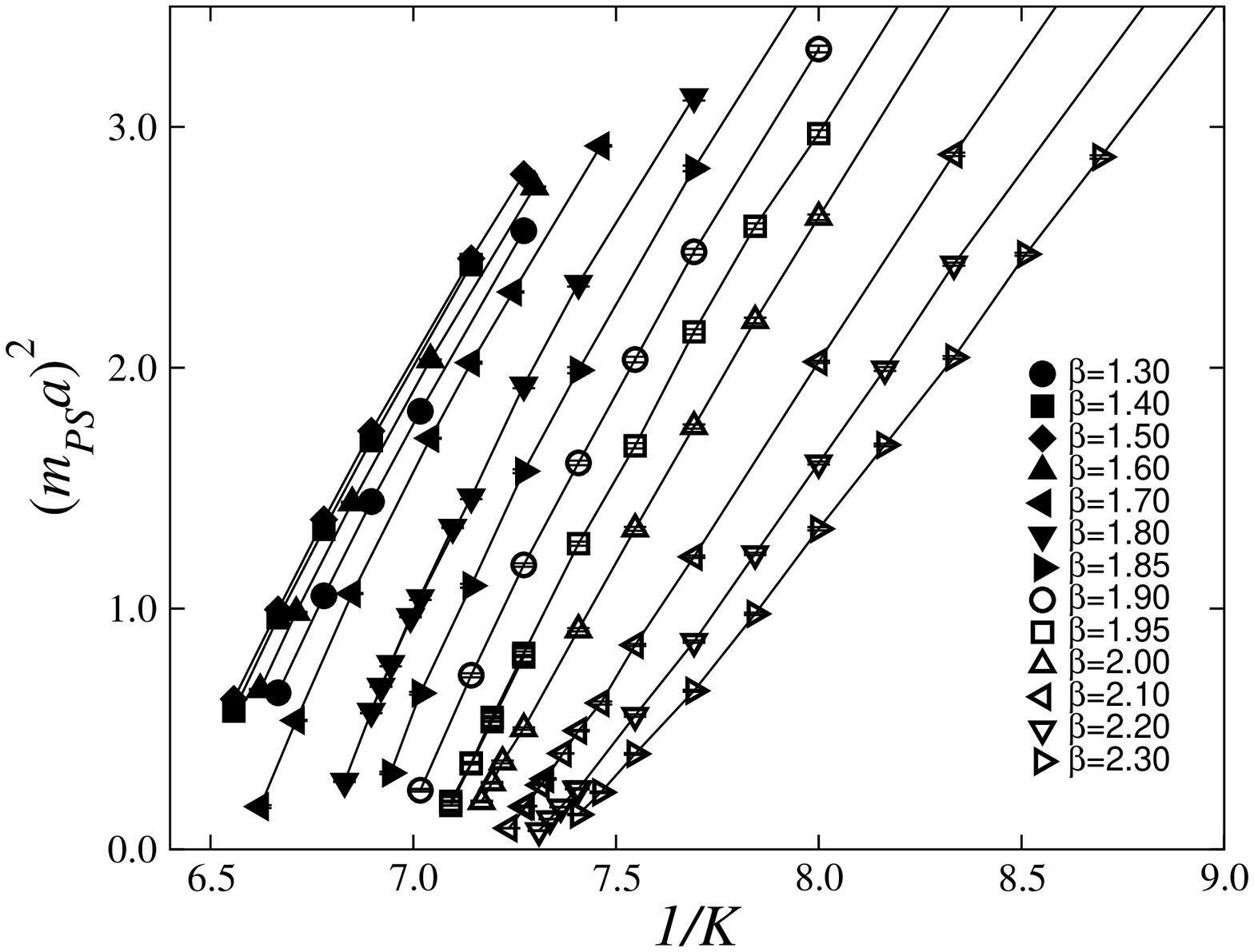} &
    \includegraphics[width=70mm]{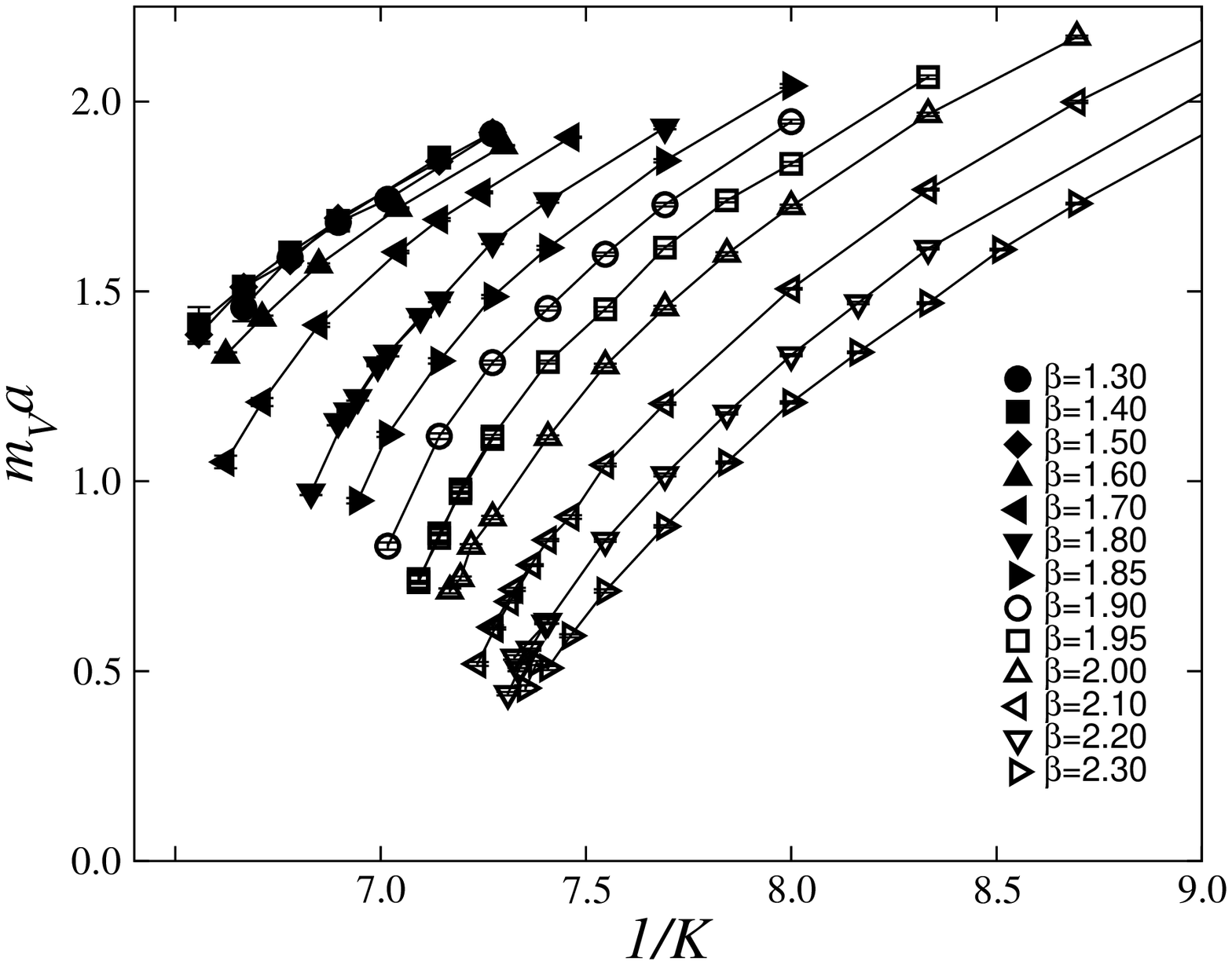}
    \end{tabular}
    \caption{Pseudo-scalar meson mass squared (left) and vector meson
   mass (right) as a function of $1/K$ for several values of $\beta$ at $T=0$.}
    \label{fig2}
  \end{center}
\end{figure}

\begin{figure}[tbp]
  \begin{center}
    \begin{tabular}{c}
    \includegraphics[width=80mm]{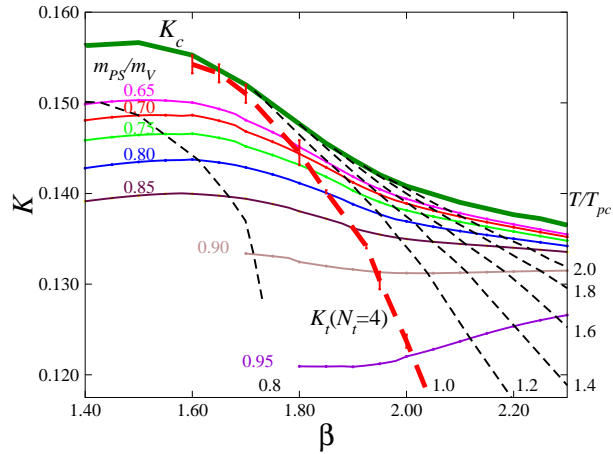}
    \end{tabular}
    \caption{
    Lines of constant $m_{\rm PS}/m_{\rm V}$ (solid lines)
    in the ($\beta,K$) plane for $m_{\rm PS}/m_{\rm V} = 0.65$,
     0.70, 0.75, 0.80, 0.85, 0.90 and 0.95.
    Dashed lines represent lines of constant $T/T_{pc}$ on $N_t=4$ lattices, 
    where $T_{pc}$ is the pseudo-critical temperature corresponding to
    $K_t(N_t=4)$ shown by the thick solid line. }
    \label{fig3}
  \end{center}
\end{figure}

For phenomenological applications, we need to investigate the
temperature dependence of thermodynamic observables on a line of
constant physics (LCP), which we determine by $m_{\rm PS}/m_{\rm V}$ 
(the ratio of pseudo-scalar and vector meson masses at $T=0$). 
For our purpose, we need LCP in a wider range of parameters 
than that studied in Ref.~\cite{cp2}. 
Therefore, we re-analyze the data for $m_{\rm PS}a$ and $m_{\rm V}a$ at
zero temperature in a wider range shown 
in Fig.~\ref{fig2}, where $a$ is the lattice spacing \cite{cp1,cp2,cp3,cp4}. 
The thin solid lines in Fig.~\ref{fig3} shows our results for LCP
corresponding to $m_{\rm PS}/m_{\rm V}=0.65$, 0.70, 0.75, 0.80, 0.85,
0.90 and 0.95. 
The bold solid line denoted as $K_c$ represents the
critical line, i.e. $m_{\rm PS}/m_{\rm V}=0$.
Our LCP's are consistent with those of \cite{cp2} in the range of
the previous study.

We also re-analyze the lines of constant $T/T_{pc}$.
The temperature $T$ is estimated by the zero-temperature vector meson
mass $m_{\rm V}a(\beta,K)$ using
\begin{eqnarray}
 \frac{T}{m_{\rm V}}(\beta,K)=\frac{1}{N_t \times m_{\rm V}a(\beta,K)}.
\end{eqnarray}
The lines of constant $T/T_{pc}$ is determined by the ratio of
${T}/{m_{\rm V}}$ to ${T_{pc}}/{m_{\rm V}}$ where
$T_{pc}/m_{\rm V}$ is obtained by $T/m_{\rm V}$ at $K_t$ on the same LCP.
We use an interpolation function,  
$ T_{pc}/m_{\rm V} = A (1+B( m_{\rm PS}/ m_{\rm V})^2)/(1+C(m_{\rm PS}/m_{\rm V})^2) $
with $A=0.2253(71)$, $B=-0.933(17)$ and $C=-0.820(39)$,
obtained in Ref.~\cite{cp2} to evaluate $T_{pc}/m_{\rm V}$ 
for each $m_{\rm PS}/m_{\rm V}$.
The bold dashed line denoted as $K_t(N_t=4)$ in Fig.~\ref{fig3} represents the
pseudo-critical line $T/T_{pc}=1$.
The thin dashed lines represent the results for 
$T/T_{pc}=0.8$, 1.2, 1.4, 1.6, 1.8, 2.0 at $N_t=4$.

We perform finite temperature simulations on a lattice with a
temporal extent $N_t=4$ and a spatial extent $N_s=16$ along the LCP's for 
$m_{\rm PS}/m_{\rm V}=0.65$ and 0.80. 
The standard hybrid Monte Carlo algorithm is employed to generate full QCD
configurations with two flavors of dynamical quarks. The length of one
trajectory is unity and the step size of 
 the molecular dynamics is tuned to
achieve an acceptance rate greater than 70\%.
Runs are carried out in the range $\beta=1.50$--2.40 at thirteen values of
$T/T_{pc}\sim 0.82$--4.0 for $m_{\rm PS}/m_{\rm V}=0.65$ and twelve values of
$T/T_{pc}\sim 0.76$--3.0 for $m_{\rm PS}/m_{\rm V}=0.80$.
Our simulation
parameters and the corresponding temperatures are summarized 
in Tab.~\ref{tab:parameter}.
Because the determination of the pseudo critical line is 
more difficult than the calculation of $T/m_{\rm V}$, the dominant source  
for the error of $T/T_{pc}$ in Table~\ref{tab:parameter}
 is the overall factor $T_{pc}/m_{\rm V}$.
The number of trajectories for each run after thermalization is
5000--6000. We measure physical quantities at every 10 trajectories.

\begin{table}[tbp]
 \begin{center}
 \caption{Simulation parameters for $m_{\rm PS}/m_{\rm V}=0.65$ (left) 
 and $m_{\rm PS}/m_{\rm V}=0.80$ (right).}
 \label{tab:parameter}
 {\renewcommand{\arraystretch}{1.2} \tabcolsep = 3mm
 \newcolumntype{a}{D{.}{.}{2}}
 \newcolumntype{b}{D{.}{.}{6}}
 \newcolumntype{d}{D{.}{.}{0}}
 \begin{tabular}{|abbd|c|abbd|}
 \cline{1-4} \cline{6-9}
 \multicolumn{1}{|c}{$\beta$} &
 \multicolumn{1}{c} {$K$}    & 
 \multicolumn{1}{c} {$T/T_{pc}$} & 
 \multicolumn{1}{c|}{Traj.} & 
 \multicolumn{1}{c} {} & 
 \multicolumn{1}{|c}{$\beta$} &
 \multicolumn{1}{c} {$K$}    & 
 \multicolumn{1}{c} {$T/T_{pc}$} & 
 \multicolumn{1}{c|}{Traj.} \\
 \cline{1-4} \cline{6-9}
 1.50 & 0.150290 & 0.82(3)  & 5000 & & 1.50 & 0.143480 & 0.76(4)  & 5500 \\
 1.60 & 0.150030 & 0.86(3)  & 5000 & & 1.60 & 0.143749 & 0.80(4)  & 6000 \\
 1.70 & 0.148086 & 0.94(3)  & 5000 & & 1.70 & 0.142871 & 0.84(4)  & 6000 \\
 1.75 & 0.146763 & 1.00(4)  & 5000 & & 1.80 & 0.141139 & 0.93(5)  & 6000 \\
 1.80 & 0.145127 & 1.07(4)  & 5000 & & 1.85 & 0.140070 & 0.99(5)  & 6000 \\
 1.85 & 0.143502 & 1.18(4)  & 5000 & & 1.90 & 0.138817 & 1.08(5)  & 6000 \\
 1.90 & 0.141849 & 1.32(5)  & 5000 & & 1.95 & 0.137716 & 1.20(6)  & 6000 \\
 1.95 & 0.140472 & 1.48(5)  & 5000 & & 2.00 & 0.136931 & 1.35(7)  & 5000 \\
 2.00 & 0.139411 & 1.67(6)  & 5000 & & 2.10 & 0.135860 & 1.69(8)  & 5000 \\
 2.10 & 0.137833 & 2.09(7)  & 5000 & & 2.20 & 0.135010 & 2.07(10) & 5000 \\
 2.20 & 0.136596 & 2.59(9)  & 5000 & & 2.30 & 0.134194 & 2.51(13) & 5000 \\
 2.30 & 0.135492 & 3.22(12) & 5000 & & 2.40 & 0.133395 & 3.01(15) & 5000 \\
 2.40 & 0.134453 & 4.02(15) & 5000 & &      &          &          &      \\
 \cline{1-4} \cline{6-9}
 \end{tabular}}
 \end{center}
\end{table}

%%%%%%%%%%%%%%%%%%%%%%%%%%%%%%%%%%%%%%%%%%%%%%%%%%%%%%%%%%%%%%%%%%%%%%
\section{Heavy quark free energies}
\label{sec:3}

A free energy of static quarks on the lattice
is described by the correlations of the  Polyakov loop:
$\Omega ( {\bf x} )  = \prod_{ \tau = 1}^{N_t} U_4 (\tau, {\bf x})$
 where the $U_\mu (\tau, {\bf x}) \in$ SU(3) is the link variable.
 With an appropriate gauge fixing,
one can define the free energy in various 
 color channels separately
\cite{Nadkarni1,Nadkarni2}:
 the color singlet $Q\overline{Q}$ channel ({\bf 1}),
 the color octet $Q\overline{Q}$ channel ({\bf 8}),
 the color anti-triplet $QQ$ channel (${\bf 3}^*$),
 and the color sextet $QQ$ channel ({\bf 6}),
 given as follows.
\begin{eqnarray}
e^{-F_{\bf 1}(r,T)/T}
 &=&
  \frac{1}{3} \langle {\rm Tr} 
\Omega^\dagger({\bf x}) \Omega ({\bf y})
\rangle
, 
\label{eq:singlet}
\\
e^{-F_{\bf 8}(r,T)/T}
 &=& 
\frac{1}{8} \langle {\rm Tr} \Omega^\dagger({\bf x})
{\rm Tr} \Omega ({\bf y})
\rangle
-
\frac{1}{24} \langle {\rm Tr} \Omega^\dagger({\bf x})
 \Omega ({\bf y}) \rangle
, \\
e^{-F_{\bf 6}(r,T)/T}
 &=& 
\frac{1}{12} \langle {\rm Tr} \Omega({\bf x})
{\rm Tr} \Omega ({\bf y})
\rangle
+
\frac{1}{12} \langle {\rm Tr} \Omega({\bf x})
 \Omega ({\bf y}) \rangle
, \\
e^{-F_{{\bf 3}^*}(r,T)/T}
 &=& 
\frac{1}{6} \langle {\rm Tr} \Omega({\bf x})
{\rm Tr} \Omega ({\bf y})
\rangle
-
\frac{1}{6} \langle {\rm Tr} \Omega({\bf x})
 \Omega ({\bf y}) \rangle
,
\label{eq:sextet}
\end{eqnarray}
where $r = |{\bf x} - {\bf y}|$.
We adopt the Coulomb gauge fixing in this study.

Above $T_{pc}$, we introduce normalized free energies 
$(V_{\bf 1}, V_{\bf 8}, V_{\bf 6}, V_{{\bf 3}^*})$
such that they vanish at large distances. 
 This is equivalent to defining the free energies by 
 dividing the right-hand sides of 
Eqs.~(\ref{eq:singlet})--(\ref{eq:sextet})
by $\langle {\rm Tr} \Omega \rangle^2$.

\begin{figure}[tbp]
  \begin{center}
    \begin{tabular}{cc}
    \includegraphics[width=80mm]{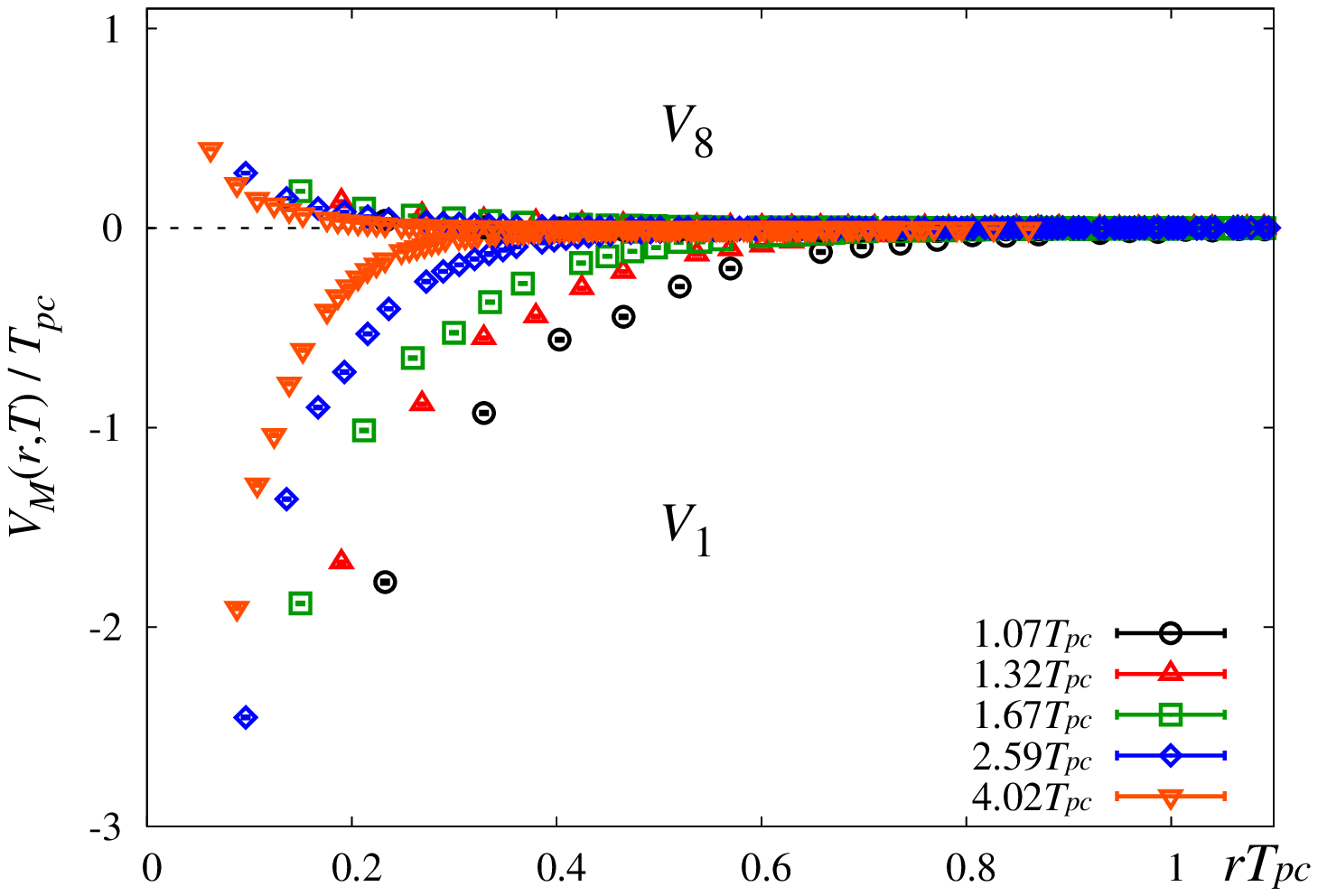} &
    \includegraphics[width=80mm]{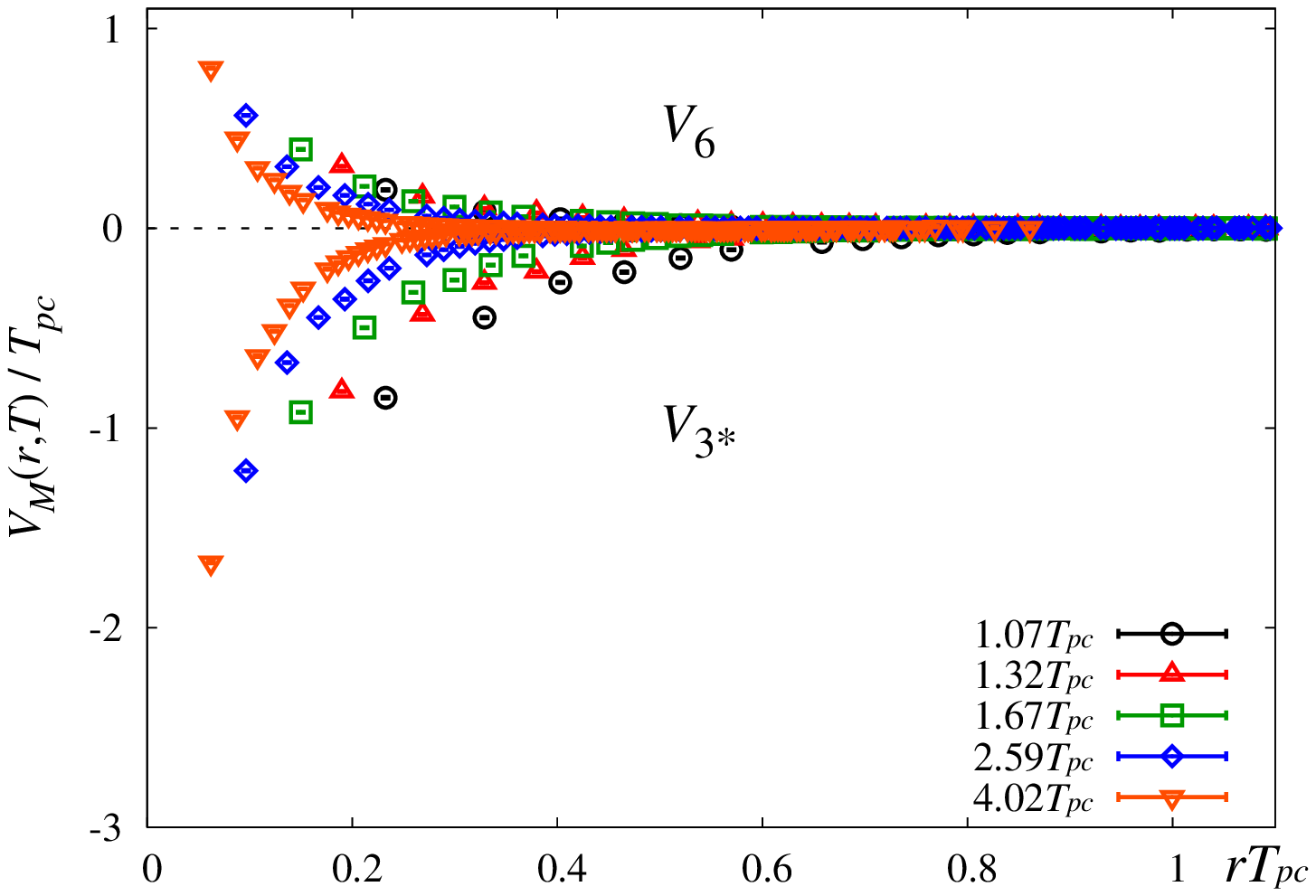}
    \end{tabular}
    \caption{
    Simulation results of the 
    normalized free energies scaled by $T_{pc}$
    for color singlet and octet $Q\overline{Q}$ channels (left)
    and color anti-triplet and sextet $QQ$ channels (right)
    at $m_{\rm PS}/m_{\rm V} = 0.65$.
        }
    \label{fig:NFE065}
  \end{center}
\end{figure}

The normalized free energies are shown in Fig.~\ref{fig:NFE065}
for color singlet and octet $Q\overline{Q}$ channels (left) and 
color anti-triplet and sextet $QQ$ channels (right)
 for  $m_{\rm PS}/m_{\rm V} = 0.65$ and $T \ge T_{pc}$.
These for $m_{\rm PS}/m_{\rm V} = 0.80$ are also shown 
in Fig.~\ref{fig:NFE080}.
Data of the normalized free energies for all temperatures above $T_{pc}$
are summarized in Appendix \ref{sec:apb}.
From these figures, we find that the inter-quark interaction is ``attractive''  
in the color singlet and anti-triplet channels 
and is ``repulsive''  in the color octet and sextet channels.
We also see that, irrespective of the channels, the inter-quark interaction becomes rapidly weak 
at long distances as $T$ increases, as expected from the Debye screening at high temperatures.
These behaviors are qualitatively similar to the case of quenched QCD in the Lorenz gauge 
as reported in Ref.~\cite{Nakamura1,Nakamura2}.

\begin{figure}[tbp]
  \begin{center}
    \begin{tabular}{cc}
    \includegraphics[width=80mm]{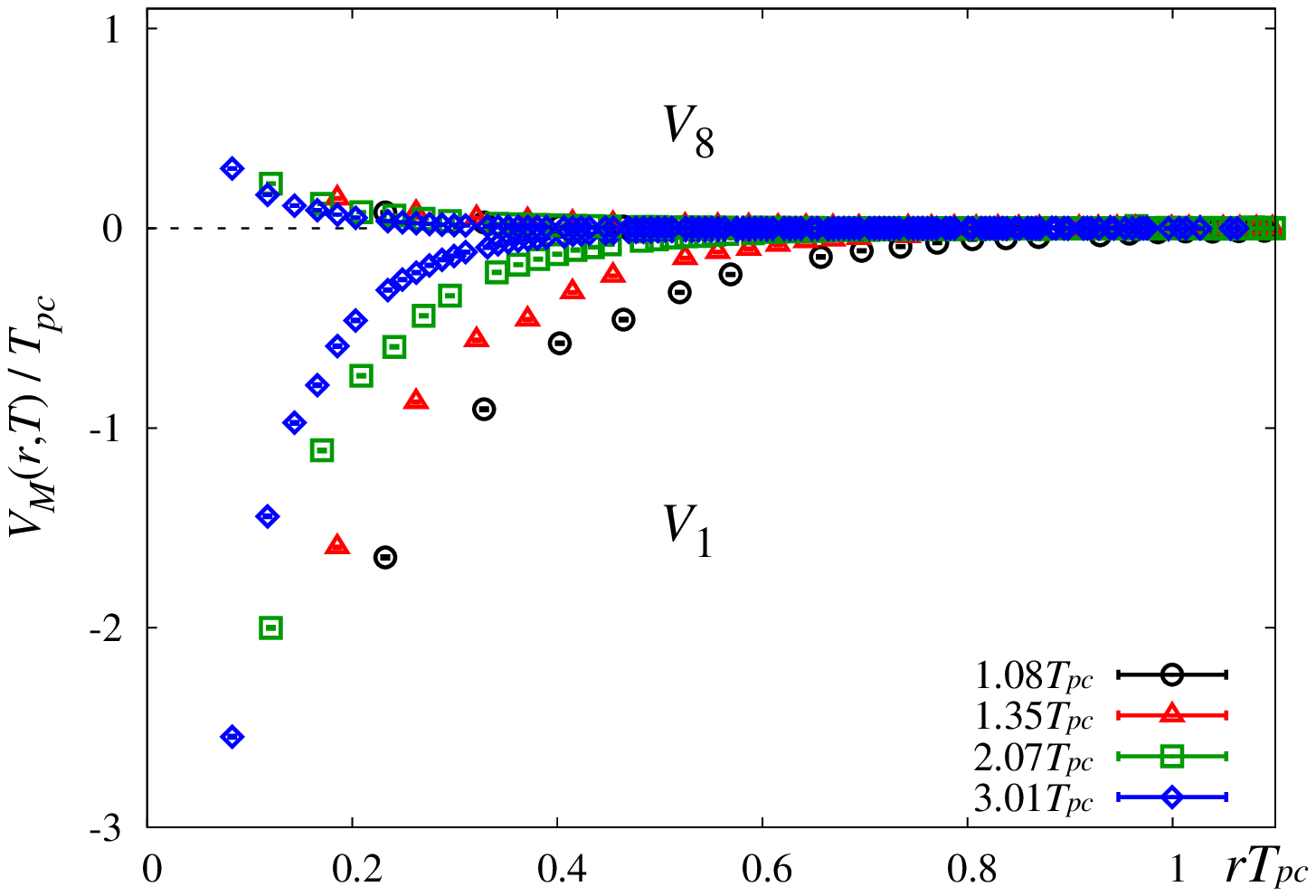} &
    \includegraphics[width=80mm]{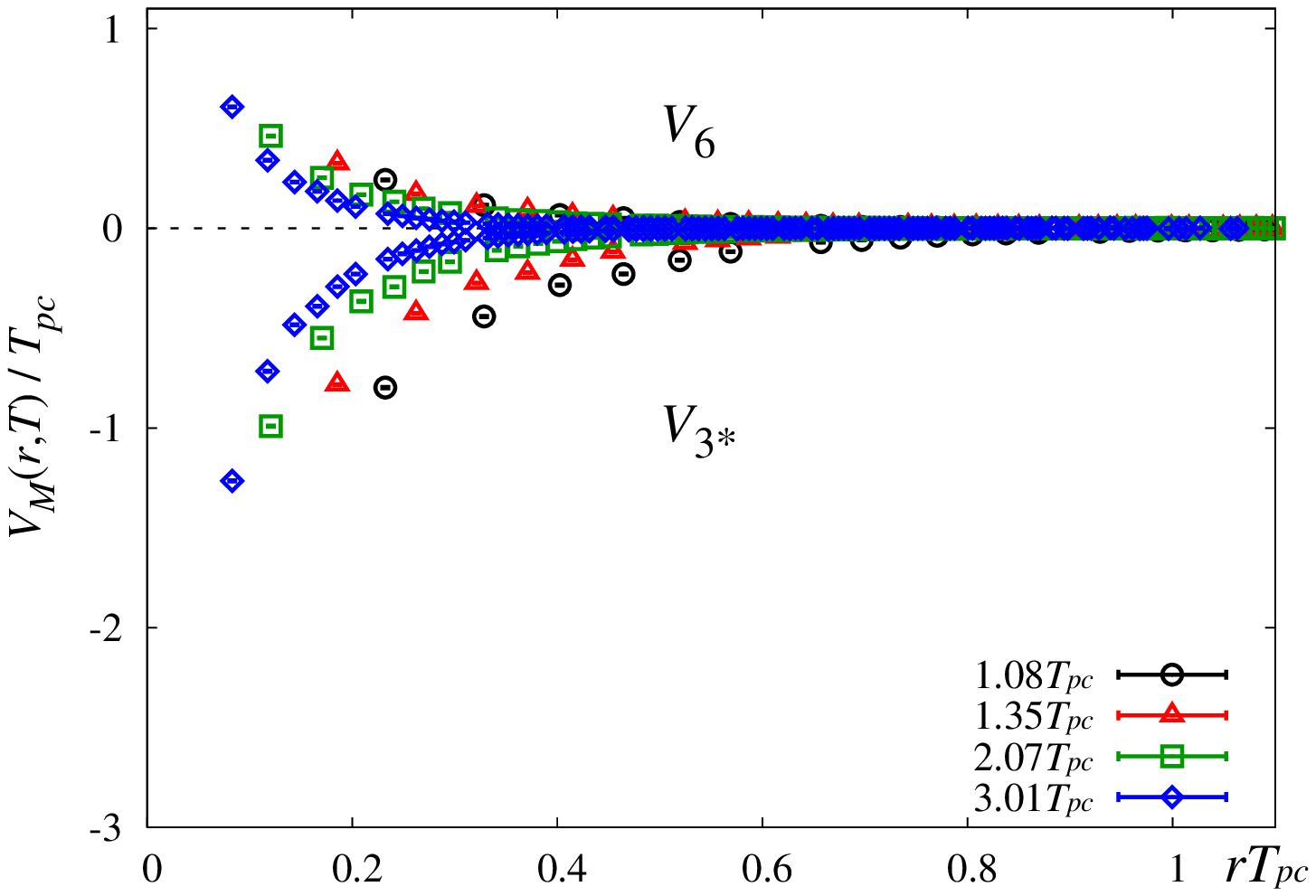}
    \end{tabular}
    \caption{
    The same figure with Fig.~\ref{fig:NFE065}
    at $m_{\rm PS}/m_{\rm V} = 0.80$.
        }
    \label{fig:NFE080}
  \end{center}
\end{figure}

To study the screening effects in each color channel more closely,
we fit the free energies by the screened Coulomb form:
\begin{equation}
V_M(r,T) = C(M) \frac{\alpha_{\rm eff}(T)}{r} e^{-m_D(T) r} ,
\label{eq:SCP}
\end{equation}
where $\alpha_{\rm eff}(T)$ and $m_D(T)$ 
are the effective running coupling and 
 Debye screening mass, respectively.
The Casimir factor $C(M) \equiv \langle \sum_{a=1}^{8} t_1^a\cdot t_2^a \rangle_M$
 for color channel $M$  is explicitly given  by 
\begin{eqnarray}
C({\bf 1})     = -\frac{4}{3}, \ \ \
C({\bf 8})     =  \frac{1}{6}, \ \ \
C({\bf 6})     =  \frac{1}{3}, \ \ \
C({\bf 3}^*)   = -\frac{2}{3},
\end{eqnarray}
for our cases.
Here, it is worth stressing that, with the improved actions we adopt,
the rotational symmetry is well restored in the heavy quark free
energies \cite{Aoki:1999ff}.
 Therefore,
 we do not need to introduce 
 terms correcting lattice artifacts at short distances 
 in Eq.~(\ref{eq:SCP}) to
fit the data shown in Fig.~\ref{fig:NFE065} and \ref{fig:NFE080}.

 The Debye screening effect is defined through
the long distance behavior 
of $V_M(r,T)$.
A discussion to determine an appropriate fit range is given
in Appendix \ref{sec:apa} and we choose it 
to be $ \sqrt{11}/4 \le rT \le 1.5$.
We also discuss the systematic errors due to a difference of 
the fit ranges in Appendix \ref{sec:apa}, and find that
the systematic errors are smaller than the statistical errors
at $T \simge 2 T_{pc}$.
The fit is performed by minimizing $\chi^2 / N_{DF}$,
where $N_{DF}=20$.
Results of the $\chi^2 / N_{DF}$ for each color channel and 
temperature are summarized in Tab.~\ref{tab:chi2}
for $m_{\rm PS}/m_{\rm V}=0.65$ (left) 
and $0.80$ (right).

\begin{table}[tbp]
 \begin{center}
 \caption{The $\chi^2 / N_{DF}$ for each color channel and 
temperature at $m_{\rm PS}/m_{\rm V}=0.65$ (left) 
and 0.80 (right).}
 \label{tab:chi2}
 {\renewcommand{\arraystretch}{1.2} \tabcolsep = 3mm
 \begin{tabular}{|c|cccc|c|c|cccc|}
 \cline{1-5} \cline{7-11}
 $T/T_{pc}$  & $M={\bf 1}$ & ${\bf 8}$ &
 ${\bf 6}$ & ${\bf 3}^*$ & \ \ \ \  &
 $T/T_{pc}$  & $M={\bf 1}$ & ${\bf 8}$ &
 ${\bf 6}$ & ${\bf 3}^*$  \\
 \cline{1-5} \cline{7-11}
 1.00 &  1.97 &  2.26 &  2.50 &  1.41 & &  1.08 &  1.25 &  1.14 &  $-$\footnote{
Fit is unstable since $V_{\bf 6}(r,T)$ at this parameter point is smaller than the statistical errors
 in the fit range.} &  0.82 \\
 1.07 &  2.06 &  0.99 &  1.04 &  1.59 & &  1.20 &  1.78 &  1.66 &  1.33 &  0.95 \\
 1.18 &  2.08 &  1.86 &  1.21 &  1.59 & &  1.35 &  1.32 &  0.76 &  2.10 &  0.81 \\
 1.32 &  1.19 &  4.02 &  1.58 &  0.85 & &  1.69 &  1.01 &  1.58 &  1.07 &  1.22 \\
 1.48 &  1.32 &  1.55 &  1.06 &  0.98 & &  2.07 &  4.07 &  1.51 &  2.28 &  2.80 \\
 1.67 &  2.96 &  1.65 &  1.98 &  2.64 & &  2.51 &  2.00 &  2.72 &  1.43 &  1.93 \\
 2.09 &  3.51 &  2.16 &  2.07 &  1.64 & &  3.01 &  2.13 &  2.60 &  1.56 &  2.08 \\
 2.59 &  1.49 &  0.98 &  1.25 &  3.55 & & & & & & \\
 3.22 &  2.05 &  1.66 &  2.14 &  1.83 & & & & & & \\
 4.02 &  2.75 &  1.51 &  1.37 &  2.73 & & & & & & \\
 \cline{1-5} \cline{7-11}
 \end{tabular}}
 \end{center}
\end{table}

\begin{figure}[tbp]
  \begin{center}
    \begin{tabular}{cc}
      \includegraphics[width=80mm]{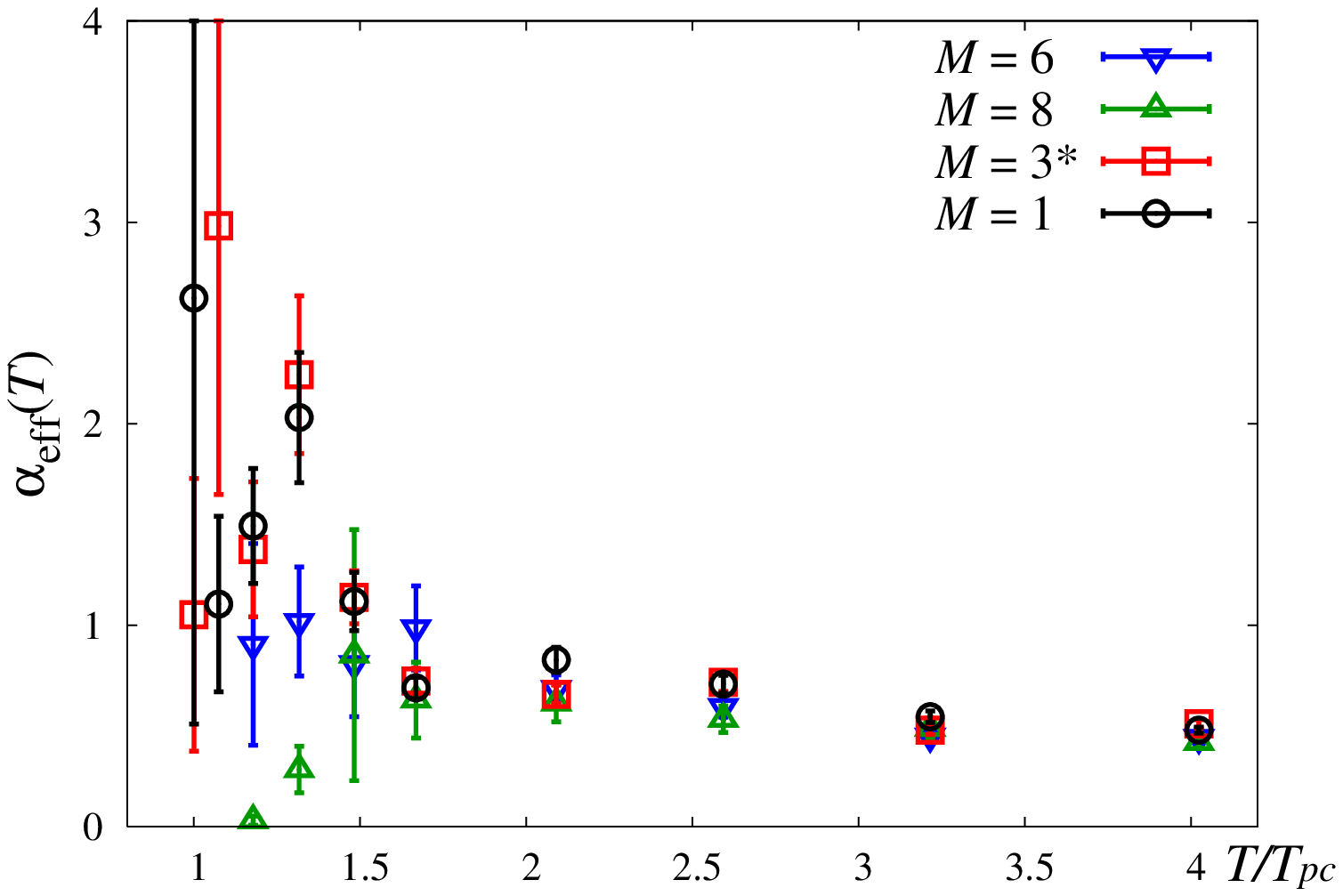} &
      \includegraphics[width=80mm]{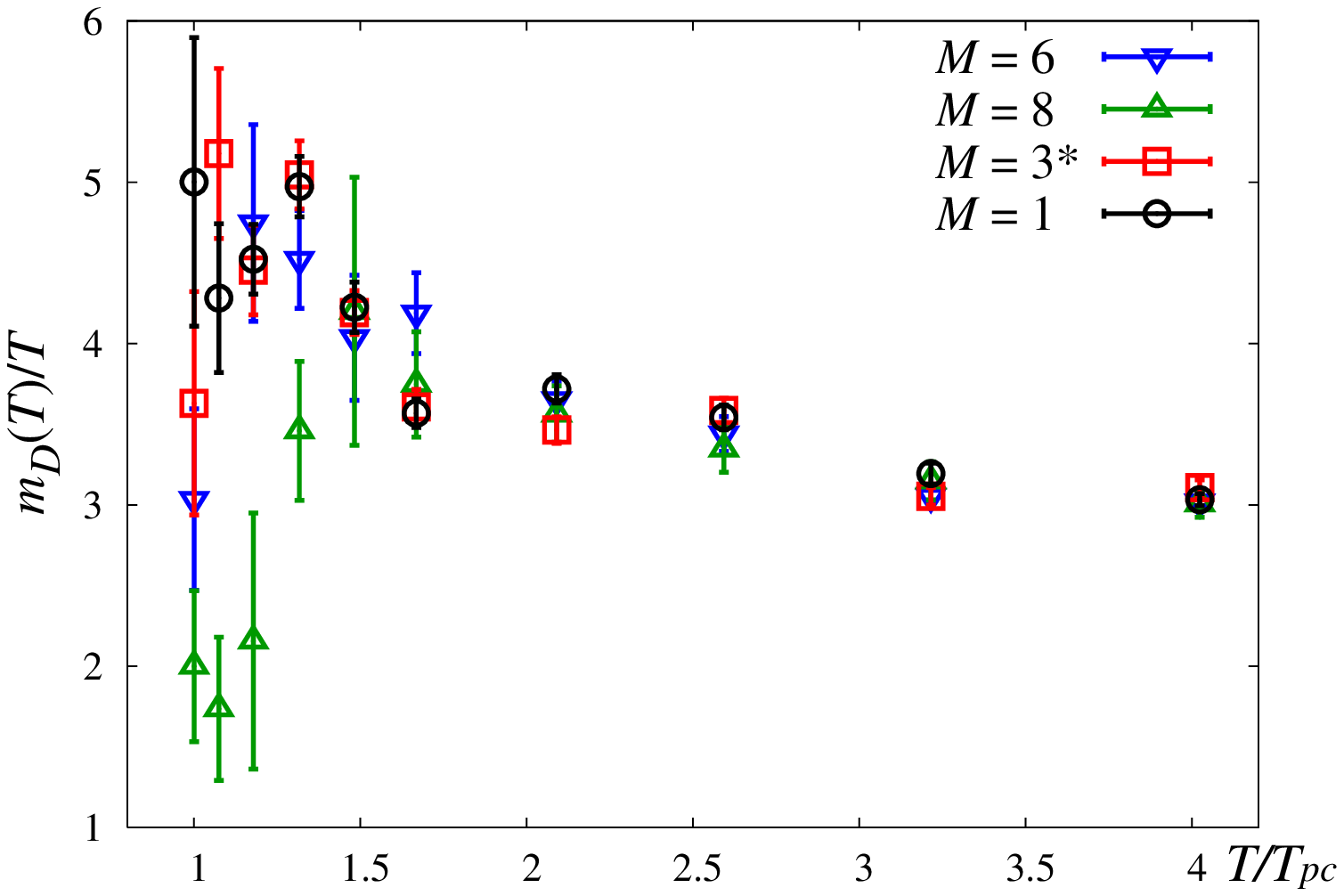} 
    \end{tabular}
    \caption{The effective running coupling $\alpha_{\rm eff}(T)$ (left) and
    Debye screening mass $m_D(T)$ (right) for each color channel
    as a function of temperature
    from the large distance behavior of the potentials
    at $m_{\rm PS}/m_{\rm V} = 0.65$.
    }
    \label{fig:para065}
  \end{center}
\end{figure}

\begin{figure}[tbp]
  \begin{center}
    \begin{tabular}{cc}
      \includegraphics[width=80mm]{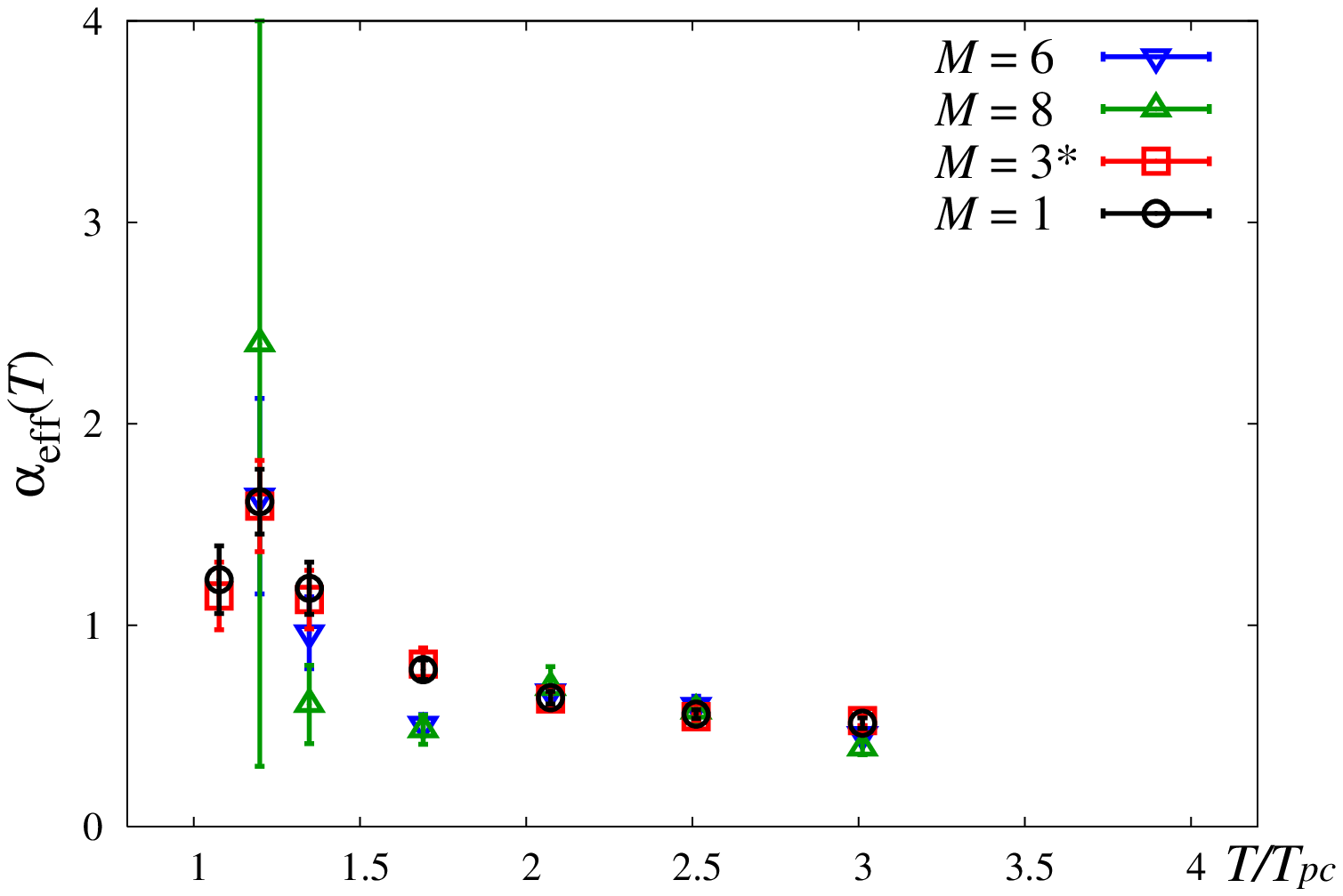} &
      \includegraphics[width=80mm]{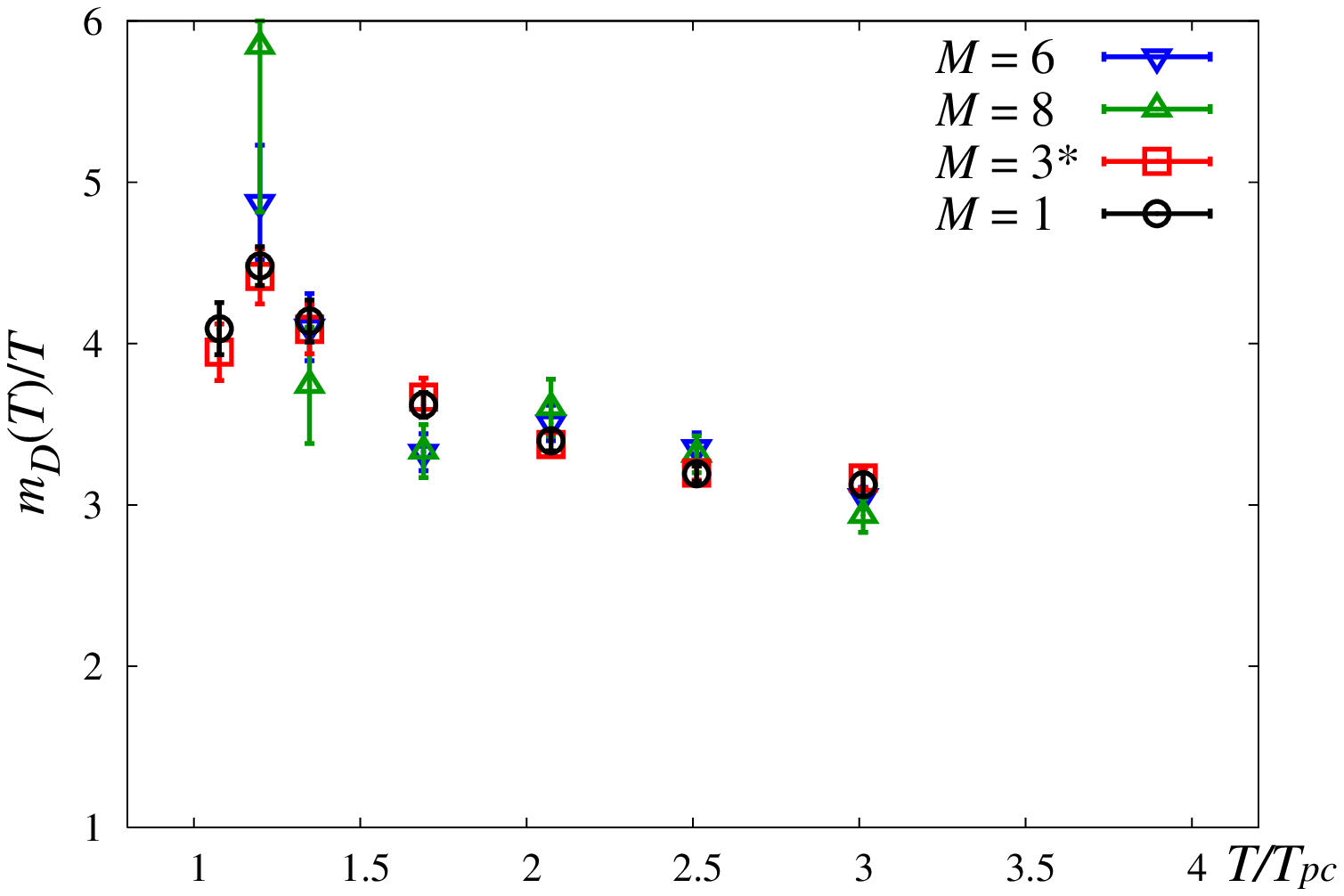} 
    \end{tabular}
    \caption{The same figures with Fig.~\ref{fig:para065}
    at $m_{\rm PS}/m_{\rm V} = 0.80$ .}
    \label{fig:para080}
  \end{center}
\end{figure}

The results of $\alpha_{\rm eff}(T)$ and $m_D(T)$ are shown
in Fig.~\ref{fig:para065} for $m_{\rm PS}/m_{\rm V} = 0.65$
and Fig.~\ref{fig:para080} for 0.80.
The numerical results are also summarized in Tab.~\ref{tab:am_065}
for $m_{\rm PS}/m_{\rm V} = 0.65$
and Tab.~\ref{tab:am_080} for 0.80.
We find that there is no significant channel dependence in  
 $\alpha_{\rm eff}(T)$ and $m_D(T)$
at sufficiently high temperatures $(T \simge 2T_{pc})$.
In other words, the channel dependence in the free energy
can be well absorbed in the kinematical Casimir factor at high temperatures, 
as first indicated in quenched studies \cite{Nakamura1,Nakamura2}.

\begin{table}[tbp]
 \begin{center}
 \caption{Results of $\alpha_{\rm eff} (T)$ and
 $m_D(T)$ at $m_{\rm PS}/m_{\rm V}=0.65$
 with statistical errors determined by a jackknife method with
 the bin-size of 100 trajectories.}
 \label{tab:am_065}
 {\renewcommand{\arraystretch}{1.2} \tabcolsep = 3mm
 \newcolumntype{.}{D{.}{.}{6}}
 \begin{tabular}{|c|....|....|}
 \hline
 \multicolumn{1}{|l|}{} &
 \multicolumn{4}{c|}{$\alpha_{\rm eff}(T)$} & 
 \multicolumn{4}{c|}{$m_D(T)$} \\
 \hline
 \multicolumn{1}{|c|}{$T/T_{pc}$} &
 \multicolumn{1}{c} {$M={\bf 1}$} & 
 \multicolumn{1}{c} {${\bf 8}$} & 
 \multicolumn{1}{c} {${\bf 6}$} & 
 \multicolumn{1}{c|}{${\bf 3}^*$} & 
 \multicolumn{1}{c} {${\bf 1}$} & 
 \multicolumn{1}{c} {${\bf 8}$} & 
 \multicolumn{1}{c} {${\bf 6}$} & 
 \multicolumn{1}{c|}{${\bf 3}^*$} \\
 \hline
 1.00 &  2.62(211) & -0.47( 21) & -0.60( 31) &  1.05( 67) &  5.00( 89) &  2.00( 46) &  3.03( 56) &  3.63( 69) \\
 1.07 &  1.11( 43) & -0.17(  6) & -0.01(  0) &  2.98(133) &  4.28( 46) &  1.74( 44) &  0.09( 40) &  5.18( 52) \\
 1.18 &  1.49( 28) &  0.03(  2) &  0.90( 50) &  1.38( 33) &  4.52( 21) &  2.16( 79) &  4.75( 60) &  4.46( 27) \\
 1.32 &  2.03( 32) &  0.28( 11) &  1.02( 27) &  2.24( 39) &  4.97( 18) &  3.46( 43) &  4.52( 30) &  5.05( 21) \\
 1.48 &  1.12( 14) &  0.85( 62) &  0.81( 26) &  1.14( 13) &  4.23( 15) &  4.20( 83) &  4.04( 38) &  4.19( 13) \\
 1.67 &  0.69(  5) &  0.63( 18) &  0.99( 20) &  0.72(  6) &  3.57(  8) &  3.75( 32) &  4.19( 25) &  3.61( 10) \\
 2.09 &  0.83(  6) &  0.61(  9) &  0.69(  6) &  0.66(  4) &  3.72(  8) &  3.56( 17) &  3.65( 11) &  3.46(  8) \\
 2.59 &  0.71(  4) &  0.53(  6) &  0.60(  5) &  0.72(  4) &  3.54(  7) &  3.35( 14) &  3.44( 10) &  3.59(  7) \\
 3.22 &  0.55(  2) &  0.49(  5) &  0.45(  2) &  0.48(  2) &  3.19(  6) &  3.15( 11) &  3.05(  6) &  3.05(  5) \\
 4.02 &  0.48(  1) &  0.42(  2) &  0.44(  1) &  0.51(  2) &  3.03(  3) &  3.01(  8) &  3.02(  5) &  3.11(  4) \\
 \hline
 \end{tabular}}
 \end{center}
\end{table}

\begin{table}[tbp]
 \begin{center}
 \caption{The same Table as Tab.~\ref{tab:am_065}
 at $m_{\rm PS}/m_{\rm V}=0.80$.}
\label{tab:am_080}
 {\renewcommand{\arraystretch}{1.2} \tabcolsep = 3mm
 \newcolumntype{.}{D{.}{.}{6}}
 \begin{tabular}{|c|....|....|}
 \hline
 \multicolumn{1}{|l|}{} &
 \multicolumn{4}{c|}{$\alpha_{\rm eff}(T)$} & 
 \multicolumn{4}{c|}{$m_D(T)$} \\
 \hline
 \multicolumn{1}{|c|}{$T/T_{pc}$} &
 \multicolumn{1}{c} {$M={\bf 1}$} & 
 \multicolumn{1}{c} {${\bf 8}$} & 
 \multicolumn{1}{c} {${\bf 6}$} & 
 \multicolumn{1}{c|}{${\bf 3}^*$} & 
 \multicolumn{1}{c} {${\bf 1}$} & 
 \multicolumn{1}{c} {${\bf 8}$} & 
 \multicolumn{1}{c} {${\bf 6}$} & 
 \multicolumn{1}{c|}{${\bf 3}^*$} \\
 \hline
 \multicolumn{1}{|c|}{1.08} &
 \multicolumn{1}{.}  { 1.23( 16)} &
 \multicolumn{1}{.}  {-0.03(  1)} &
 \multicolumn{1}{c}  {$-$} &
 \multicolumn{1}{.|}  { 1.15( 16)} &
 \multicolumn{1}{.}  { 4.09( 16)} &
 \multicolumn{1}{.}  { 0.63( 39)} &
 \multicolumn{1}{c}  {$-$} &
 \multicolumn{1}{.|}  { 3.95( 17)} \\
 1.20 &  1.61( 16) &  2.40(209) &  1.64( 48) &  1.59( 22) &  4.48( 12) &  5.84(102) &  4.87( 35) &  4.42( 17) \\
 1.35 &  1.18( 12) &  0.61( 19) &  0.96( 17) &  1.13( 14) &  4.14( 12) &  3.74( 36) &  4.10( 20) &  4.09( 15) \\
 1.69 &  0.78(  4) &  0.48(  7) &  0.51(  5) &  0.81(  8) &  3.62(  7) &  3.33( 16) &  3.33( 11) &  3.67( 11) \\
 2.07 &  0.64(  3) &  0.69( 10) &  0.67(  6) &  0.63(  3) &  3.40(  6) &  3.60( 17) &  3.51( 10) &  3.37(  6) \\
 2.51 &  0.56(  2) &  0.57(  5) &  0.60(  4) &  0.55(  1) &  3.19(  4) &  3.31( 11) &  3.36(  9) &  3.20(  4) \\
 3.01 &  0.51(  2) &  0.39(  3) &  0.46(  2) &  0.53(  2) &  3.13(  6) &  2.93( 10) &  3.05(  5) &  3.17(  6) \\
 \hline
 \end{tabular}}
 \end{center}
\end{table}

%%%%%%%%%%%%%%%%%%%%%%%%%%%%%%%%%%%%%%%%%%%%%%%%%%%%%%%%%%%%%%%%%%%%%%
\subsection{Debye mass on the lattice and that in perturbative theory}
\label{sec:PT}    

 Let us first compare the Debye mass on the lattice
  with that calculated in the thermal perturbation theory.
The  2-loop running coupling is given by
\begin{eqnarray}
g^{-2}_{2 {\rm l}} (\mu) 
=  \beta_0 \ln \left( \frac{\mu}{\Lambda} \right)^2 + 
\frac{\beta_1}{\beta_0} 
\ln \ln \left( \frac{\mu}{\Lambda} \right)^2
,
\label{eq:RC}
\end{eqnarray}
where  the argument  in the logarithms can be written as 
$ \mu / \Lambda = (\mu/T)(T/T_{pc})(T_{pc}/\Lambda)$
  where we adopt 
$\Lambda= \Lambda_{\overline{MS}}^{N_f=2} \simeq 261$ MeV \cite{Gockeler:2005rv} 
and $T_{pc} \simeq 171$ MeV \cite{cp1}.
We assume that the renormalization point $\mu$ is 
in the range $\mu = \pi T$ to $3 \pi T$.
Therefore, $g_{\rm 2l}$ can be viewed as a function of $T/T_{pc}$. 
With $g_{\rm 2l}$, the Debye screening mass  is given by
\begin{eqnarray}
\frac{m_D^{\rm LO}(T)}{T} = \sqrt{ 1 + \frac{N_f}{6} } \  g_{\rm 2l} (\mu) ,
\end{eqnarray} 
in the leading-order (LO) thermal perturbation theory, neglecting the effects of quark masses.

\begin{figure}[tbp]
  \begin{center}
    \begin{tabular}{cc}
      \includegraphics[width=80mm]{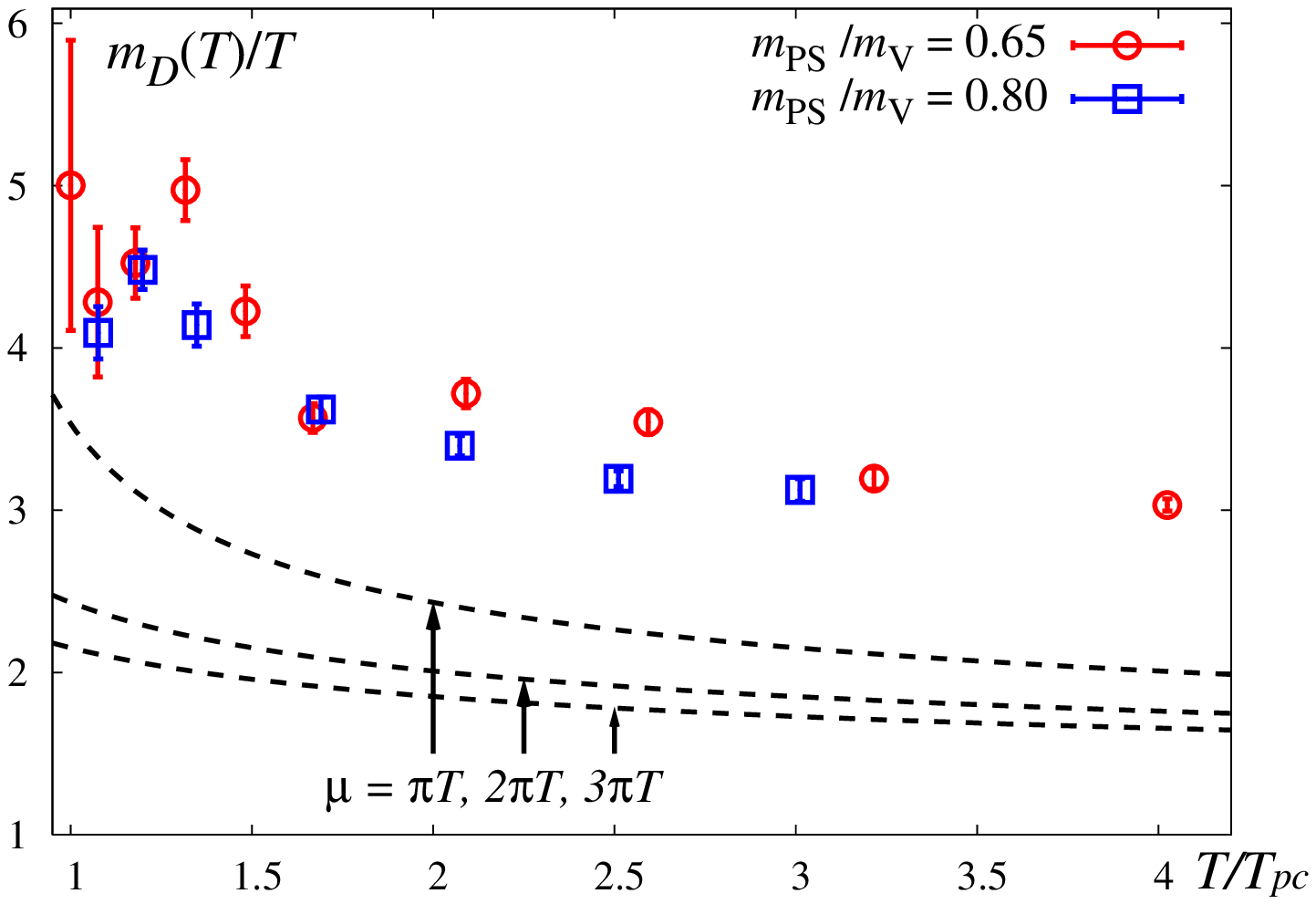} &
      \includegraphics[width=80mm]{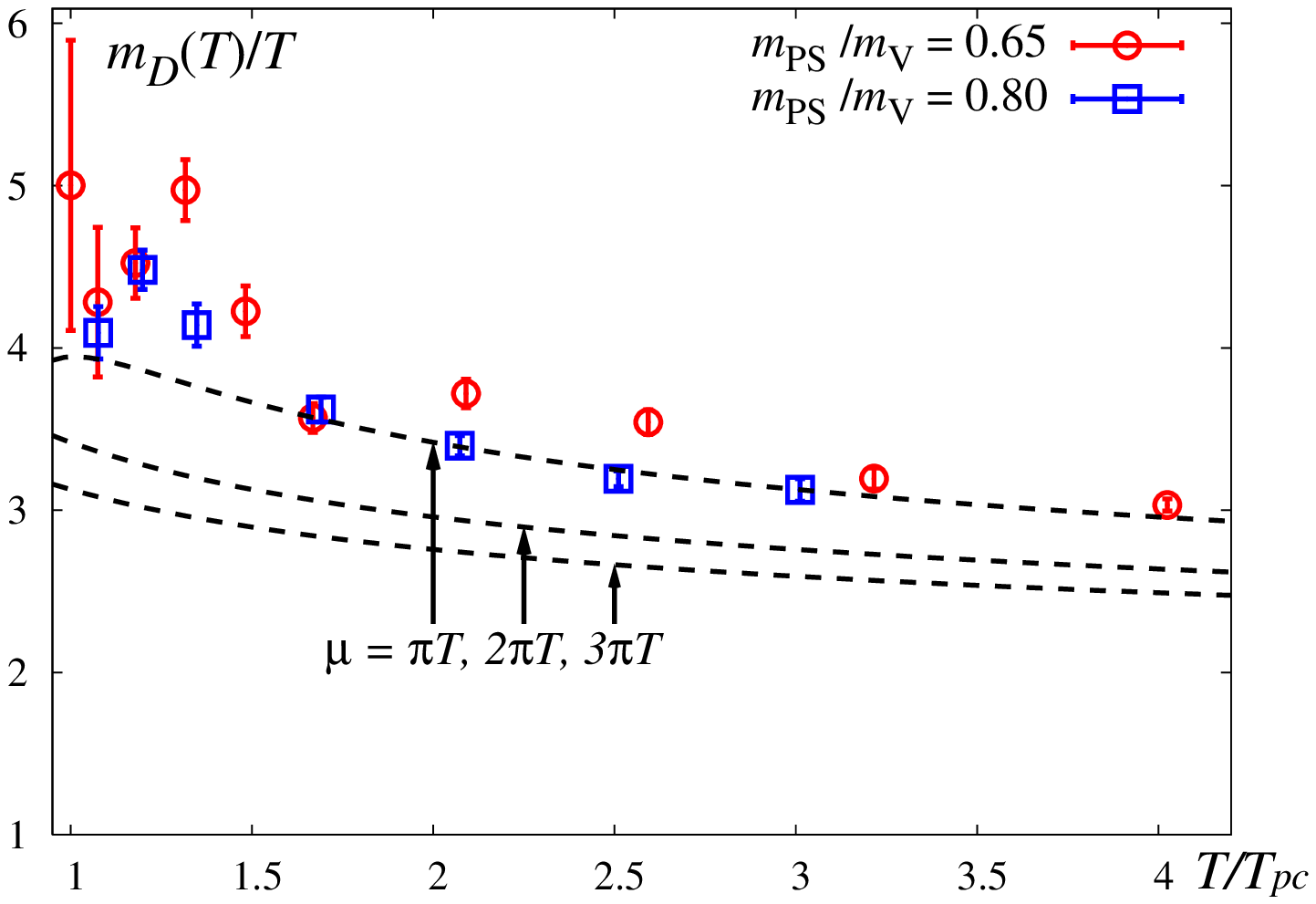} 
    \end{tabular}
    \caption{The Debye screening masses $m_D(T)$ 
    at $m_{\rm PS}/m_{\rm V} = 0.65$ and 0.80 in the color singlet channel
    together with that calculated in the leading-order (left)
    and next-to-leading-order (right) thermal perturbation theory 
    shown by the dashed lines. $\mu$ is the renormalization point
    chosen at $\mu = \pi T, \ 2\pi T, \ 3\pi T$.
    }
    \label{fig:pQCD}
  \end{center}
\end{figure}

In Fig.~\ref{fig:pQCD} (left) we compare $m_D(T)$ 
 in the color singlet channel
with ${m_D^{\rm LO}(T)}/{T}$
for $\mu = \pi T$, $2\pi T$ and $3\pi T$.
We find that the leading-order screening mass 
$m_D^{\rm LO}(T)$
  does not reproduce  the  simulation results at all.
Similar discrepancy has been observed also in quenched QCD  \cite{Nakamura:2003pu}
   and in full QCD with staggered quarks \cite{Kaczmarek:2005ui} .

To study higher-order contributions in 
the thermal perturbation theory,
we test the Debye mass  in the  next-to-leading-order 
calculated by the hard thermal loop resummation method \cite{Rebhan:1993az},
\begin{eqnarray}
\frac{m_D^{\rm NLO}(T)}{T} = 
\sqrt{ 1 + \frac{N_f}{6} } \
g_{\rm 2l}(\mu) \left[ 1 + 
g_{\rm 2l}(\mu) \frac{3}{2 \pi} \sqrt{\frac{1}{1+ N_f/6}}
\left(
\ln \frac{2 m_D^{\rm LO}}{m_{\rm mag}} - \frac{1}{2}
\right)
+ \mathcal{O}(g^2)
\right]
.
\label{eq:m_D_NLO}
\end{eqnarray}
 Here $m_{\rm mag}(T) = C_m \, g_{\rm 2l}^2(\mu) \, T$  denotes 
 the magnetic screening mass.
The factor $C_m$ cannot be determined within the perturbation theory 
due to the infrared problem.
In this study, we adopt $C_m \simeq 0.482$ obtained 
by a quenched lattice simulation
 \cite{Nakamura:2003pu}
 as a typical value \footnote{
 When we fit $C_m$ from our simulation results 
 $m_D(T=4.02T_{pc})$ with $\mu=2\pi T$, we obtain $C_m \simeq 0.33$.}.
 In Fig.~\ref{fig:pQCD} (right), our simulation results 
 for $m_D$ are compared with 
 $m_D^{\rm NLO}(T)$ shown by the  dashed lines.
 The values of $m_D^{\rm NLO}$ are larger than $m_D^{\rm LO}$ 
 by approximately 50 \% 
  and  agree better with the simulation results.

%%%%%%%%%%%%%%%%%%%%%%%%%%%%%%%%%%%%%%%%%%%%%%%%%%%%%%%%%%%%%%%%%%%%%%
\subsection{Phenomenological relation 
between $\alpha_{\rm eff}$ and $m_D$}

So far, we have fitted the free energies on the lattice treating 
 $\alpha_{\rm eff}$ and $m_D$ as independent parameters.
In this subsection, we test an ansatz inspired from the leading-order
perturbation theory:
\begin{eqnarray}
\frac{m_D(T)}{T} = \sqrt{1 + \frac{N_f}{6} } \ g_{\rm eff} (T) ,
\label{eq:m_eff}
\end{eqnarray}
with $g_{\rm eff} (T) \equiv \sqrt{4 \pi \alpha_{\rm eff}(T) } $.
Therefore, if this relation holds, we expect that the ratio
\begin{eqnarray}
R(T) \equiv  
\frac{( 1 + N_f / 6 )^{-1/2} \ m_D(T)/T}{\sqrt{4 \pi \alpha_{\rm eff}(T)}}
,
\end{eqnarray}
should be close to unity ($R(T) \sim 1$).

\begin{figure}[tbp]
  \begin{center}
     \includegraphics[width=80mm]{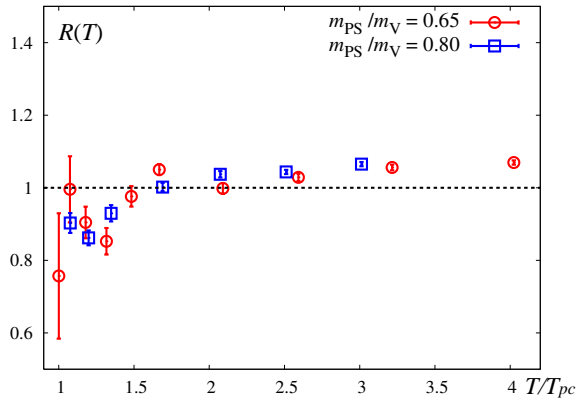}
     \caption{
     The ratio $R(T)$ in the text which is 
      supposed to be close to unity if  
      $m_D(T) = \sqrt{1 + N_f/6 } \, g_{\rm eff} (T) \, T$ 
      holds.
     The results 
        are for the color singlet channel. 
        }
     \label{fig:P_ratio}
 \end{center}
\end{figure}

In Fig.~\ref{fig:P_ratio},  our simulation results of $R(T)$ 
for the singlet channel are shown as a function of $T/T_{pc}$.
We find that $R(T)$ is consistent  with unity 
 for $ T \simge 1.5 T_{pc}$ within 10\% accuracy.
This is a non-trivial observation particularly near $T_{pc}$
 and  suggests that the major part of the higher-order effects
  and non-perturbative effects of $m_D(T)$ can be 
   expressed by the effective running coupling $g_{\rm eff}(T)$. 
We note that a similar effective coupling at $T=0$ defined through the lattice
potential was introduced to improve the lattice perturbation theory \cite{Lepage:1992xa}.

%%%%%%%%%%%%%%%%%%%%%%%%%%%%%%%%%%%%%%%%%%%%%%%%%%%%%%%%%%%%%%%%%%%%%%
\subsection{Comparison with the staggered quark action}

\begin{figure}[tbp]
  \begin{center}
    \begin{tabular}{cc}
      \includegraphics[width=80mm]{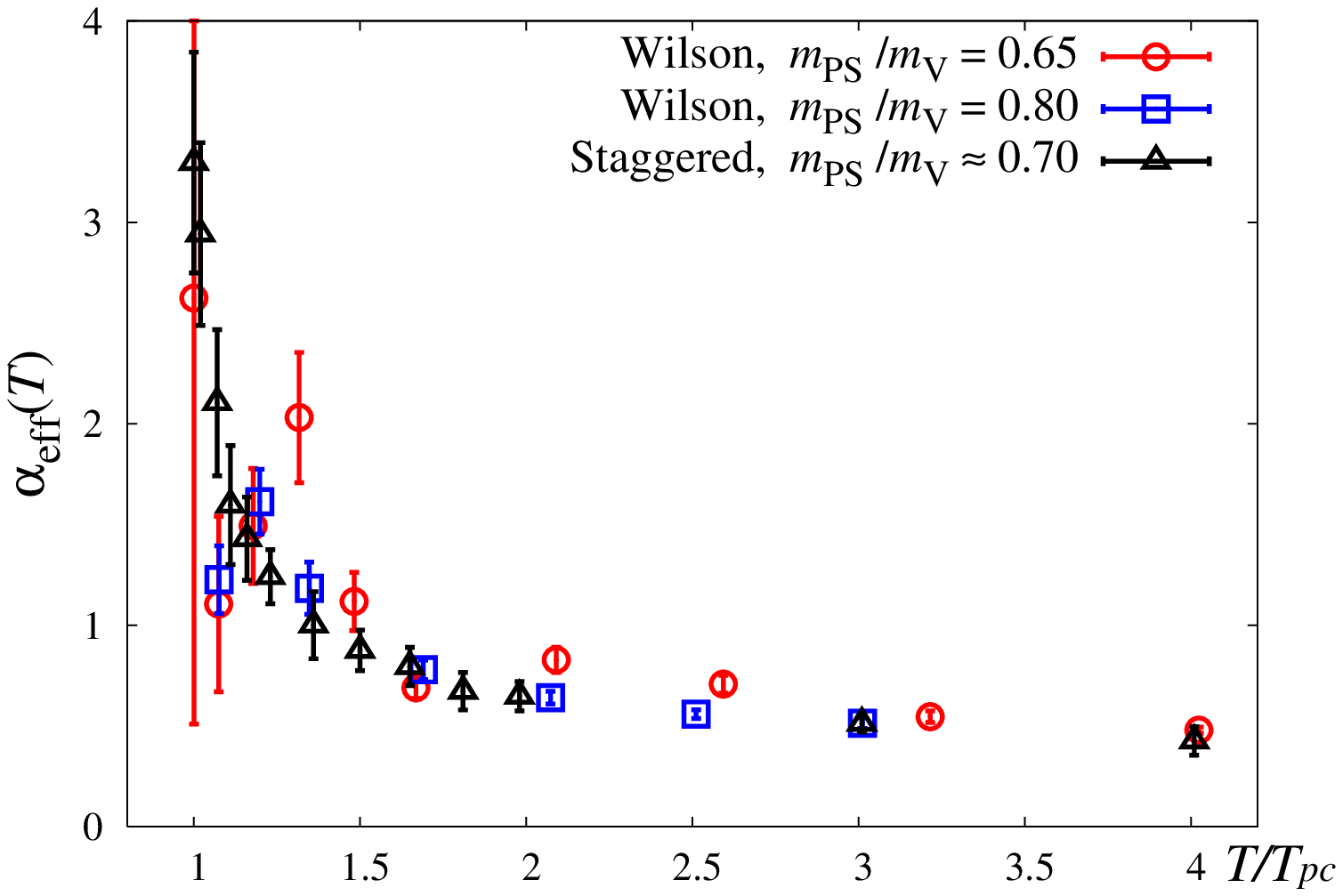} &
      \includegraphics[width=80mm]{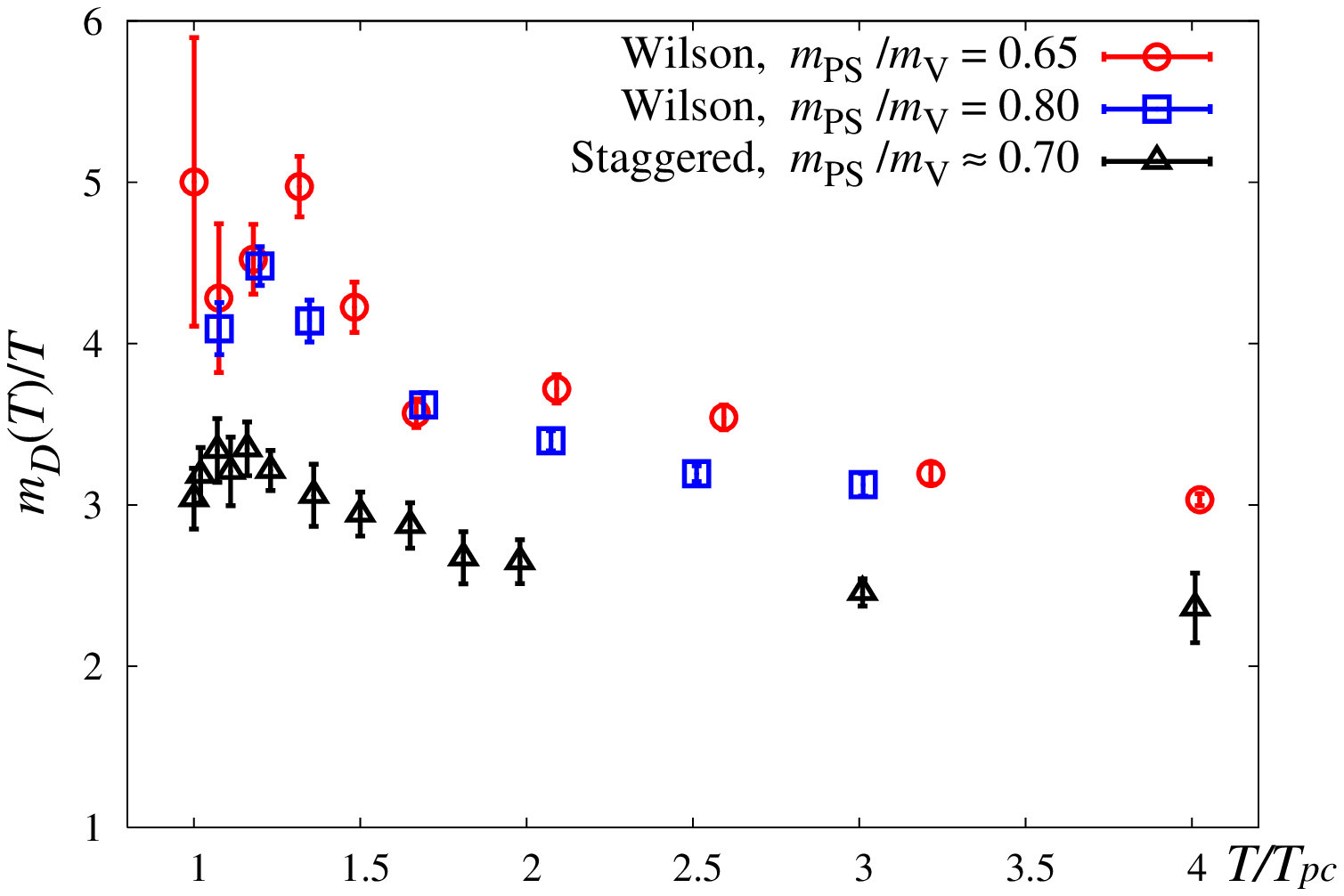} 
    \end{tabular}
    \caption{Comparison of the $\alpha_{\rm eff}(T)$(left) 
    and $m_D(T)$(right) between
    the results of the Wilson quark action and 
    staggered quark action.
    }
    \label{fig:KS}
  \end{center}
\end{figure}

\begin{figure}[tbp]
   \begin{center}
     \begin{tabular}{cc}
     \includegraphics[width=80mm]{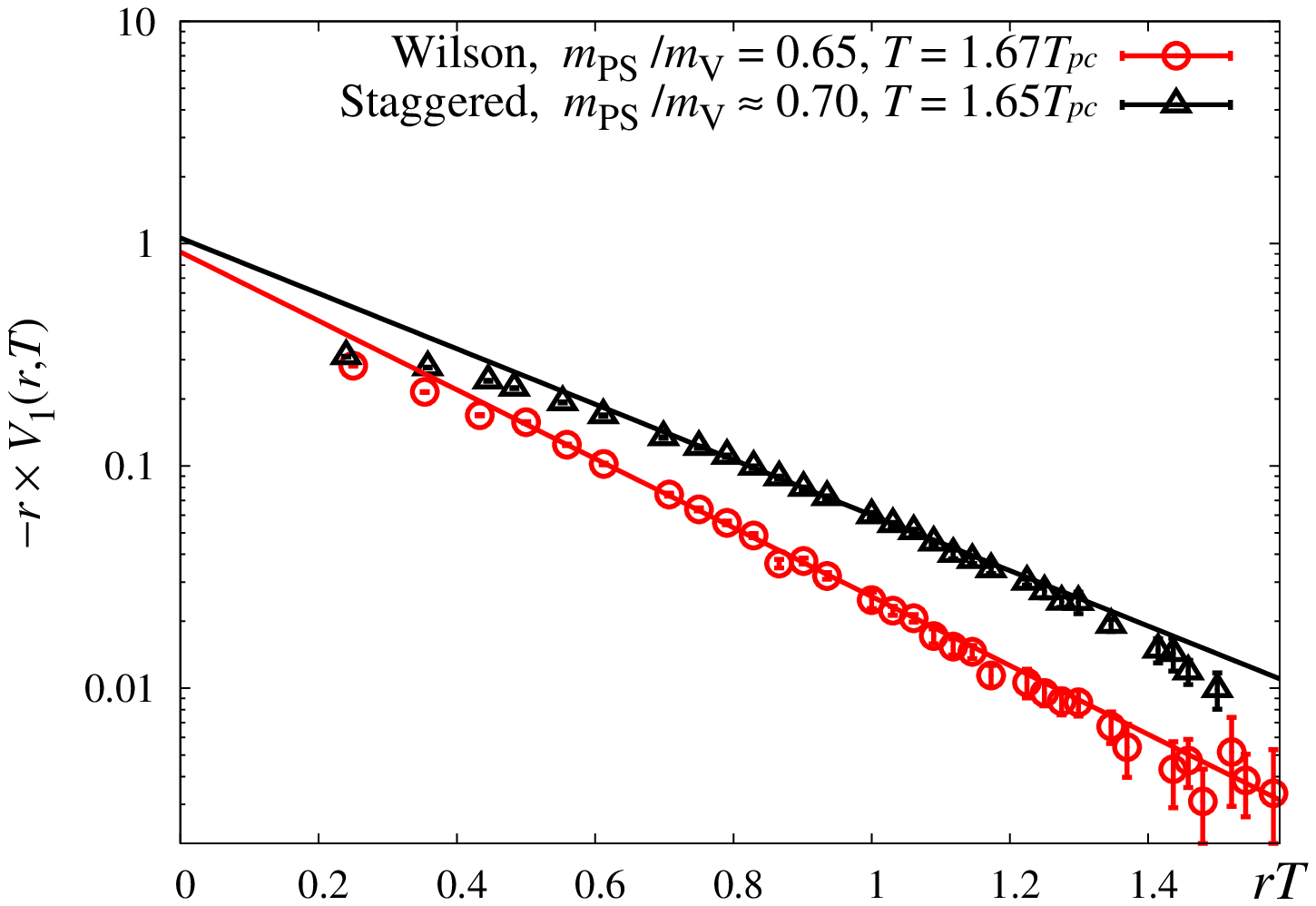} &
     \includegraphics[width=80mm]{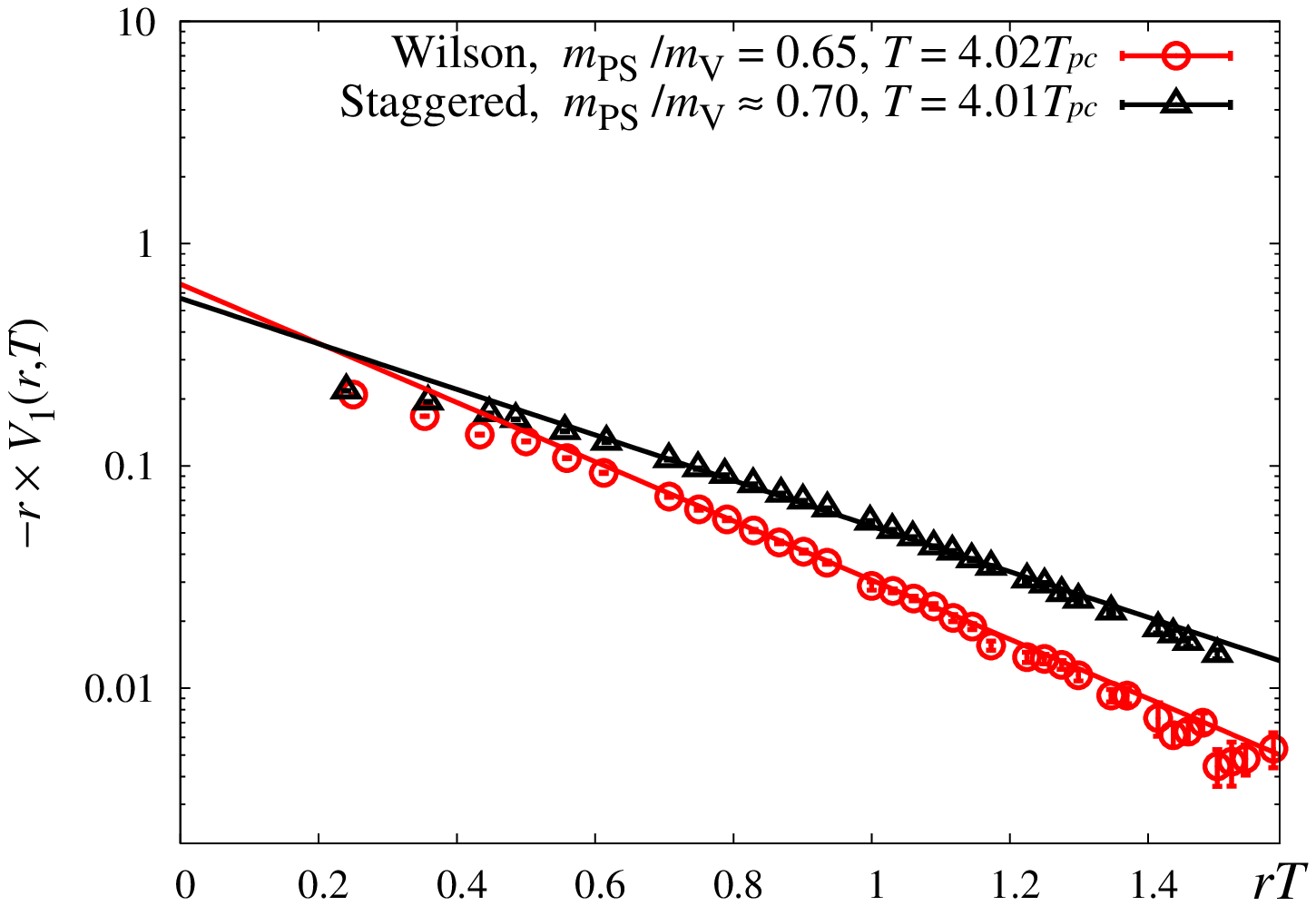}
     \end{tabular}
     \caption{Comparison of the normalized free energies, 
     $-r \times V_{\bf 1}(r)$, between the Wilson quark action
     (present work)
     and the staggered quark action in Ref.~\cite{Kaczmarek:2005ui}
 as a function of $rT$ at $T \simeq 1.7 T_{pc}$ (left) 
 and $\simeq 4.0 T_{pc}$ (right)
 with the log scale.  Straight lines show the fits based on the 
 screened Coulomb form in the range 
 $\sqrt{11}/4 \le rT \le 1.5$ (circles) and 
   $0.8 - 1.0 \simle rT $ (triangles).
 }
     \label{fig:rV}
   \end{center}
\end{figure}

Finally, we compare the results of  $\alpha_{\rm eff}(T)$ and $m_D(T)$
obtained by the Wilson quark action (present work)
with those by an improved staggered quark action
 on a  $16^3 \times 4$ lattice
 at $m_{\rm PS}/m_{\rm V} \simeq 0.70$ \cite{Kaczmarek:2005ui,Doring}.
The comparison is shown in
 Fig.~\ref{fig:KS} for  $\alpha_{\rm eff}(T)$ (left panel)
and $m_D(T)$ (right panel).
Although  $\alpha_{\rm eff}(T)$ does not show significant
 difference between
the two actions, $m_D(T)$ in the Wilson quark action is
systematically higher than that of the staggered quark action 
by about 20\% 
even at $T= 4 T_{pc}$.
This discrepancy can be seen directly from the normalized free energies
 in  Fig. \ref{fig:rV} where $-r \times V_{\bf 1}(r)$
 is shown as a function of $rT$ at $T \simeq 1.7 T_{pc}$ (left) 
 and $\simeq 4.0 T_{pc}$ (right).
 The circles (triangles) are the results of the 
  Wilson quark action (staggered quark action). The data for
  the staggered quark action is taken from Ref.~\cite{Kaczmarek:2005ui}.
 The straight lines are the fits 
 with the screened Coulomb form, Eq.~(\ref{eq:SCP}),
  in the range $\sqrt{11}/4 \le rT \le 1.5$ (circles) and 
   $0.8 - 1.0 \simle rT $ (triangles).
 The intercepts of the lines with the vertical axis
 and the slopes of the lines correspond to
  $\alpha_{\rm eff}(T)$ and $m_D(T)$, respectively.
There is an obvious difference in the slope in 
  different quark actions,
   which may be regarded 
   as systematic errors due to the lattice discretization.
 This discrepancy should be further investigated
  at smaller lattice spacing,
 i.e. larger lattice size in the temporal direction.

%%%%%%%%%%%%%%%%%%%%%%%%%%%%%%%%%%%%%%%%%%%%%%%%%%%%%%%%%%%%%%%%%%%%%%
\subsection{Force between heavy quarks}

\begin{figure}[tbp]
   \begin{center}
     \begin{tabular}{cc}
     \includegraphics[width=80mm]{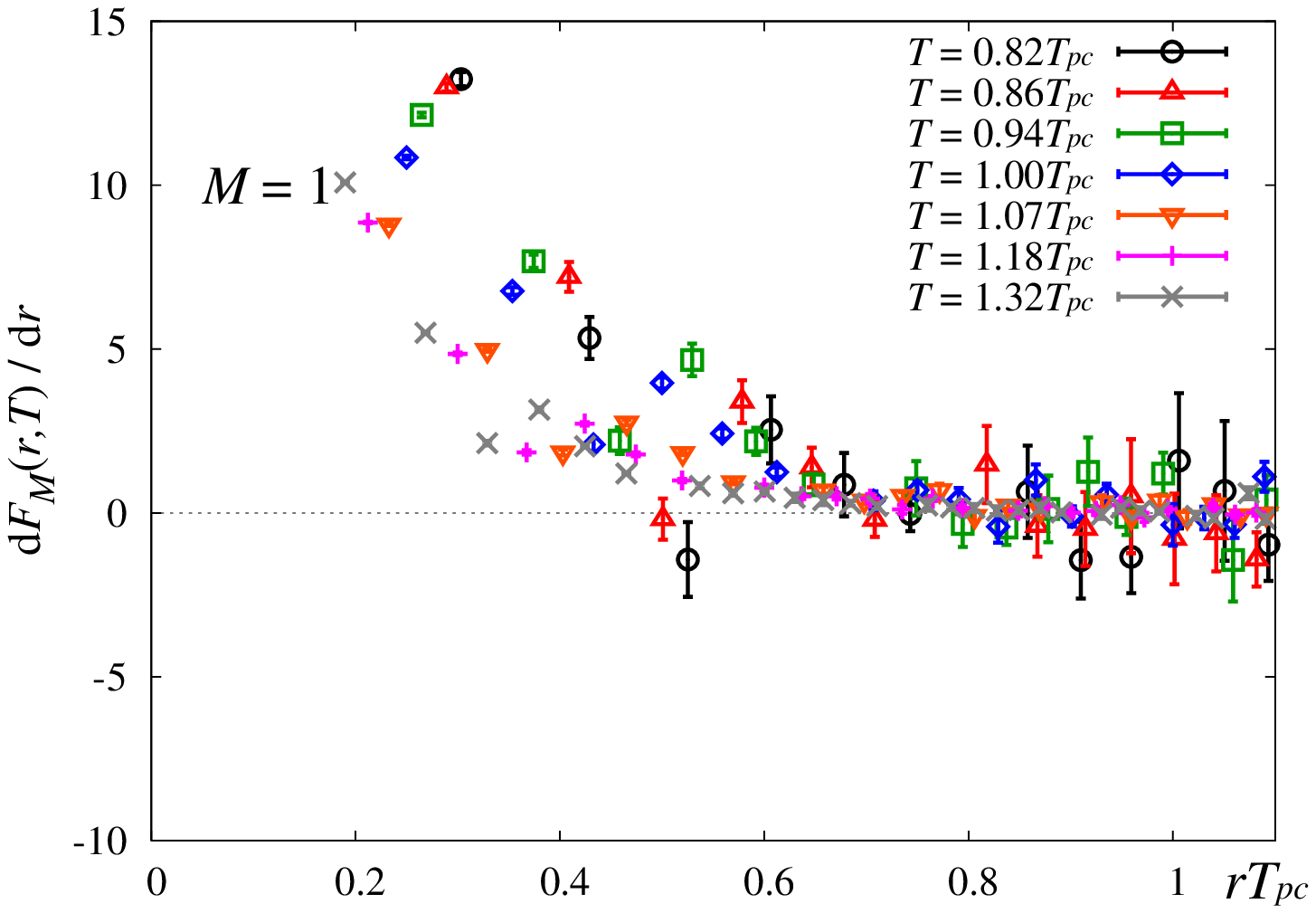} &
     \includegraphics[width=80mm]{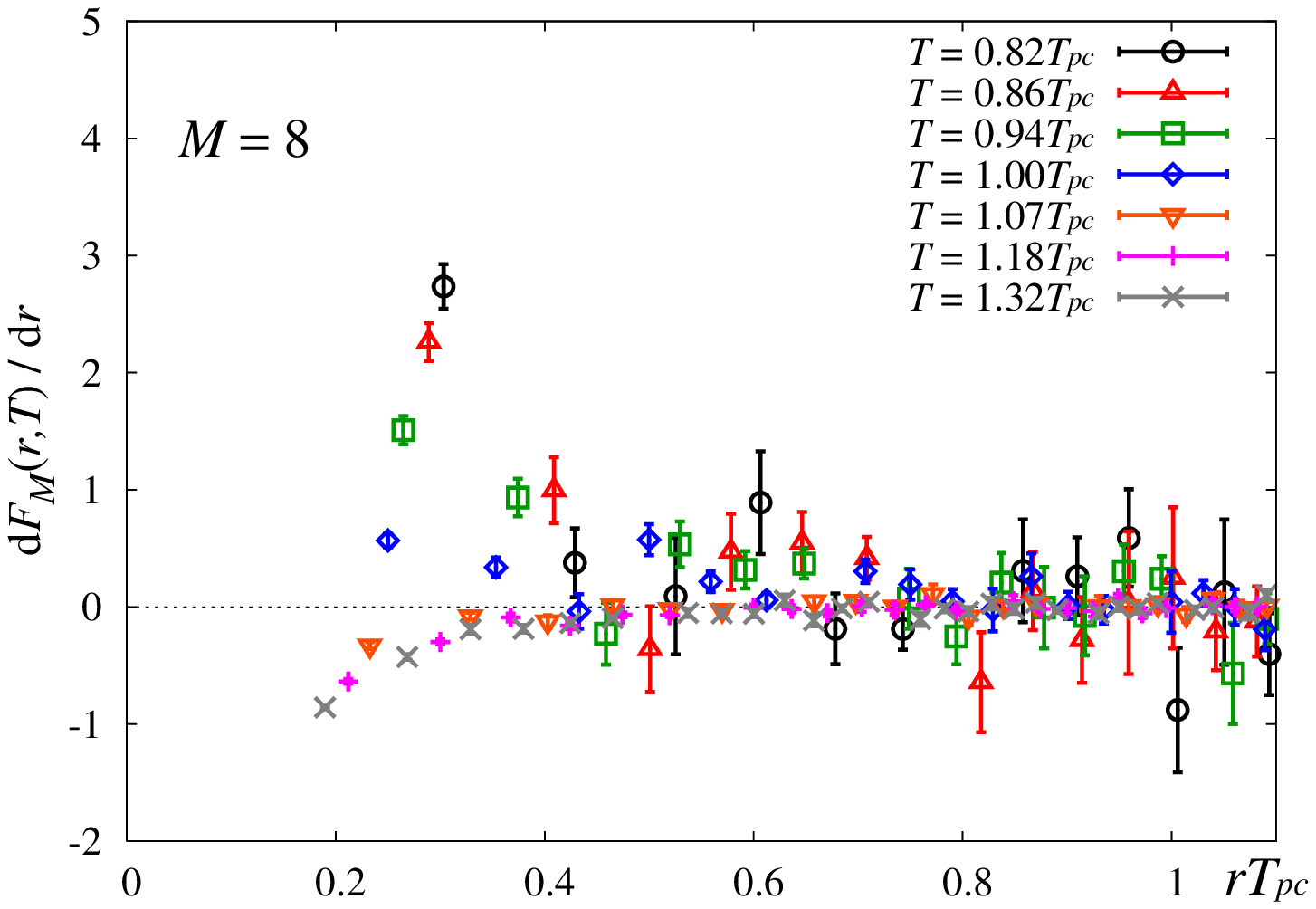}
     \end{tabular}
     \caption{Results of 
     $dF_M(r)/dr$ scaled by $T_{pc}$
     for color singlet (left) and octet (right) $Q\overline{Q}$ channels
     at $m_{\rm PS}/m_{\rm V} = 0.65$.
         }
     \label{fig:force1_065}
   \end{center}
\end{figure}

\begin{figure}[tbp]
   \begin{center}
     \begin{tabular}{cc}
     \includegraphics[width=80mm]{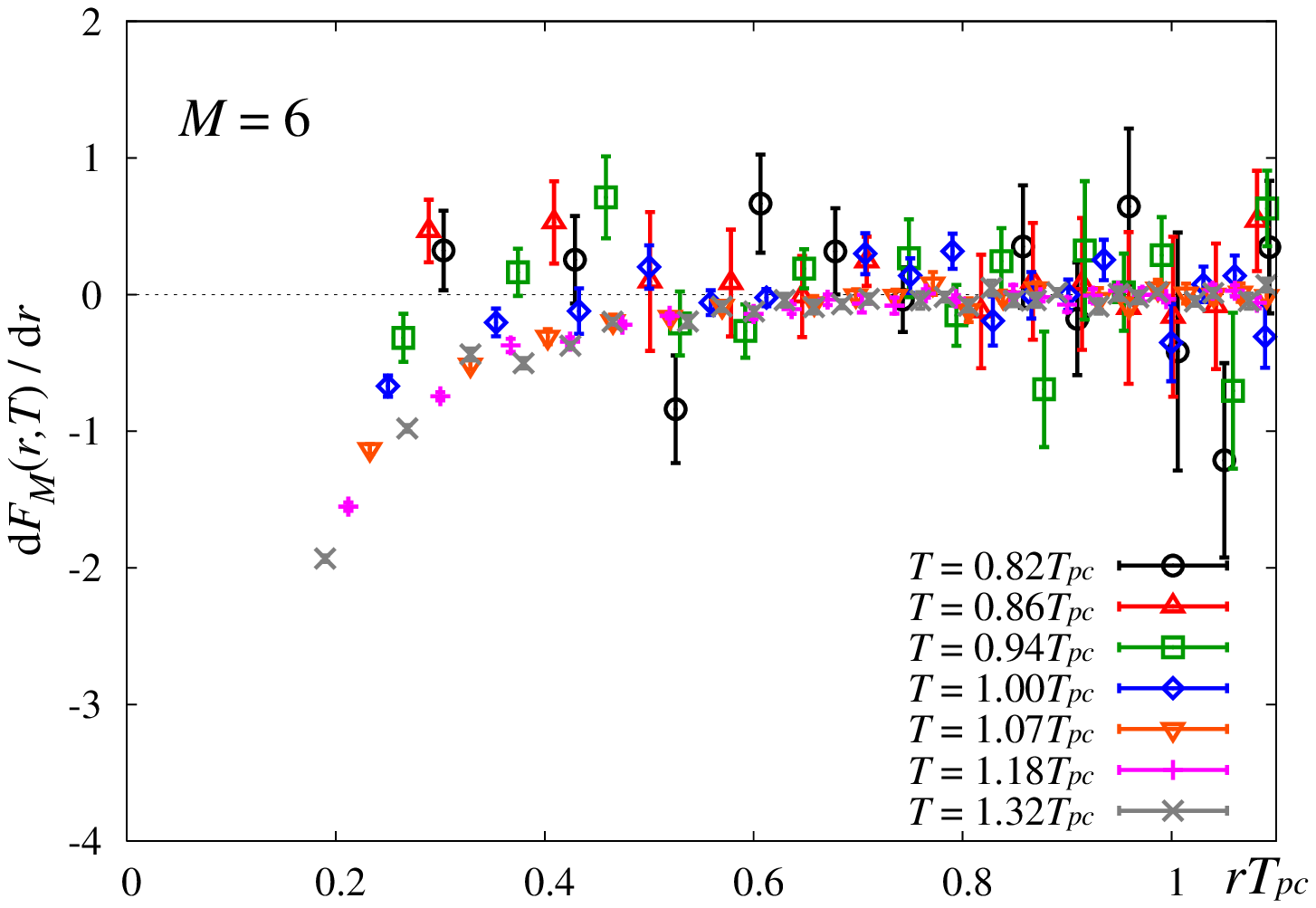} &
     \includegraphics[width=80mm]{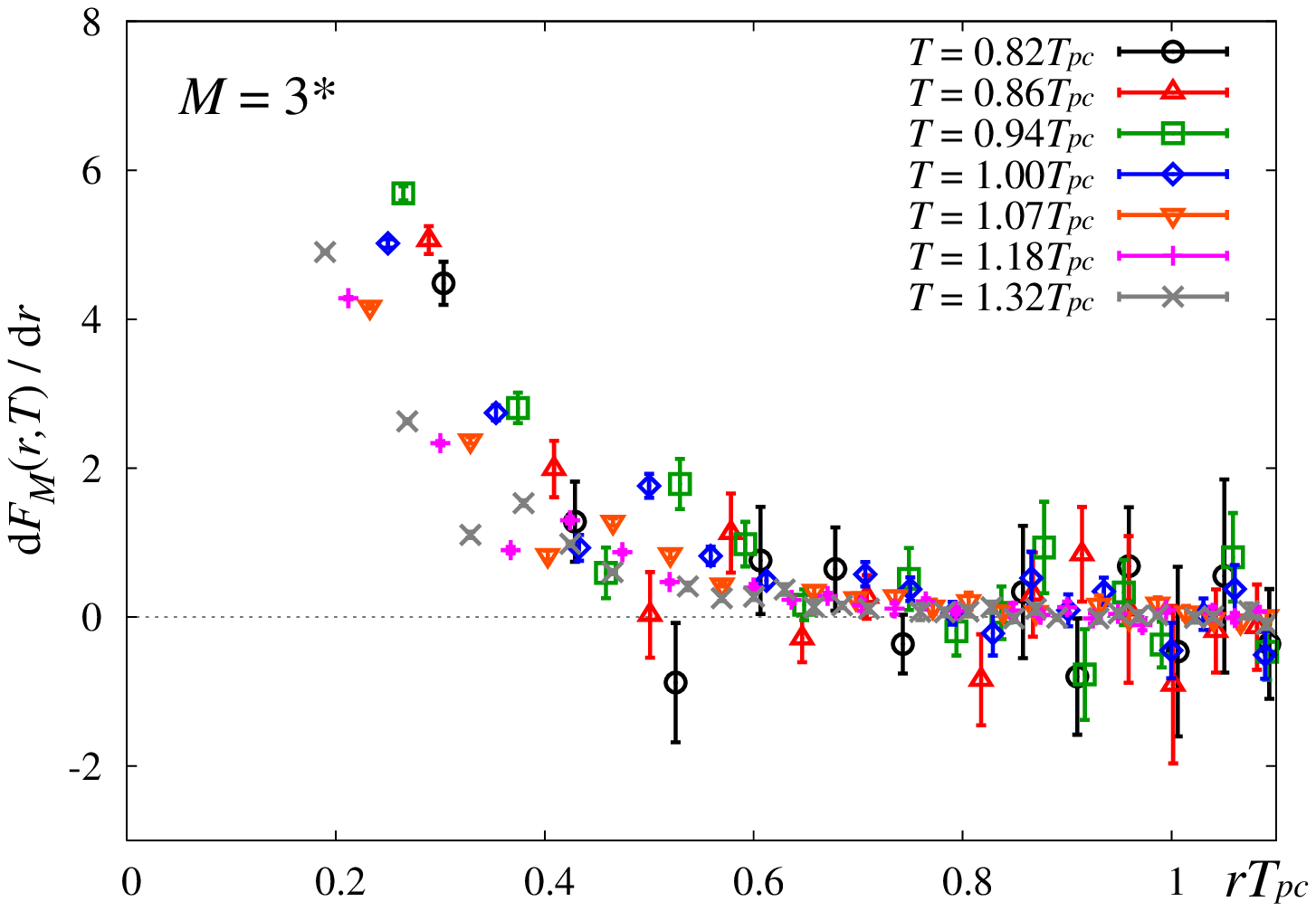}
     \end{tabular}
     \caption{Results of 
     $dF_M(r)/dr$ scaled by $T_{pc}$
     for color sextet (left) and anti-triplet (right) $QQ$ channels
     at $m_{\rm PS}/m_{\rm V} = 0.65$.
         }
     \label{fig:force2_065}
   \end{center}
\end{figure}

\begin{figure}[tbp]
   \begin{center}
     \begin{tabular}{cc}
     \includegraphics[width=80mm]{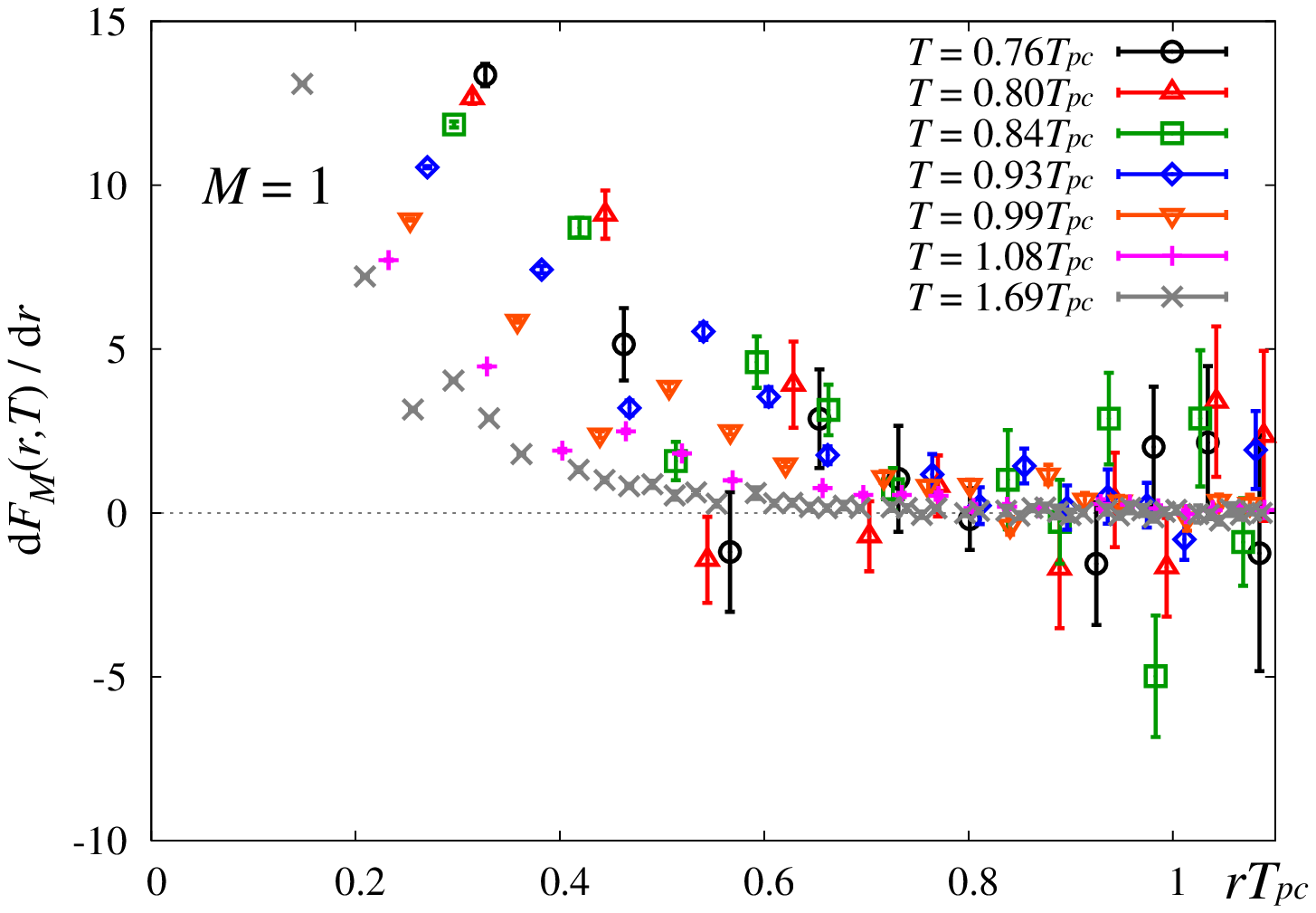} &
     \includegraphics[width=80mm]{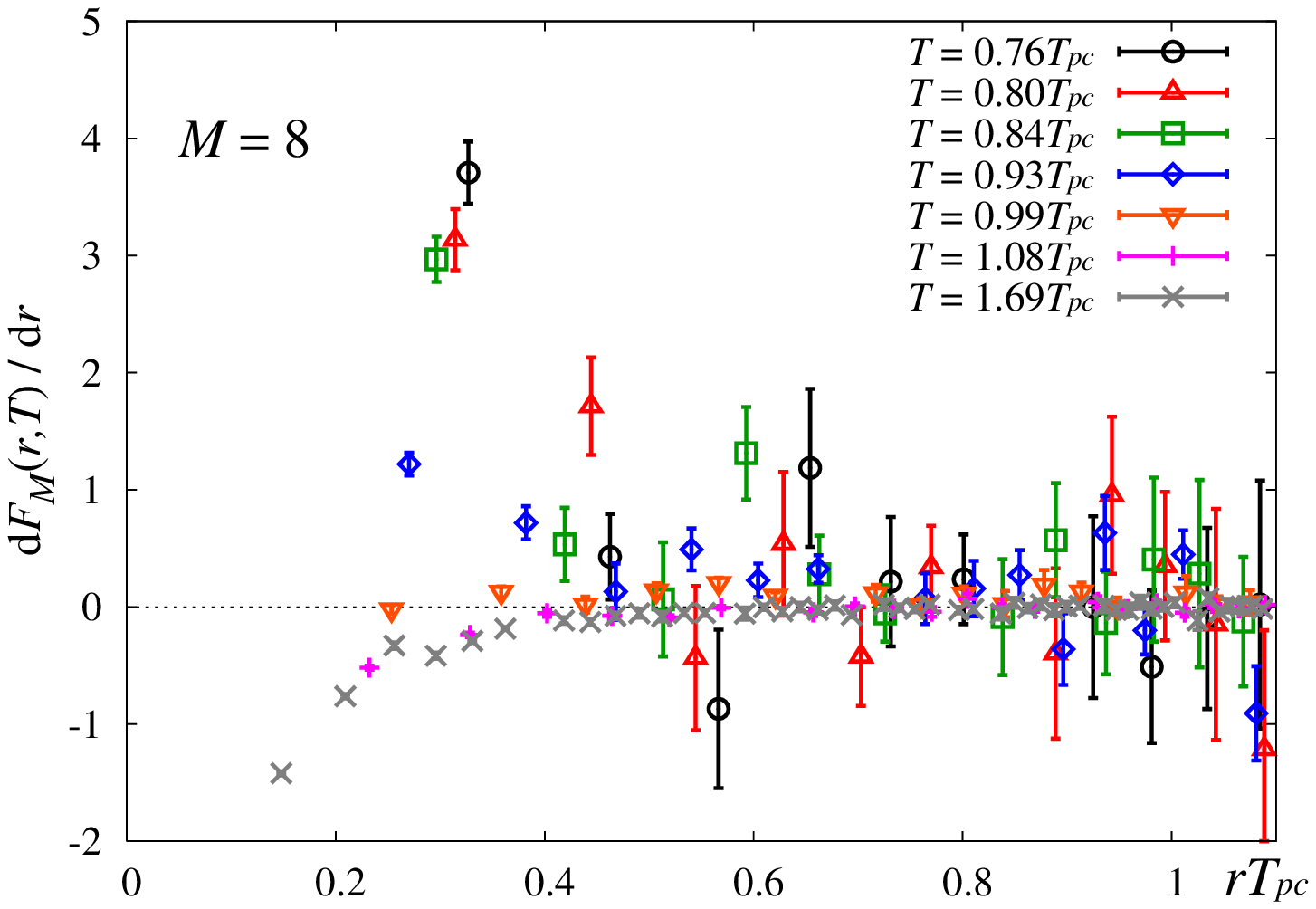}
     \end{tabular}
     \caption{
     The same as Fig.~\ref{fig:force1_065}
     at $m_{\rm PS}/m_{\rm V} = 0.80$.
         }
     \label{fig:force1_080}
   \end{center}
\end{figure}

\begin{figure}[tbp]
   \begin{center}
     \begin{tabular}{cc}
     \includegraphics[width=80mm]{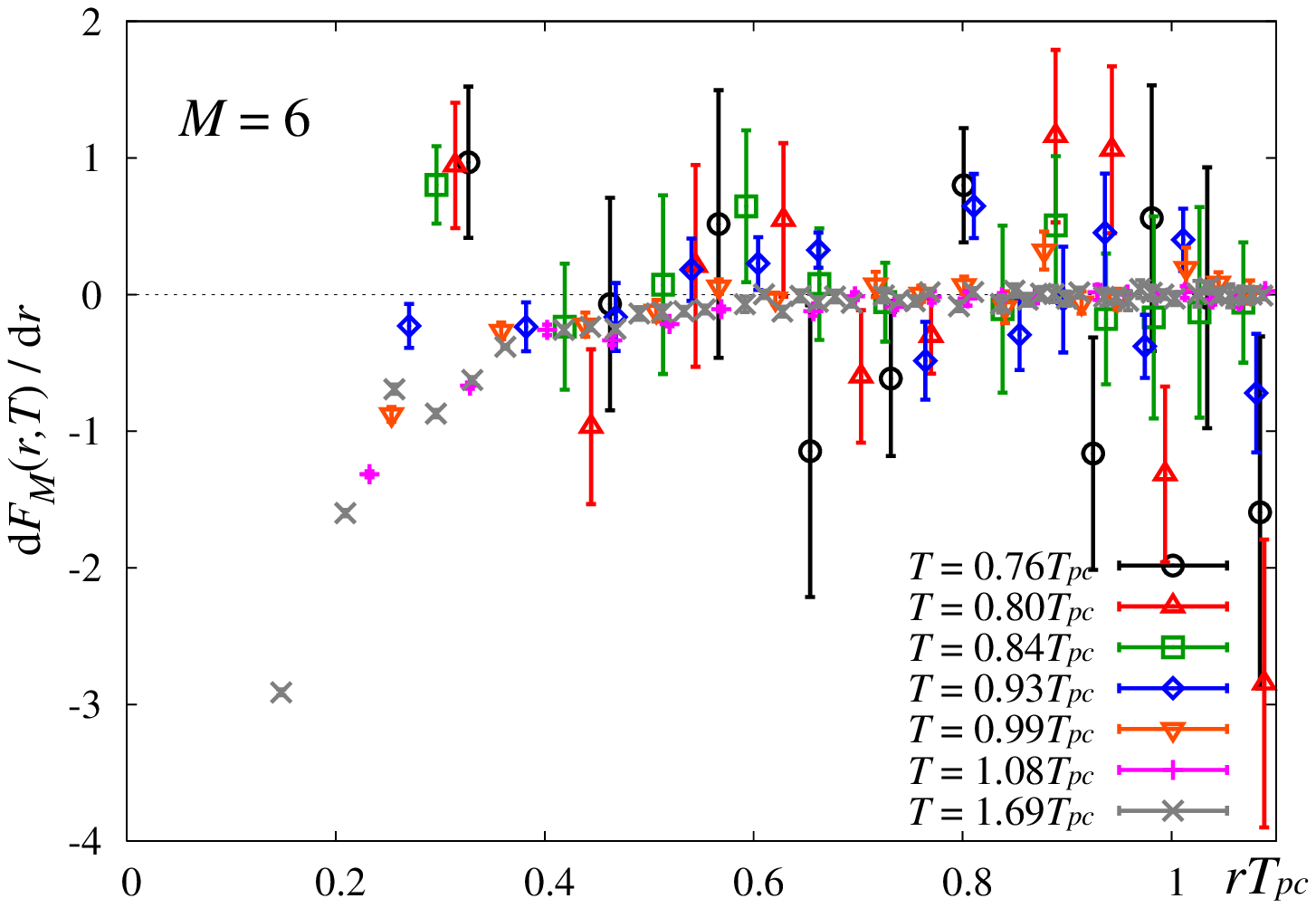} &
     \includegraphics[width=80mm]{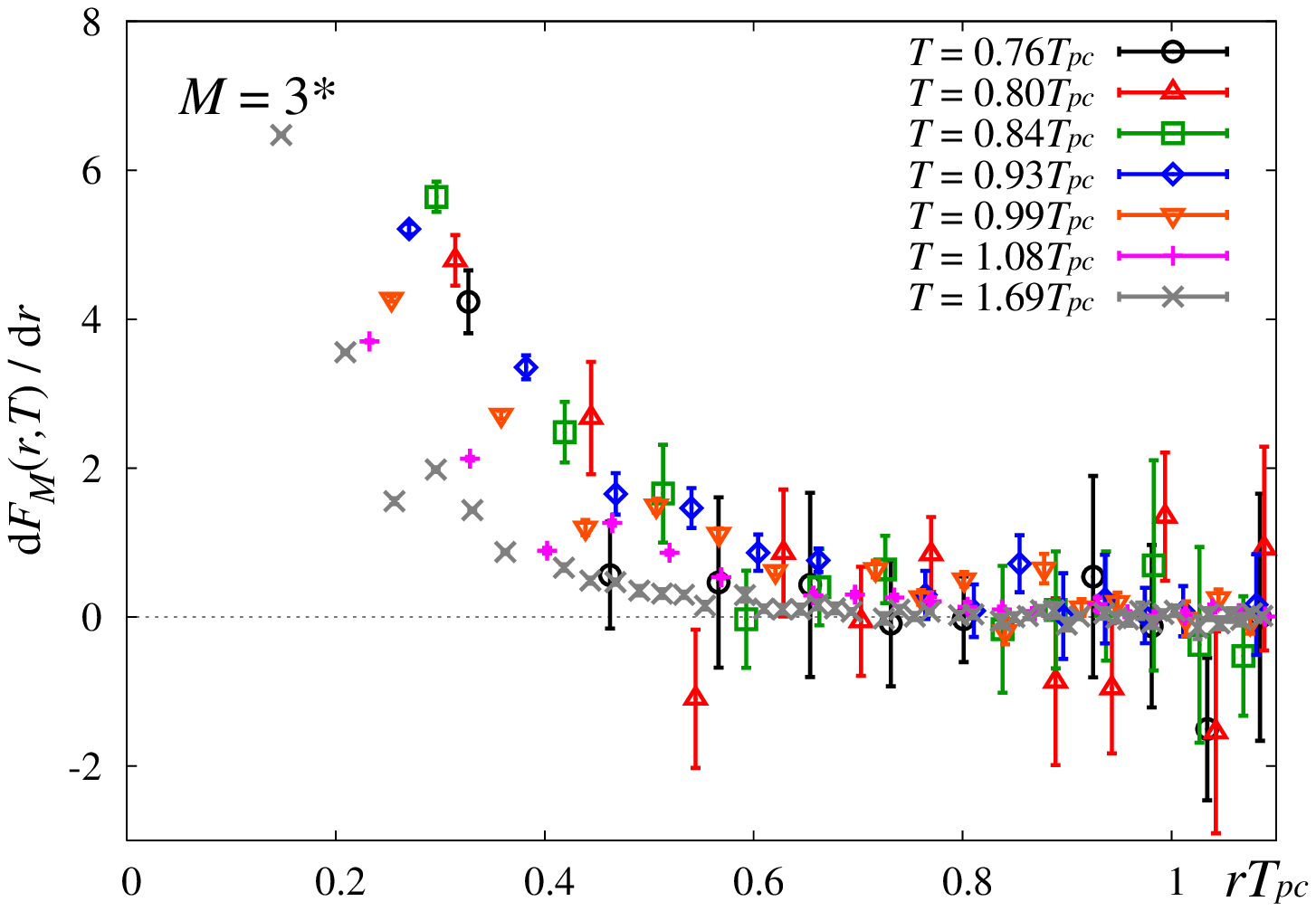}
     \end{tabular}
     \caption{
     The same as  Fig.~\ref{fig:force2_065}
     at $m_{\rm PS}/m_{\rm V} = 0.80$.
         }
     \label{fig:force2_080}
   \end{center}
\end{figure}

To make a direct comparison of the free energies below and above
$T_{pc}$, we study the ``force'' between heavy quarks defined by
$dF_M(r,T)/dr$ without introducing the subtraction of 
$\langle {\rm Tr}\Omega \rangle^2$. 
  In Fig.~\ref{fig:force1_065}, such forces are shown
  for color singlet (left) and octet (right) $Q \overline{Q}$ channels,
and in Fig.~\ref{fig:force2_065},
for color sextet (left) and anti-triplet (right) $QQ$ channels
at $m_{\rm PS}/m_{\rm V} = 0.65$. 
Results at $m_{\rm PS}/m_{\rm V} = 0.80$ are shown 
in Fig.~\ref{fig:force1_080} and \ref{fig:force2_080}.

  In the color singlet and anti-triplet channels,
   there is always an ``attraction'' both
  for $T < T_{pc}$ and for  $T > T_{pc}$, and the
  attraction is stronger as we decrease the temperature.
  The signal for the color sextet channel below
  $T_{pc}$ is rather weak and we cannot make a definite
   statement from our data.
  On the other hand, the force in the color octet channel
   is clearly ``attractive'' (``repulsive'')  at low (high) $T$.
   Namely, the simple Casimir scaling law, which worked
   well at high $T$, does not hold below  $T_{pc}$.
In Ref.~\cite{Jahn:2004qr}, it has be suggested that
the Polyakov loop correlation in the octet channel is saturated by the
color-singlet intermediate states at very low temperature.
If this is true, the attraction in the color octet channel found here
is explained by the attraction in the singlet channel.

%%%%%%%%%%%%%%%%%%%%%%%%%%%%%%%%%%%%%%%%%%%%%%%%%%%%%%%%%%%%%%%%%%%%%%%%%%
\section{Spatial Wilson Loop}
\label{sec:4}

In previous quenched studies \cite{bali,karr,karsch,boyd},
the Wilson loop in spatial direction is
found to show non-vanishing spatial string tension $\sigma_s$
even at $T>T_{pc}$, which is called the spatial confinement.
We study this phenomenon with the existence of dynamical quarks in the system.
Although quarks are expected to decouple from the
spatial observables for $T \gg T_{pc}$ due to dimensional reduction
and thus do not affect $\sigma_s$ in the high temperature limit,
it is not obvious whether the same is true near $T_{pc}$.

\begin{figure}[tbp]
  \begin{center}
    \begin{tabular}{cc}
    \includegraphics[width=80mm]{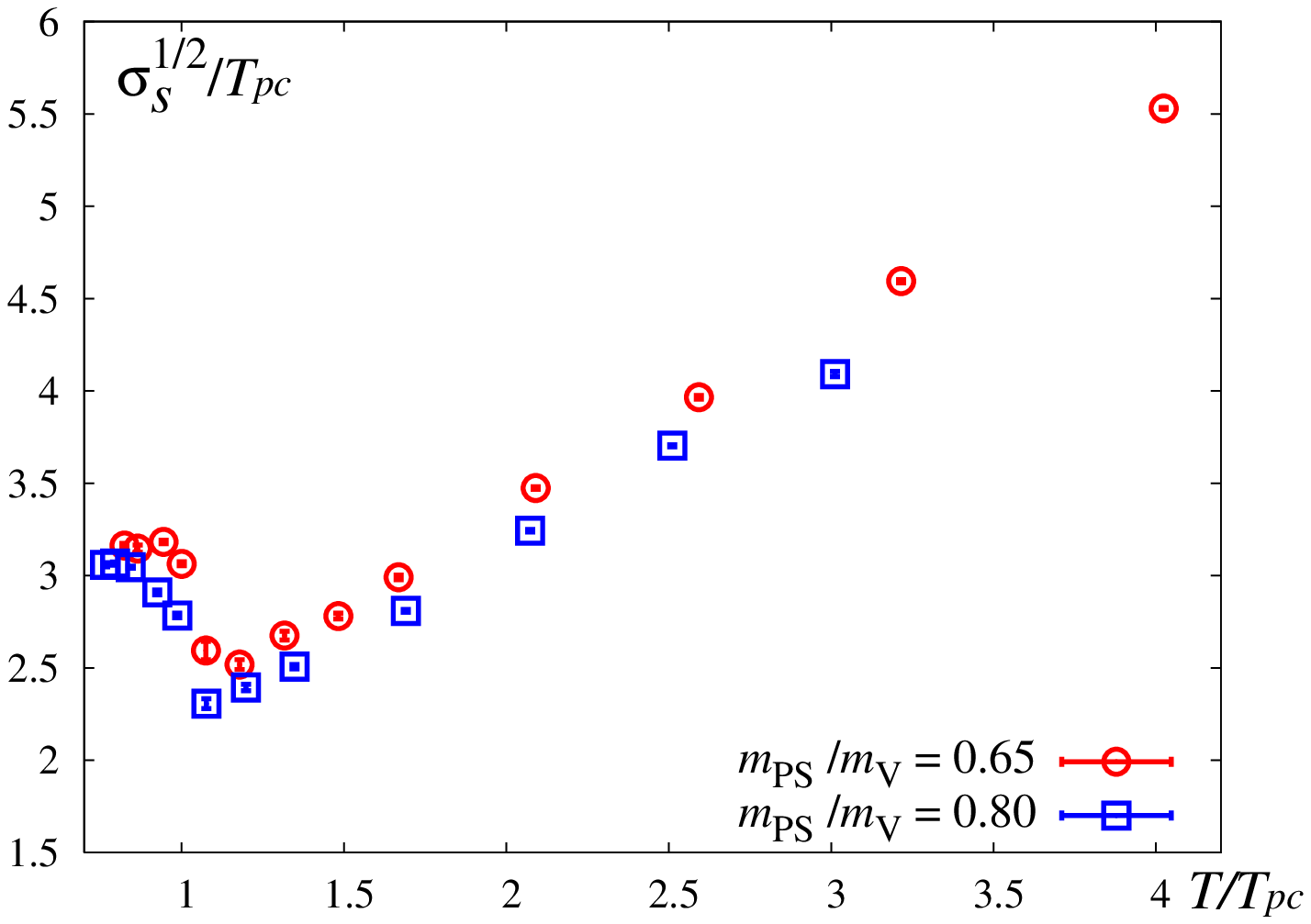} &
    \includegraphics[width=80mm]{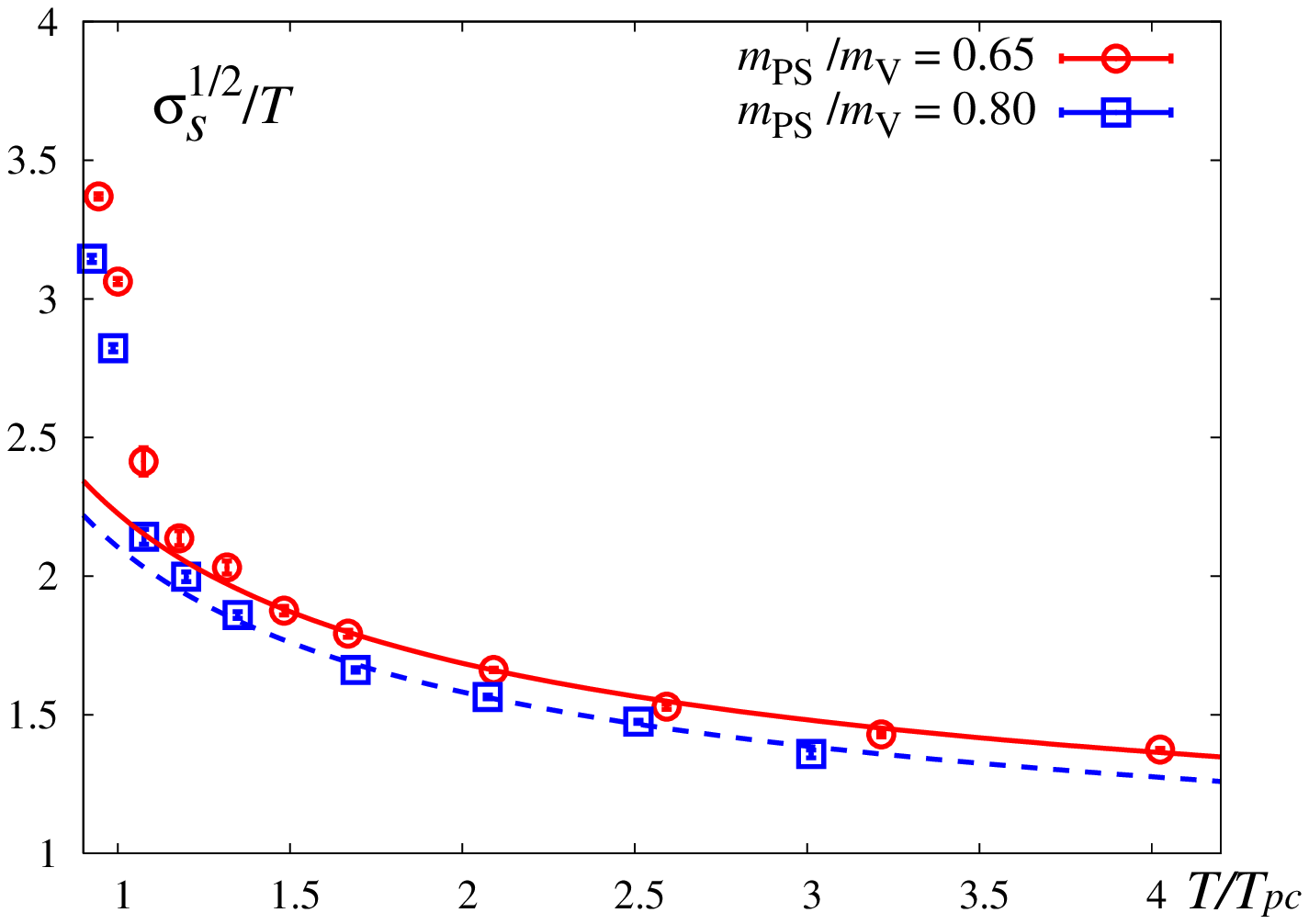}
    \end{tabular}
    \caption{The square root of the spatial string tension over $T_{pc}$ (left) 
    and $T$ (right) as a function of $T/T_{pc}$ at $m_{\rm PS}/m_{\rm V}=0.65$
    and 0.80. 
    The solid and dashed lines in the right panel express the fit results
    based on Eq.~(\ref{eq:fit}) at $m_{\rm PS}/m_{\rm V}=0.65$ and 0.80, 
    respectively.}
    \label{fig:S1}
  \end{center}
\end{figure}

We evaluate $\sigma_s$ assuming a simplest ansatz
for the spatial Wilson loop $W(I,J)$ with the size $I\times J $:
\begin{eqnarray}
 -\ln W(I,J) = \sigma_s IJ + \sigma_{p}(2I+2J) + C_w,
\end{eqnarray}
where $\sigma_s$, $\sigma_{p}$ and $C_w$ are fit parameters.
The results of $\sqrt{\sigma_s(T)}/T_{pc}$ are shown in Fig. \ref{fig:S1} (left)
as a function of $T/T_{pc}$.
We find that $\sqrt{\sigma_s(T)}/T_{pc}$ approaches to a constant
 as $T$ decreases below $T_{pc}$,
  while it increases linearly as $T$ above $T_{pc}$.
 Similar behavior has been observed in the quenched case \cite{bali}.

 Let us first make a phenomenological parameterization of 
 the spatial string tension:
\begin{eqnarray}
 {\sqrt{\sigma_s(T)}} &=& c \, g_{\rm 2l}^2(\mu ) \, T,
 \label{eq:fit}
\end{eqnarray}
 motivated by the dimensional reduction at high temperature
with $g_{\rm 2l}^2(\mu)$ being the two-loop running coupling
 of two-flavor QCD defined in Eq.~(\ref{eq:RC}). 
Our results for $\sqrt{\sigma_s(T)}/T$ are shown in Fig.~\ref{fig:S1}
(right). The solid and dashed lines in the figure are the results of
a fit of our data in the range $T > 1.3 T_{pc}$ to the formula
Eq.~(\ref{eq:fit}). Here, we fix $\mu = 2 \pi T$ and take $c$ and
$\Lambda/T_{pc}$ as fitting parameters to obtain 
\begin{eqnarray}
 && c=0.694(12), \ \ \Lambda/T_{pc} = 0.866(48) 
 \ \ {\rm for} \ \ m_{\rm PS}/m_{\rm V} = 0.65  , 
\\
 && c=0.639(34), \ \ \Lambda/T_{pc} = 0.92(14) 
 \ \ {\rm for} \ \ m_{\rm PS}/m_{\rm V} = 0.80  .
\end{eqnarray}
Similar numbers have been observed in a quenched simulation \cite{boyd},
$(c, \Lambda/T_{pc}) =(0.566(13), 0.653(57))$,
and in a $2+1$ flavor simulation with staggered quark \cite{Umeda:2006xi},
$(c,\Lambda/T_{pc}) =(0.587(41), 0.72(17))$. (Here we rescaled the previous
 results of $\Lambda$ by $2 \pi$ 
 to compare at the common value of $\mu = 2 \pi T$.)

Let us now make an alternative comparison of our data with 
 a recent  prediction by the parameter-free three-dimensional $(3d)$ effective
theory \cite{Laine:2005ai,ls}, which gives 
\begin{eqnarray}
\sqrt{\sigma_s} = 0.553(1) \, g_{\rm M}^2 , 
\label{eq:3d}
\end{eqnarray}
where the coefficient 0.553(1) expresses a non-perturbative contribution
 determined by $3d$ quenched
 lattice simulations, and $g_{\rm M}$ is a dimensionful $3d$ gauge coupling
defined through the four-dimensional running coupling, $g(\mu)$, as 
\begin{eqnarray}
g_{\rm M}^2 =  g_{\rm 2l}^2(\mu) \, T 
\left[ 1 + C_1 \, g_{\rm 2l}^2(\mu) 
+ C_2 \, g_{\rm 2l}^4(\mu) + \cdots \right] .
\label{eq:gm}
\end{eqnarray}
 Known coefficients $C_1$ and $C_2$
 represent the next-to-leading (NLO) and next-to-next-leading (NNLO)
 contributions in $3d$ effective theory ~\cite{Laine:2005ai}. 
We choose $\mu/T=2 \pi $, which is 
in the range 6.0--7.0
assumed in the study of Ref.~\cite{Laine:2005ai}.
 We take the same value for 
 $\Lambda/T_{pc}= \Lambda^{N_f=2}_{\overline{MS}}/T_{pc}$ 
 as that in Section \ref{sec:PT}.

\begin{figure}[tbp]
  \begin{center}
    \includegraphics[width=80mm]{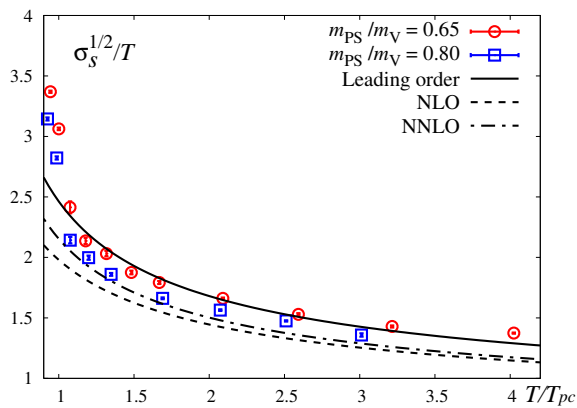}
    \caption{Comparison of our results and predictions of 
    the three-dimensional effective theory with contributions
    up to leading order (solid), NLO (dashed) and 
    NNLO (dash-dotted line), respectively.} 
    \label{fig:S2}
  \end{center}
\end{figure}

Figure \ref{fig:S2} shows the results of the spatial string tension,
together with the prediction from the $3d$ effective theory in the
leading order, NLO  and NNLO.
We observe that convergence of the series is rather well
and the NNLO result is quite consistent with our lattice data.
Similar agreement in quenched QCD  has been reported in Ref.~\cite{Laine:2005ai}.

%%%%%%%%%%%%%%%%%%%%%%%%%%%%%%%%%%%%%%%%%%%%%%%%%%%%%%%%%%%%%%%%%%%%%%
\section{Conclusions}
\label{sec:5}

 In the past few years, many 
 of the lattice QCD simulations at finite temperature and density have been  
 performed using staggered quark actions. However, 
 to control the lattice artifacts, comparison with other quark actions
 are indispensable.  This motivates us to 
 carry out a systematic study with an improved Wilson quark action.
As a first step, we performed simulations of $N_f=2$ QCD on an 
$N_s^3 \times N_t=16^3 \times 4$ lattice.
We have identified the lines of constant physics and studied the
temperature-dependence of
various quantities at $m_{\rm PS}/m_{\rm V}=0.65$ and 0.80
in the range $T/T_{pc}\sim 0.76$--4.0.

 We found that, at $T \ge T_{pc}$,
  the free energies of $QQ$ and $Q \overline{Q}$ 
 normalized to be zero at large separation show attraction (repulsion)
in the color singlet and anti-triplet channels (color octet and sextet
channels).
We fitted the free energy data in each channel by the screened Coulomb
form, Eq.~(\ref{eq:SCP}), which consists of the Casimir factor, the effective
coupling $\alpha_{\rm eff}(T)$ and the Debye screening mass $m_D(T)$. 
We found that free energies in different channels can be fitted for
$T \simge 2 T_{pc}$ in terms of universal $\alpha_{\rm eff}(T)$ and
$m_D(T)$ while all the channel dependence can be absorbed by the
Casimir factor.
The magnitude and the $T$-dependence of
 $m_D(T)$ is consistent
 with the next-to-leading order calculation in thermal
 perturbation theory.
Moreover it is also well
 approximated by the leading order form
with an ``effective'' running coupling defined 
from  $\alpha_{\rm eff}(T)$.

 By comparing our results with the improved Wilson quark action and 
  those with the improved staggered quark action,
 we found that $\alpha_{\rm eff}(T)$ does not show appreciable difference
  while $m_D(T)$ in the Wilson quark action is 
 larger than that of the staggered quark action by 20\%.
To draw a definite conclusion, however, simulations with smaller
lattice spacings, i.e., larger lattice sizes in the temporal direction
(such as $N_t=6$ or larger) at smaller quark masses are required.

We also discussed the force between heavy quarks
to make a direct comparison of the free energies 
below and above $T_{pc}$.
In the color octet channel, we find
 that the force becomes attractive below $T_{pc}$
 against the simple Casimir scaling law.
This may be related to the suggestion that
the Polyakov loop correlation in the octet channel
is not independent from that in the singlet channel
in the low temperature limit \cite{Jahn:2004qr}.
Moreover,  spatial string tension $\sigma_s(T)$ in the high 
temperature phase was investigated.
The result of the spatial string tension  
in the quark-gluon plasma shows a behavior consistent with 
${\sqrt{\sigma_s(T)}} = c \, g_{\rm 2l}^2(2\pi T) \, T$ 
and agrees well
 with a parameter-free prediction of the NNLO three-dimensional 
 effective theory.

 Using our configurations,  
 we are currently studying
 the critical temperature ($T_{pc}$ in the chiral limit),
  the chiral and isospin susceptibility across the 
   phase transition, the effect of the chemical potential
    on the heavy-quark free energies,  and 
 the equation of state at finite temperature and density.
 Some of the results are 
  reported in Ref.~\cite{ejiri,maezawa}.

%%%%%%%%%%%%%%%%%%%%%%%%%%%%%%%%%%%%%%%%%%%%%%%%%%%%%%%%%%%%%%%%%%%%%%
\section*{Acknowledgements}
We would like to thank K.-I.~Ishikawa and the members of the CP-PACS 
Collaboration for providing us with the basic code 
for generating the configurations.
We also thank O.~Kaczmarek for providing us the data
from a staggered quark action.
We are grateful to the Yukawa Institute for Theoretical Physics 
at Kyoto University
for discussions during the YITP workshops YITP-W-06-07 and YKIS2006. 
This work is in part supported 
by Grants-in-Aid of the Japanese Ministry
of Education, Culture, Sports, Science and Technology, 
(Nos.~13135204, 15540251, 17340066, 18540253, 18740134). 
SE is supported by the Sumitomo Foundation (No.~050408),
and YM is supported by the Japan Society for the Promotion
of Science for Young Scientists. 
This work is in part supported 
also by the Large-Scale Numerical Simulation
Projects of ACCC, Univ.~of Tsukuba,
and by the Large Scale Simulation Program of High Energy
Accelerator Research Organization 
No.06-19 (FY2006) of KEK.
%%%%%%%%%%%%%%%%%%%%%%%%%%%%%%%%%%%%%%%%%%%%%%%%%%%%%%%%%%%%%%%%%%%%%%

%%%%%%%%%%%%%%%%%%%%%%%%%%%%%%%%%%%%%%%%%%%%%%%%%%%%%%%%%%%%%%%%%%%%%%
\appendix

\section{Fit range and systematic errors}
\label{sec:apa}

\begin{figure}[tbp]
  \begin{center}
    \begin{tabular}{cc}
    \includegraphics[width=80mm]{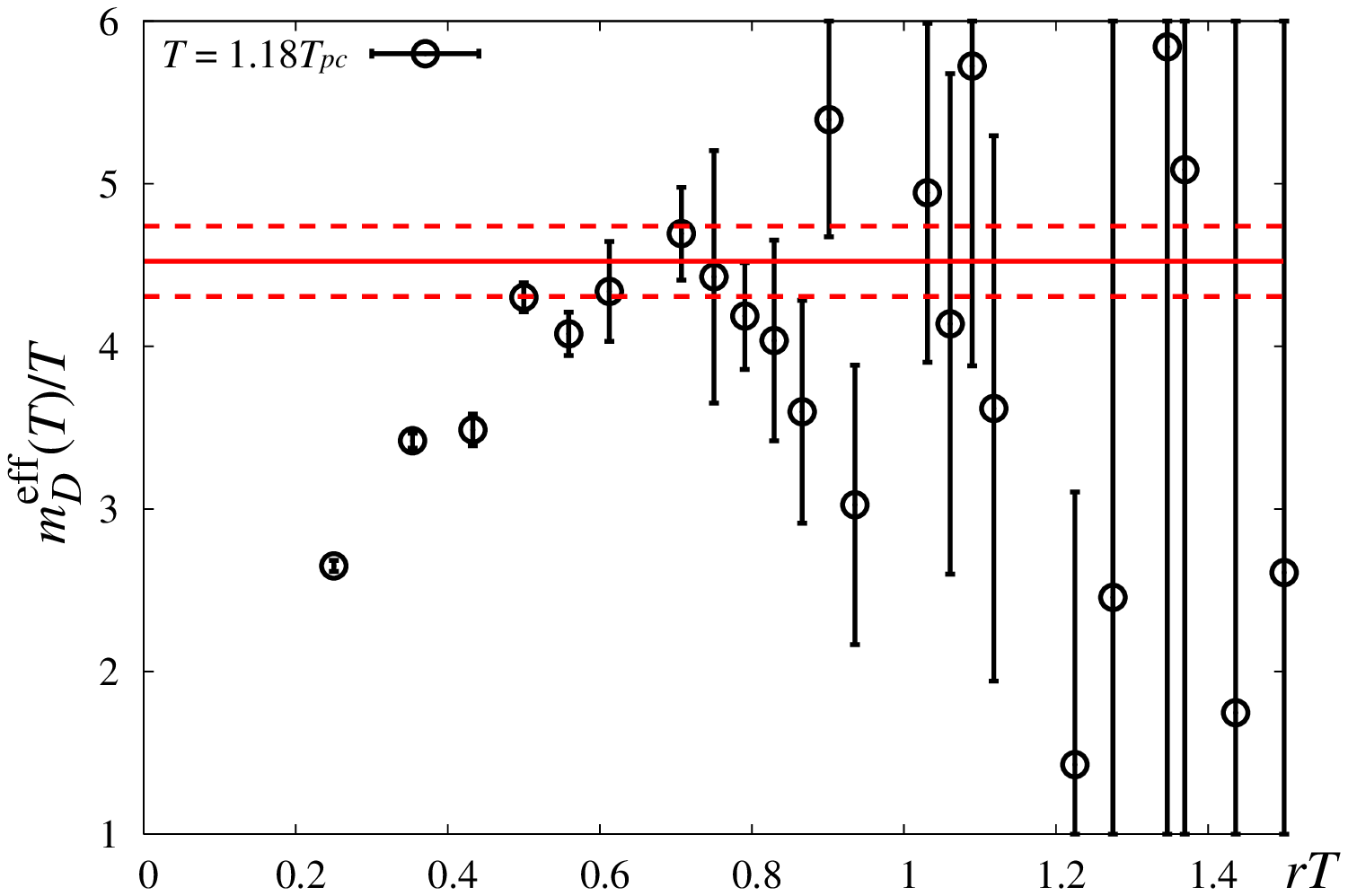} &
    \includegraphics[width=80mm]{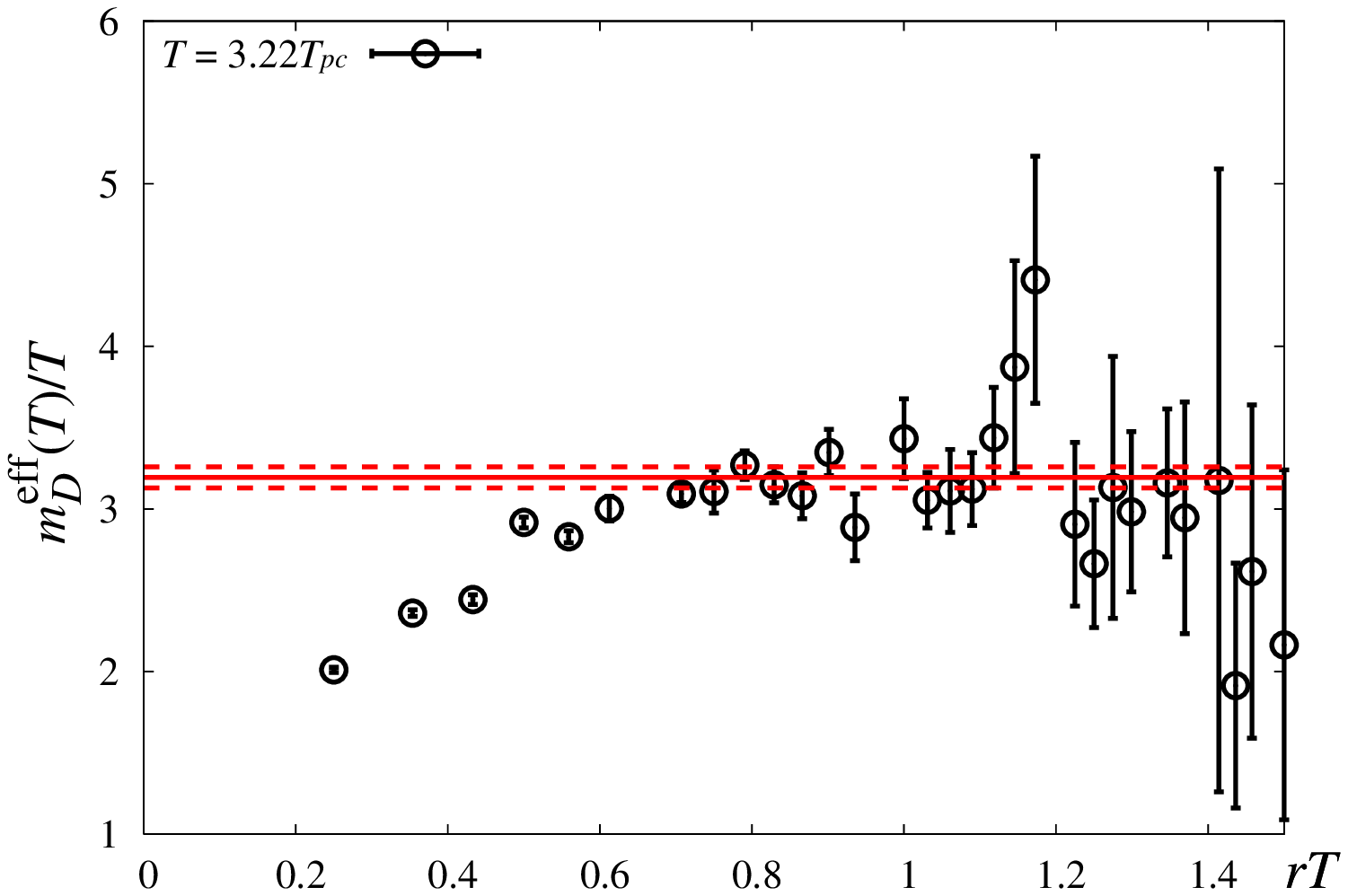}
    \end{tabular}
    \caption{The effective Debye mass defined by 
    a ratio of normalized free energies, Eq.~(\ref{eq:mDeff}),
    for color singlet channel at $T=1.18T_{pc}$ (left) 
    and $3.22 T_{pc}$ (right)
    for $m_{\rm PS}/m_{\rm V} = 0.65$. 
    The fit results (statistical errors) are also given 
    as solid lines (dashed lines).
        }
    \label{fig:mDeff}
  \end{center}
\end{figure}

In order to determine  the appropriate fit range
 for free energies by
the screened Coulomb form, Eq.~(\ref{eq:SCP}), 
we estimate the effective Debye mass
from the ratio of normalized free energies:
\begin{eqnarray}
 m_D^{\rm eff} (T;r) 
 = \frac{1}{\Delta r} \log \frac{V_M(r)}{V_M(r+ \Delta r)}
 - \frac{1}{\Delta r} \log \left[ 1 + \frac{\Delta r}{r} \right]
.
\label{eq:mDeff}
\end{eqnarray}
The effective Debye mass is  shown in Fig.~\ref{fig:mDeff}
for color singlet channel at $T=1.18T_{pc}$ (left) 
and $3.22 T_{pc}$ (right)
for $m_{\rm PS}/m_{\rm V} = 0.65$.
Since the plateaus of $m_D^{\rm eff} (T;r)$ are found 
at $rT > 0.8$,
 we choose the fit range to be $ \sqrt{11}/4 \le rT \le 1.5$.
 The solid lines in Fig.~\ref{fig:mDeff} are the results
  of the fit with the statistical errors indicated by the dashed lines.

\begin{figure}[tbp]
  \begin{center}
    \begin{tabular}{cc}
    \includegraphics[width=80mm]{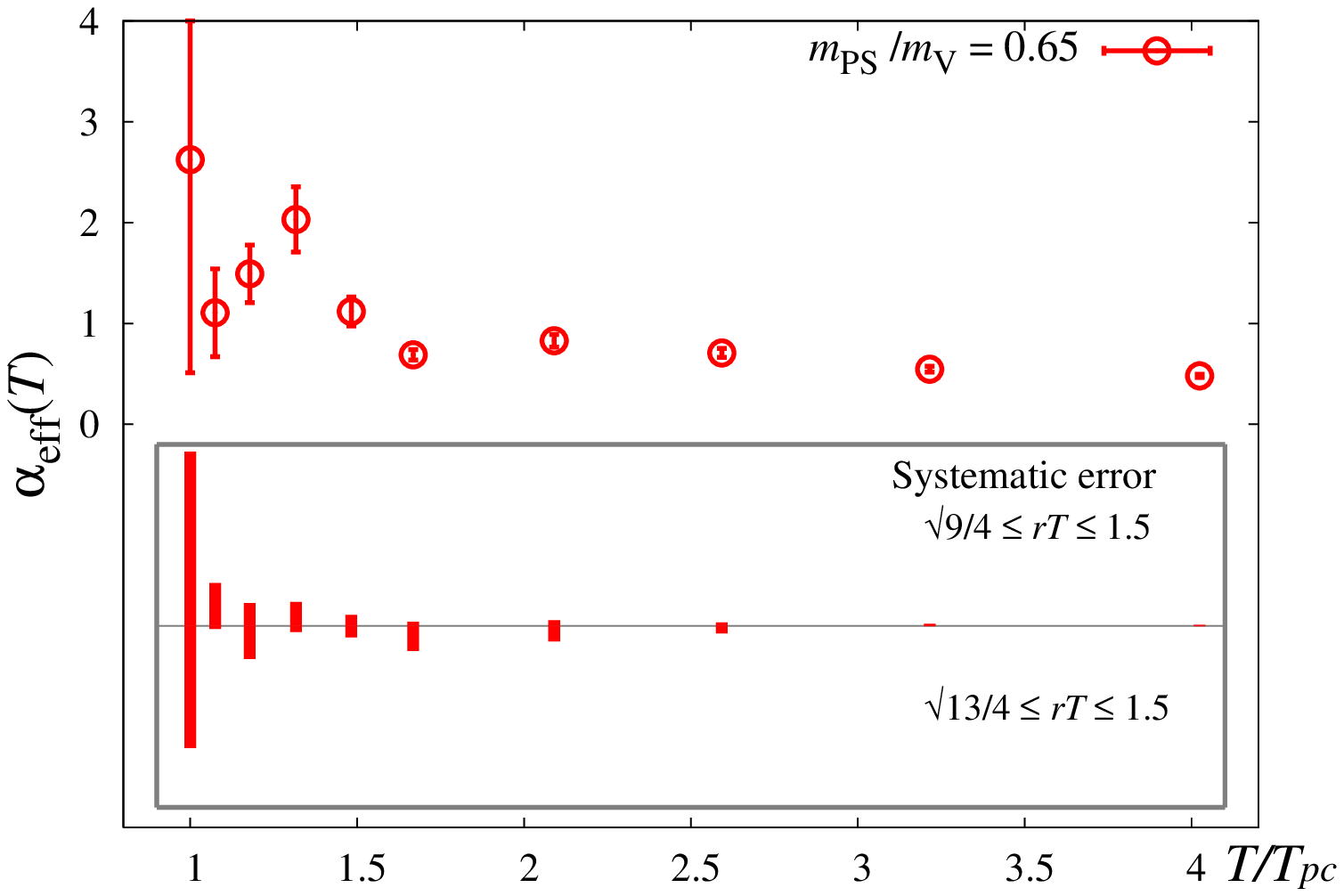} &
    \includegraphics[width=80mm]{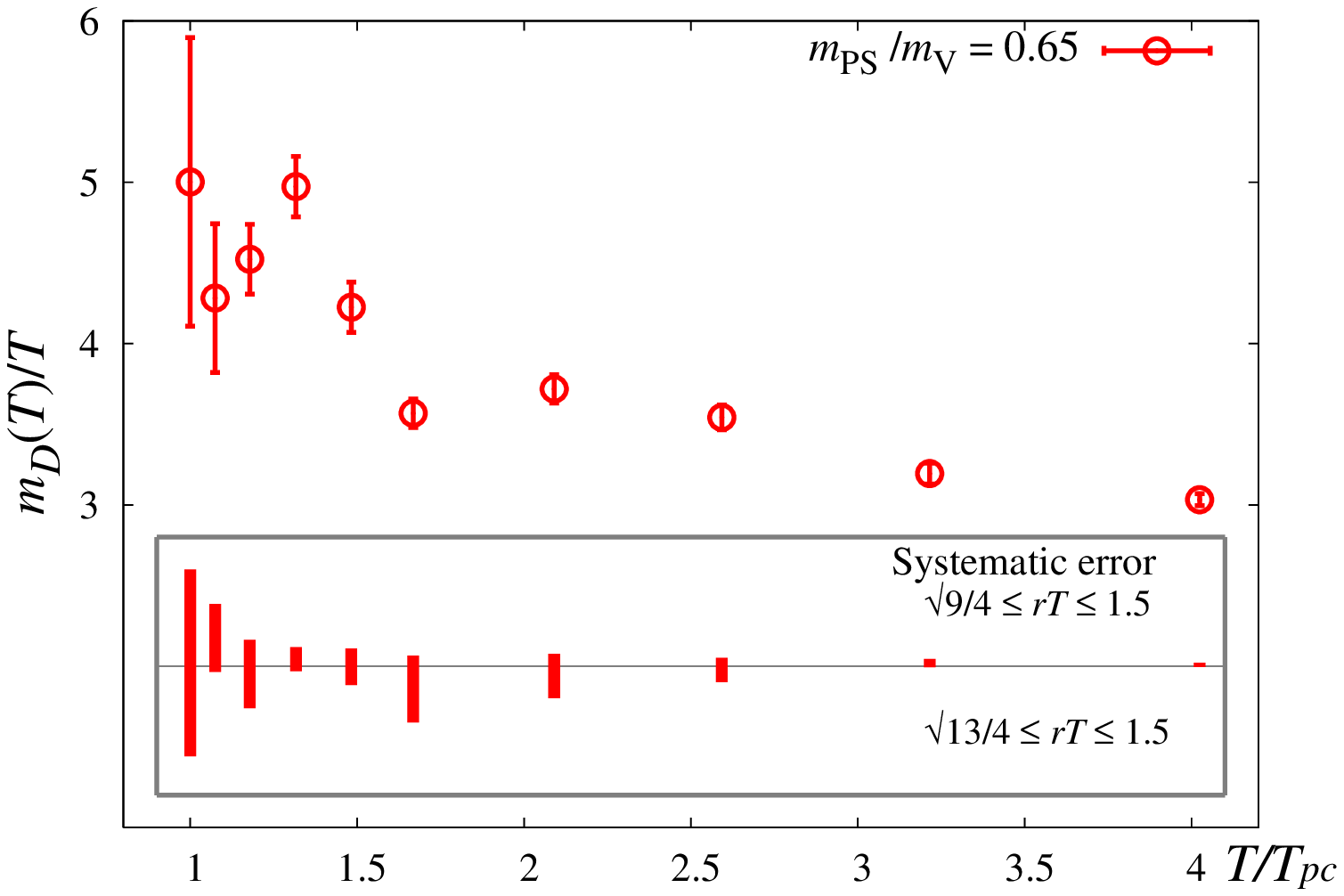}
    \end{tabular}
    \caption{The effective running coupling (left) and
    Debye screening mass (right) for color singlet channel
    with the systematic errors due to the difference of 
    the fit range defined in text
    at $m_{\rm PS}/m_{\rm V} = 0.65$.
        }
    \label{fig:sys065}
  \end{center}
\end{figure}

\begin{figure}[tbp]
  \begin{center}
    \begin{tabular}{cc}
    \includegraphics[width=80mm]{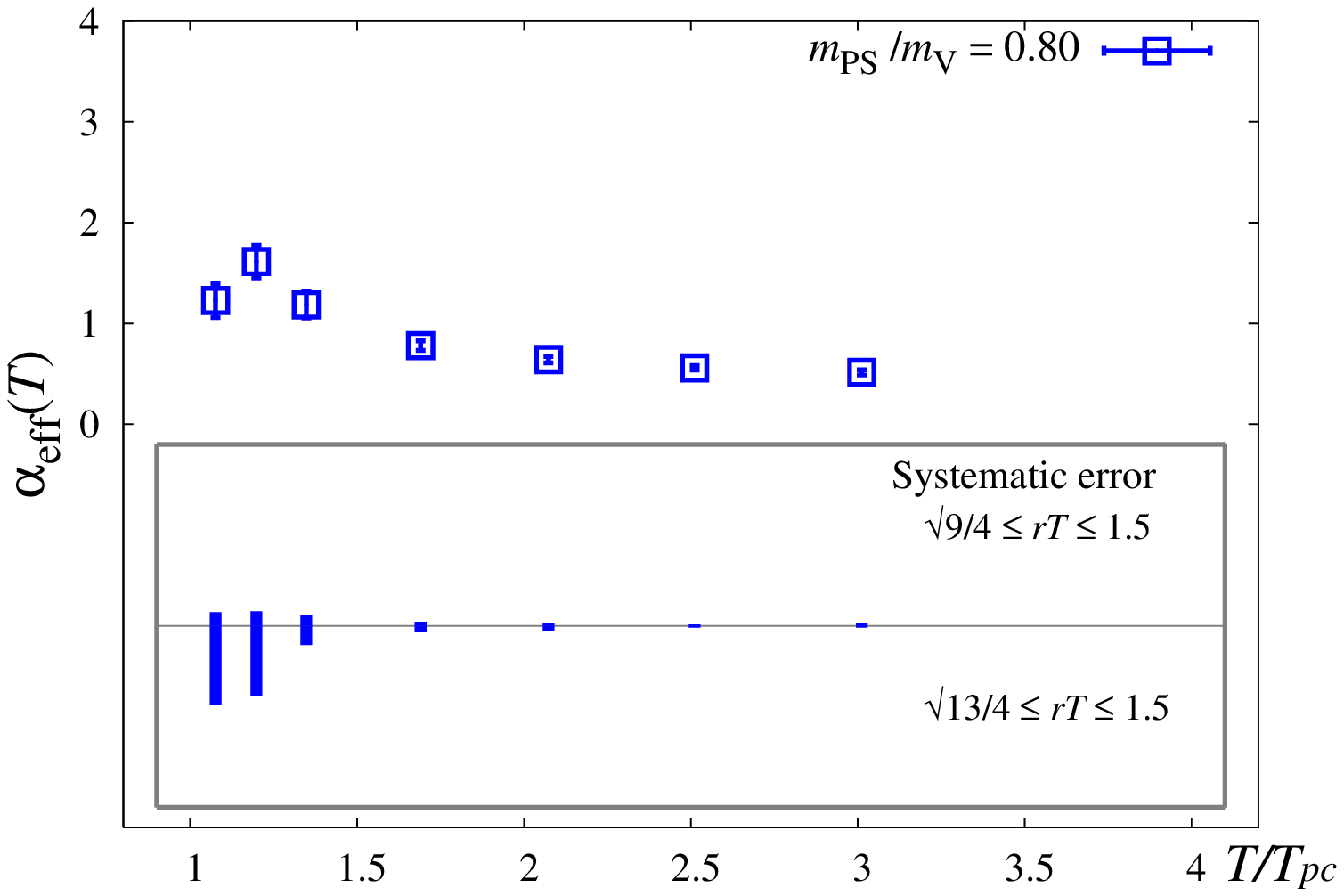} &
    \includegraphics[width=80mm]{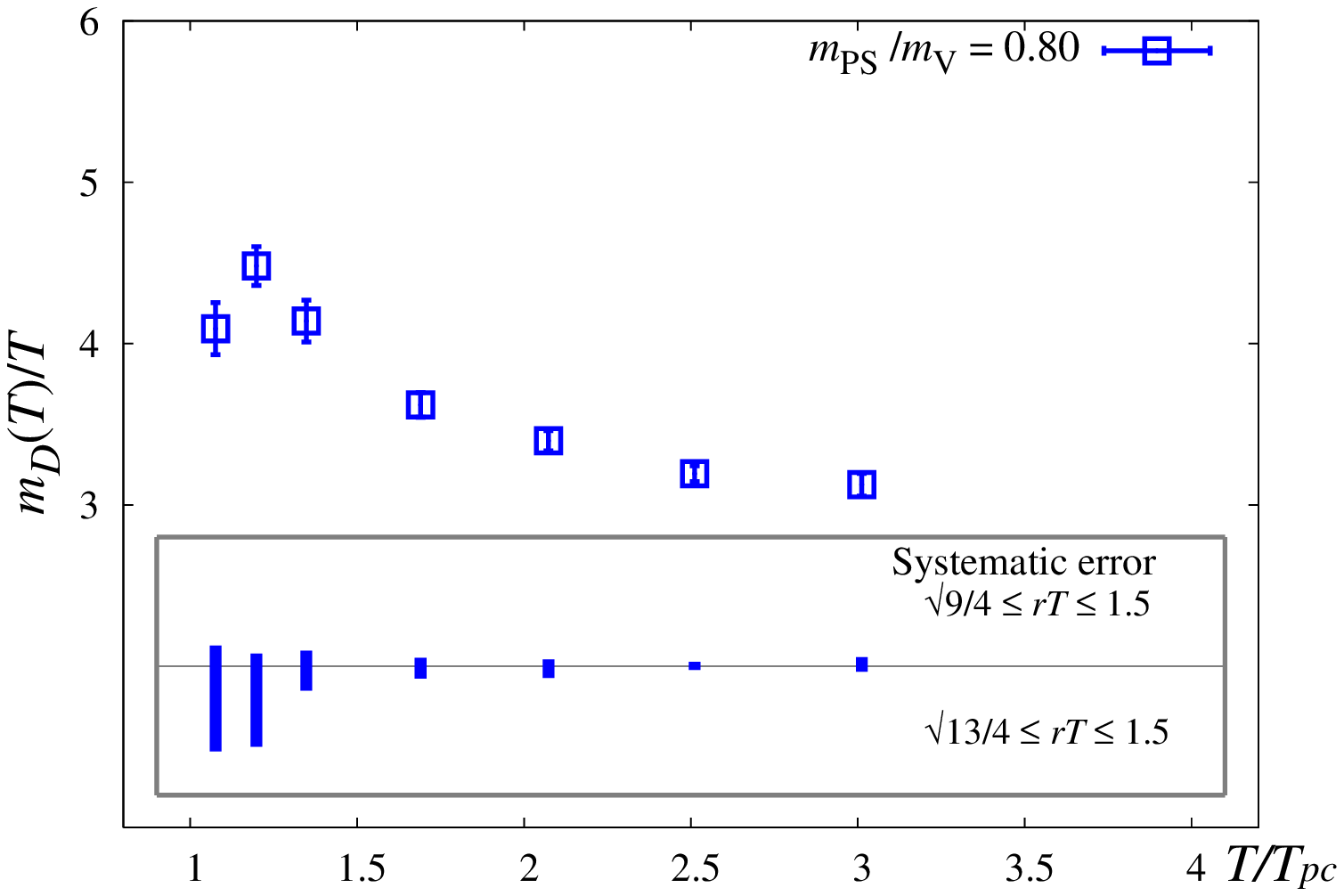}
    \end{tabular}
    \caption{The same figures with Fit.~\ref{fig:sys065}
    at $m_{\rm PS}/m_{\rm V} = 0.80$
        }
    \label{fig:sys080}
  \end{center}
\end{figure}

In order to investigate  
systematic errors due to  a dependence on fit ranges,
we consider two other fit ranges:
 $ \sqrt{9}/4 \le rT \le 1.5$
and $ \sqrt{13}/4 \le rT \le 1.5$. 
The systematic errors due to the 
 difference of the lower end of the fit range
 are shown in Fig.~\ref{fig:sys065}
for $\alpha_{\rm eff}(T)$ (left) and $m_D(T)$ (right)
in the  color singlet channel at $m_{\rm PS}/m_{\rm V} = 0.65$.
The similar plots at $m_{\rm PS}/m_{\rm V} = 0.80$ are shown
in Fig.~\ref{fig:sys080}.
We find that, for $T \simle 2T_{pc}$, 
the systematic errors amount to at most the twice of the statistical errors,
i.e. $2 \sigma$.
On the other hand, for $T \simge 2T_{pc}$,
the systematic errors are less than 10\% 
of the statistical errors.

%%%%%%%%%%%%%%%%%%%%%%%%%%%%%%%%%%%%%%%%%%%%%%%%%%%%%%%%%%%%%%%%%%%%%%
\clearpage
\section{Data lists of normalized free energies}
\label{sec:apb}

In this Appendix, we give data of the normalized free energies
for all color channels and temperatures above $T_{pc}$ 
as a function of distance ($rT$).
The normalized free energies are measured at every ten trajectories
for each quark mass and temperature summarized in Tab.~\ref{tab:parameter}.
The values in parenthesis of the normalized free energies 
express the statistical errors determined by a jackknife method
with the bin-size of 100 trajectories.

 \begin{table}[htbp]
  \begin{center}
  \caption{Data lists of normalized free energies for all color channels at $T=
 1.00
 T_{pc}$ (left) and $
 1.07
 T_{pc}$ (right) for $m_{\rm PS}/m_{\rm V}=0.65$
 as a function of $rT$}
 \label{tab:NFE065_1}
 {\renewcommand{\arraystretch}{1.2} \tabcolsep = 1mm
 \newcolumntype{.}{D{.}{.}{8}}
\begin{tabular}{|c|....|....|}
 \hline
 \multicolumn{1}{|l|}{} &
 \multicolumn{4}{c|}{$T= 1.00T_{pc}$, $m_{\rm PS} / m_{\rm V} = 0.65$}
  & 
 \multicolumn{4}{c|}{$T= 1.07T_{pc}$, $m_{\rm PS} / m_{\rm V} = 0.65$}
 \\ \hline
 \multicolumn{1}{|c|}{$rT$} &
 \multicolumn{1}{c} {$V_{\bf 1}$} & 
 \multicolumn{1}{c} {$V_{\bf 8}$} & 
 \multicolumn{1}{c} {$V_{\bf 6}$} & 
 \multicolumn{1}{c|}{$V_{{\bf 3}^*}$} & 
 \multicolumn{1}{c} {$V_{\bf 1}$} & 
 \multicolumn{1}{c} {$V_{\bf 8}$} & 
 \multicolumn{1}{c} {$V_{\bf 6}$} & 
 \multicolumn{1}{c|}{$V_{{\bf 3}^*}$} \\
 \hline
  0.250 &    -2.3976( 132)  &    -0.1786(  59)  &     0.0466(  78)  &    -1.0760(  69)  &    -1.6513(  71)  &     0.0337(  27)  &     0.1796(  31)  &    -0.7895(  34)  \\
  0.354 &    -1.2753( 110)  &    -0.1199(  51)  &    -0.0228(  56)  &    -0.5567(  58)  &    -0.8627(  51)  &     0.0041(  18)  &     0.0783(  24)  &    -0.4158(  28)  \\
  0.433 &    -0.7376( 104)  &    -0.0931(  60)  &    -0.0390(  75)  &    -0.3389(  72)  &    -0.5205(  46)  &    -0.0013(  22)  &     0.0430(  28)  &    -0.2525(  27)  \\
  0.500 &    -0.5979(  95)  &    -0.0956(  74)  &    -0.0471(  90)  &    -0.2766(  90)  &    -0.4137(  52)  &    -0.0085(  25)  &     0.0251(  26)  &    -0.2042(  28)  \\
  0.559 &    -0.3640(  72)  &    -0.0618(  39)  &    -0.0352(  42)  &    -0.1726(  54)  &    -0.2728(  33)  &    -0.0077(  11)  &     0.0152(  13)  &    -0.1388(  19)  \\
  0.612 &    -0.2347(  64)  &    -0.0503(  38)  &    -0.0384(  36)  &    -0.1287(  54)  &    -0.1884(  32)  &    -0.0087(  13)  &     0.0076(  15)  &    -0.0997(  21)  \\
\hline
  \end{tabular}}
  \end{center}
 \end{table}
 
 \begin{table}[htbp]
  \begin{center}
  \caption{Data lists of normalized free energies for all color channels at $T=
 1.18
 T_{pc}$ (left) and $
 1.32
 T_{pc}$ (right) for $m_{\rm PS}/m_{\rm V}=0.65$
 as a function of $rT$}
 \label{tab:NFE065_2}
 {\renewcommand{\arraystretch}{1.2} \tabcolsep = 1mm
 \newcolumntype{.}{D{.}{.}{8}}
 \begin{tabular}{|c|....|....|}
 \hline
 \multicolumn{1}{|l|}{} &
 \multicolumn{4}{c|}{$T= 1.18T_{pc}$, $m_{\rm PS} / m_{\rm V} = 0.65$}
  & 
 \multicolumn{4}{c|}{$T= 1.32T_{pc}$, $m_{\rm PS} / m_{\rm V} = 0.65$}
 \\ \hline
 \multicolumn{1}{|c|}{$rT$} &
 \multicolumn{1}{c} {$V_{\bf 1}$} & 
 \multicolumn{1}{c} {$V_{\bf 8}$} & 
 \multicolumn{1}{c} {$V_{\bf 6}$} & 
 \multicolumn{1}{c|}{$V_{{\bf 3}^*}$} & 
 \multicolumn{1}{c} {$V_{\bf 1}$} & 
 \multicolumn{1}{c} {$V_{\bf 8}$} & 
 \multicolumn{1}{c} {$V_{\bf 6}$} & 
 \multicolumn{1}{c|}{$V_{{\bf 3}^*}$} \\
 \hline
  0.250 &    -1.3844(  36)  &     0.0866(  16)  &     0.2260(  22)  &    -0.6711(  18)  &    -1.2727(  31)  &     0.1011(  11)  &     0.2358(  17)  &    -0.6208(  17)  \\
  0.354 &    -0.7238(  28)  &     0.0392(   9)  &     0.1103(  14)  &    -0.3518(  15)  &    -0.6704(  25)  &     0.0498(   9)  &     0.1204(  11)  &    -0.3280(  16)  \\
  0.433 &    -0.4460(  27)  &     0.0220(  12)  &     0.0677(  16)  &    -0.2183(  17)  &    -0.4185(  22)  &     0.0303(   9)  &     0.0755(  12)  &    -0.2075(  15)  \\
  0.500 &    -0.3569(  32)  &     0.0177(  14)  &     0.0497(  15)  &    -0.1749(  20)  &    -0.3363(  28)  &     0.0229(  12)  &     0.0586(  13)  &    -0.1649(  16)  \\
  0.559 &    -0.2411(  22)  &     0.0109(   8)  &     0.0351(   8)  &    -0.1196(  13)  &    -0.2291(  17)  &     0.0165(   6)  &     0.0415(   6)  &    -0.1128(  10)  \\
  0.612 &    -0.1725(  22)  &     0.0083(   8)  &     0.0266(  10)  &    -0.0861(  12)  &    -0.1664(  17)  &     0.0123(   5)  &     0.0300(   8)  &    -0.0825(  11)  \\
\hline
  \end{tabular}}
  \end{center}
 \end{table}
 
 \begin{table}[htbp]
  \begin{center}
  \caption{Data lists of normalized free energies for all color channels at $T=
 1.48
 T_{pc}$ (left) and $
 1.67
 T_{pc}$ (right) for $m_{\rm PS}/m_{\rm V}=0.65$
 as a function of $rT$}
 \label{tab:NFE065_3}
 {\renewcommand{\arraystretch}{1.2} \tabcolsep = 1mm
 \newcolumntype{.}{D{.}{.}{8}}
 \begin{tabular}{|c|....|....|}
 \hline
 \multicolumn{1}{|l|}{} &
 \multicolumn{4}{c|}{$T= 1.48T_{pc}$, $m_{\rm PS} / m_{\rm V} = 0.65$}
  & 
 \multicolumn{4}{c|}{$T= 1.67T_{pc}$, $m_{\rm PS} / m_{\rm V} = 0.65$}
 \\ \hline
 \multicolumn{1}{|c|}{$rT$} &
 \multicolumn{1}{c} {$V_{\bf 1}$} & 
 \multicolumn{1}{c} {$V_{\bf 8}$} & 
 \multicolumn{1}{c} {$V_{\bf 6}$} & 
 \multicolumn{1}{c|}{$V_{{\bf 3}^*}$} & 
 \multicolumn{1}{c} {$V_{\bf 1}$} & 
 \multicolumn{1}{c} {$V_{\bf 8}$} & 
 \multicolumn{1}{c} {$V_{\bf 6}$} & 
 \multicolumn{1}{c|}{$V_{{\bf 3}^*}$} \\
 \hline
  0.250 &    -1.1929(  27)  &     0.1064(  12)  &     0.2382(  16)  &    -0.5850(  14)  &    -1.1278(  24)  &     0.1107(   9)  &     0.2366(  12)  &    -0.5522(  12)  \\
  0.354 &    -0.6369(  22)  &     0.0549(   7)  &     0.1237(  10)  &    -0.3118(  12)  &    -0.6082(  22)  &     0.0580(   6)  &     0.1257(   7)  &    -0.2983(  11)  \\
  0.433 &    -0.4032(  25)  &     0.0350(   7)  &     0.0781(   9)  &    -0.1979(  12)  &    -0.3905(  20)  &     0.0374(   6)  &     0.0813(  10)  &    -0.1925(  11)  \\
  0.500 &    -0.3218(  24)  &     0.0277(   8)  &     0.0643(   9)  &    -0.1595(  15)  &    -0.3144(  22)  &     0.0288(   6)  &     0.0643(   8)  &    -0.1554(  13)  \\
  0.559 &    -0.2268(  19)  &     0.0192(   4)  &     0.0437(   6)  &    -0.1110(  10)  &    -0.2227(  17)  &     0.0214(   4)  &     0.0471(   5)  &    -0.1105(   9)  \\
  0.612 &    -0.1644(  17)  &     0.0144(   5)  &     0.0318(   7)  &    -0.0817(   9)  &    -0.1665(  14)  &     0.0162(   4)  &     0.0348(   5)  &    -0.0828(   8)  \\
\hline
  \end{tabular}}
  \end{center}
 \end{table}
 
 \begin{table}[htbp]
  \begin{center}
  \caption{Data lists of normalized free energies for all color channels at $T=
 2.09
 T_{pc}$ (left) and $
 2.59
 T_{pc}$ (right) for $m_{\rm PS}/m_{\rm V}=0.65$
 as a function of $rT$}
 \label{tab:NFE065_4}
 {\renewcommand{\arraystretch}{1.2} \tabcolsep = 1mm
 \newcolumntype{.}{D{.}{.}{8}}
 \begin{tabular}{|c|....|....|}
 \hline
 \multicolumn{1}{|l|}{} &
 \multicolumn{4}{c|}{$T= 2.09T_{pc}$, $m_{\rm PS} / m_{\rm V} = 0.65$}
  & 
 \multicolumn{4}{c|}{$T= 2.59T_{pc}$, $m_{\rm PS} / m_{\rm V} = 0.65$}
 \\ \hline
 \multicolumn{1}{|c|}{$rT$} &
 \multicolumn{1}{c} {$V_{\bf 1}$} & 
 \multicolumn{1}{c} {$V_{\bf 8}$} & 
 \multicolumn{1}{c} {$V_{\bf 6}$} & 
 \multicolumn{1}{c|}{$V_{{\bf 3}^*}$} & 
 \multicolumn{1}{c} {$V_{\bf 1}$} & 
 \multicolumn{1}{c} {$V_{\bf 8}$} & 
 \multicolumn{1}{c} {$V_{\bf 6}$} & 
 \multicolumn{1}{c|}{$V_{{\bf 3}^*}$} \\
 \hline
  0.250 &    -1.0261(  19)  &     0.1090(   6)  &     0.2275(  10)  &    -0.5062(  10)  &    -0.9461(  18)  &     0.1062(   4)  &     0.2181(   7)  &    -0.4681(   9)  \\
  0.354 &    -0.5610(  16)  &     0.0585(   4)  &     0.1228(   6)  &    -0.2767(   8)  &    -0.5241(  15)  &     0.0574(   3)  &     0.1190(   5)  &    -0.2592(   8)  \\
  0.433 &    -0.3661(  16)  &     0.0386(   4)  &     0.0803(   6)  &    -0.1808(   8)  &    -0.3467(  13)  &     0.0381(   3)  &     0.0789(   5)  &    -0.1724(   7)  \\
  0.500 &    -0.2964(  15)  &     0.0302(   5)  &     0.0634(   7)  &    -0.1457(   8)  &    -0.2781(  16)  &     0.0309(   4)  &     0.0634(   5)  &    -0.1370(   8)  \\
  0.559 &    -0.2132(  12)  &     0.0218(   3)  &     0.0468(   4)  &    -0.1055(   6)  &    -0.2044(  11)  &     0.0223(   2)  &     0.0466(   4)  &    -0.1013(   6)  \\
  0.612 &    -0.1607(  12)  &     0.0166(   3)  &     0.0349(   3)  &    -0.0801(   6)  &    -0.1562(  11)  &     0.0171(   2)  &     0.0356(   3)  &    -0.0771(   5)  \\
\hline
  \end{tabular}}
  \end{center}
 \end{table}
 
 \begin{table}[htbp]
  \begin{center}
  \caption{Data lists of normalized free energies for all color channels at $T=
 3.22
 T_{pc}$ (left) and $
 4.02
 T_{pc}$ (right) for $m_{\rm PS}/m_{\rm V}=0.65$
 as a function of $rT$}
 \label{tab:NFE065_5}
 {\renewcommand{\arraystretch}{1.2} \tabcolsep = 1mm
 \newcolumntype{.}{D{.}{.}{8}}
 \begin{tabular}{|c|....|....|}
 \hline
 \multicolumn{1}{|l|}{} &
 \multicolumn{4}{c|}{$T= 3.22T_{pc}$, $m_{\rm PS} / m_{\rm V} = 0.65$}
  & 
 \multicolumn{4}{c|}{$T= 4.02T_{pc}$, $m_{\rm PS} / m_{\rm V} = 0.65$}
 \\ \hline
 \multicolumn{1}{|c|}{$rT$} &
 \multicolumn{1}{c} {$V_{\bf 1}$} & 
 \multicolumn{1}{c} {$V_{\bf 8}$} & 
 \multicolumn{1}{c} {$V_{\bf 6}$} & 
 \multicolumn{1}{c|}{$V_{{\bf 3}^*}$} & 
 \multicolumn{1}{c} {$V_{\bf 1}$} & 
 \multicolumn{1}{c} {$V_{\bf 8}$} & 
 \multicolumn{1}{c} {$V_{\bf 6}$} & 
 \multicolumn{1}{c|}{$V_{{\bf 3}^*}$} \\
 \hline
  0.250 &    -0.8865(  16)  &     0.1026(   5)  &     0.2090(   7)  &    -0.4398(   8)  &    -0.8365(  16)  &     0.0986(   4)  &     0.1994(   6)  &    -0.4159(   8)  \\
  0.354 &    -0.4962(  15)  &     0.0566(   3)  &     0.1158(   5)  &    -0.2461(   8)  &    -0.4731(  15)  &     0.0547(   3)  &     0.1116(   4)  &    -0.2352(   7)  \\
  0.433 &    -0.3332(  15)  &     0.0380(   3)  &     0.0773(   4)  &    -0.1649(   8)  &    -0.3194(  15)  &     0.0362(   2)  &     0.0749(   4)  &    -0.1591(   7)  \\
  0.500 &    -0.2681(  12)  &     0.0303(   2)  &     0.0622(   4)  &    -0.1329(   7)  &    -0.2579(  12)  &     0.0294(   3)  &     0.0602(   4)  &    -0.1280(   6)  \\
  0.559 &    -0.2000(  10)  &     0.0225(   2)  &     0.0463(   3)  &    -0.0993(   5)  &    -0.1937(  10)  &     0.0224(   1)  &     0.0454(   3)  &    -0.0961(   5)  \\
  0.612 &    -0.1556(  11)  &     0.0175(   2)  &     0.0360(   2)  &    -0.0772(   5)  &    -0.1518(  10)  &     0.0172(   2)  &     0.0353(   3)  &    -0.0752(   5)  \\
\hline
  \end{tabular}}
  \end{center}
 \end{table}
 
 \begin{table}[htbp]
  \begin{center}
  \caption{Data lists of normalized free energies for all color channels at $T=
 1.08
 T_{pc}$ (left) and $
 1.20
 T_{pc}$ (right) for $m_{\rm PS}/m_{\rm V}=0.80$
 as a function of $rT$}
 \label{tab:NFE080_1}
 {\renewcommand{\arraystretch}{1.2} \tabcolsep = 1mm
 \newcolumntype{.}{D{.}{.}{8}}
 \begin{tabular}{|c|....|....|}
 \hline
 \multicolumn{1}{|l|}{} &
 \multicolumn{4}{c|}{$T= 1.08T_{pc}$, $m_{\rm PS} / m_{\rm V} = 0.80$}
  & 
 \multicolumn{4}{c|}{$T= 1.20T_{pc}$, $m_{\rm PS} / m_{\rm V} = 0.80$}
 \\ \hline
 \multicolumn{1}{|c|}{$rT$} &
 \multicolumn{1}{c} {$V_{\bf 1}$} & 
 \multicolumn{1}{c} {$V_{\bf 8}$} & 
 \multicolumn{1}{c} {$V_{\bf 6}$} & 
 \multicolumn{1}{c|}{$V_{{\bf 3}^*}$} & 
 \multicolumn{1}{c} {$V_{\bf 1}$} & 
 \multicolumn{1}{c} {$V_{\bf 8}$} & 
 \multicolumn{1}{c} {$V_{\bf 6}$} & 
 \multicolumn{1}{c|}{$V_{{\bf 3}^*}$} \\
 \hline
  0.250 &    -1.5314(  50)  &     0.0740(  18)  &     0.2251(  23)  &    -0.7414(  28)  &    -1.3096(  35)  &     0.1048(  12)  &     0.2435(  15)  &    -0.6396(  18)  \\
  0.354 &    -0.8417(  43)  &     0.0276(  14)  &     0.1075(  16)  &    -0.4103(  25)  &    -0.7118(  31)  &     0.0524(   9)  &     0.1263(  10)  &    -0.3491(  15)  \\
  0.433 &    -0.5350(  35)  &     0.0112(  14)  &     0.0618(  15)  &    -0.2644(  23)  &    -0.4546(  25)  &     0.0322(   8)  &     0.0795(  10)  &    -0.2238(  13)  \\
  0.500 &    -0.4250(  36)  &     0.0081(  14)  &     0.0469(  18)  &    -0.2128(  19)  &    -0.3705(  25)  &     0.0250(  10)  &     0.0644(  13)  &    -0.1813(  16)  \\
  0.559 &    -0.2980(  26)  &     0.0042(   9)  &     0.0298(  10)  &    -0.1484(  17)  &    -0.2579(  20)  &     0.0170(   7)  &     0.0436(   7)  &    -0.1277(  11)  \\
  0.612 &    -0.2145(  24)  &     0.0001(   8)  &     0.0199(   9)  &    -0.1087(  15)  &    -0.1898(  18)  &     0.0124(   5)  &     0.0317(   6)  &    -0.0950(  10)  \\
\hline
  \end{tabular}}
  \end{center}
 \end{table}
 
 \begin{table}[htbp]
  \begin{center}
  \caption{Data lists of normalized free energies for all color channels at $T=
 1.35
 T_{pc}$ (left) and $
 1.69
 T_{pc}$ (right) for $m_{\rm PS}/m_{\rm V}=0.80$
 as a function of $rT$}
 \label{tab:NFE080_2}
 {\renewcommand{\arraystretch}{1.2} \tabcolsep = 1mm
 \newcolumntype{.}{D{.}{.}{8}}
 \begin{tabular}{|c|....|....|}
 \hline
 \multicolumn{1}{|l|}{} &
 \multicolumn{4}{c|}{$T= 1.35T_{pc}$, $m_{\rm PS} / m_{\rm V} = 0.80$}
  & 
 \multicolumn{4}{c|}{$T= 1.69T_{pc}$, $m_{\rm PS} / m_{\rm V} = 0.80$}
 \\ \hline
 \multicolumn{1}{|c|}{$rT$} &
 \multicolumn{1}{c} {$V_{\bf 1}$} & 
 \multicolumn{1}{c} {$V_{\bf 8}$} & 
 \multicolumn{1}{c} {$V_{\bf 6}$} & 
 \multicolumn{1}{c|}{$V_{{\bf 3}^*}$} & 
 \multicolumn{1}{c} {$V_{\bf 1}$} & 
 \multicolumn{1}{c} {$V_{\bf 8}$} & 
 \multicolumn{1}{c} {$V_{\bf 6}$} & 
 \multicolumn{1}{c|}{$V_{{\bf 3}^*}$} \\
 \hline
  0.250 &    -1.1850(  27)  &     0.1111(  11)  &     0.2410(  14)  &    -0.5807(  14)  &    -1.0540(  23)  &     0.1117(   7)  &     0.2324(  11)  &    -0.5205(  13)  \\
  0.354 &    -0.6453(  21)  &     0.0573(   7)  &     0.1282(  10)  &    -0.3159(  11)  &    -0.5794(  20)  &     0.0602(   5)  &     0.1268(   7)  &    -0.2858(  10)  \\
  0.433 &    -0.4158(  22)  &     0.0378(   8)  &     0.0830(  10)  &    -0.2033(  10)  &    -0.3787(  19)  &     0.0389(   5)  &     0.0824(   7)  &    -0.1869(  11)  \\
  0.500 &    -0.3392(  22)  &     0.0299(   7)  &     0.0660(   7)  &    -0.1657(  12)  &    -0.3048(  18)  &     0.0312(   5)  &     0.0661(   6)  &    -0.1504(   9)  \\
  0.559 &    -0.2379(  15)  &     0.0209(   5)  &     0.0471(   6)  &    -0.1172(   7)  &    -0.2213(  17)  &     0.0226(   3)  &     0.0481(   4)  &    -0.1095(   8)  \\
  0.612 &    -0.1767(  15)  &     0.0160(   4)  &     0.0361(   5)  &    -0.0873(   8)  &    -0.1674(  14)  &     0.0172(   3)  &     0.0365(   4)  &    -0.0827(   7)  \\
\hline
  \end{tabular}}
  \end{center}
 \end{table}
 
 \begin{table}[htbp]
  \begin{center}
  \caption{Data lists of normalized free energies for all color channels at $T=
 2.07
 T_{pc}$ (left) and $
 2.51
 T_{pc}$ (right) for $m_{\rm PS}/m_{\rm V}=0.80$
 as a function of $rT$}
 \label{tab:NFE080_3}
 {\renewcommand{\arraystretch}{1.2} \tabcolsep = 1mm
 \newcolumntype{.}{D{.}{.}{8}}
 \begin{tabular}{|c|....|....|}
 \hline
 \multicolumn{1}{|l|}{} &
 \multicolumn{4}{c|}{$T= 2.07T_{pc}$, $m_{\rm PS} / m_{\rm V} = 0.80$}
  & 
 \multicolumn{4}{c|}{$T= 2.51T_{pc}$, $m_{\rm PS} / m_{\rm V} = 0.80$}
 \\ \hline
 \multicolumn{1}{|c|}{$rT$} &
 \multicolumn{1}{c} {$V_{\bf 1}$} & 
 \multicolumn{1}{c} {$V_{\bf 8}$} & 
 \multicolumn{1}{c} {$V_{\bf 6}$} & 
 \multicolumn{1}{c|}{$V_{{\bf 3}^*}$} & 
 \multicolumn{1}{c} {$V_{\bf 1}$} & 
 \multicolumn{1}{c} {$V_{\bf 8}$} & 
 \multicolumn{1}{c} {$V_{\bf 6}$} & 
 \multicolumn{1}{c|}{$V_{{\bf 3}^*}$} \\
 \hline
  0.250 &    -0.9651(  22)  &     0.1079(   6)  &     0.2230(   9)  &    -0.4778(  11)  &    -0.8995(  18)  &     0.1029(   4)  &     0.2105(   6)  &    -0.4463(   9)  \\
  0.354 &    -0.5365(  19)  &     0.0589(   4)  &     0.1218(   6)  &    -0.2650(   9)  &    -0.5053(  16)  &     0.0571(   3)  &     0.1173(   5)  &    -0.2503(   8)  \\
  0.433 &    -0.3561(  19)  &     0.0387(   5)  &     0.0814(   6)  &    -0.1766(   9)  &    -0.3385(  14)  &     0.0384(   3)  &     0.0787(   4)  &    -0.1674(   7)  \\
  0.500 &    -0.2864(  19)  &     0.0308(   4)  &     0.0640(   6)  &    -0.1415(   9)  &    -0.2739(  12)  &     0.0309(   3)  &     0.0635(   5)  &    -0.1355(   6)  \\
  0.559 &    -0.2114(  16)  &     0.0229(   3)  &     0.0477(   4)  &    -0.1044(   7)  &    -0.2049(  11)  &     0.0231(   2)  &     0.0475(   3)  &    -0.1012(   5)  \\
  0.612 &    -0.1630(  14)  &     0.0176(   2)  &     0.0367(   4)  &    -0.0809(   7)  &    -0.1581(  10)  &     0.0180(   2)  &     0.0369(   3)  &    -0.0784(   5)  \\
\hline
  \end{tabular}}
  \end{center}
 \end{table}
 
 \begin{table}[htbp]
  \begin{center}
  \caption{Data lists of normalized free energies for all color channels at $T=
 3.01
 T_{pc}$ for $m_{\rm PS}/m_{\rm V}=0.80$
 as a function of $rT$}
 \label{tab:NFE080_4}
 {\renewcommand{\arraystretch}{1.2} \tabcolsep = 1mm
 \newcolumntype{.}{D{.}{.}{8}}
 \begin{tabular}{|c|....|}
 \hline
 \multicolumn{1}{|l|}{} &
 \multicolumn{4}{c|}{$T= 3.01T_{pc}$, $m_{\rm PS} / m_{\rm V} = 0.80$}
 \\ \hline
 \multicolumn{1}{|c|}{$rT$} &
 \multicolumn{1}{c} {$V_{\bf 1}$} & 
 \multicolumn{1}{c} {$V_{\bf 8}$} & 
 \multicolumn{1}{c} {$V_{\bf 6}$} & 
 \multicolumn{1}{c|}{$V_{{\bf 3}^*}$} 
   \\
 \hline
  0.250 &    -0.8454(  16)  &     0.0994(   4)  &     0.2020(   7)  &    -0.4200(   9)  \\
  0.354 &    -0.4790(  14)  &     0.0558(   3)  &     0.1131(   4)  &    -0.2376(   7)  \\
  0.433 &    -0.3233(  14)  &     0.0378(   3)  &     0.0768(   4)  &    -0.1604(   6)  \\
  0.500 &    -0.2608(  14)  &     0.0301(   3)  &     0.0615(   4)  &    -0.1295(   7)  \\
  0.559 &    -0.1959(  12)  &     0.0227(   2)  &     0.0461(   3)  &    -0.0971(   6)  \\
  0.612 &    -0.1533(  11)  &     0.0176(   2)  &     0.0361(   3)  &    -0.0761(   6)  \\
\hline
  \end{tabular}}
  \end{center}
 \end{table}

%%%%%%%%%%%%%%%%%%%%%%%%%%%%%%%%%%%%%%%%%%%%%%%%%%%%%%%%%%%%%%%%%%%%%%


\clearpage
\begin{thebibliography}{99}

\bibitem{YHM}
See, e.g., K. Yagi, T. Hatsuda, and Y. Miake, 
Quark-Gluon Plasma (Cambridge University 
Press, Cambridge, 2005).

\bibitem{cp1}
A. Ali Khan {\it et al.}  (CP-PACS Collaboration),
Phys.  Rev.  D {\bf 63} 034502 (2001).

\bibitem{cp2}
A. Ali Khan {\it et al.} (CP-PACS Collaboration), 
Phys.  Rev.  D {\bf 64} 074510 (2001).

\bibitem{Iwasaki}
  Y. Iwasaki, K. Kanaya, S. Kaya and T. Yoshie,
  Phys.  Rev.  Lett.   {\bf 78}, 179 (1997).

\bibitem{Pisarski}
  R. D. Pisarski and F. Wilczek,
  Phys.  Rev.  D {\bf 29}, 338 (1984).

\bibitem{Rajagopal}
  K. Rajagopal and F. Wilczek,
  Nucl.  Phys.  B {\bf 399}, 395 (1993).

\bibitem{cp3}
 A. Ali Khan {\it et al.}
 (CP-PACS Collaboration),  Phys. Rev. Lett. {\bf 85}, 4674 (2000).

\bibitem{cp4}
 A. Ali Khan {\it et al.}
 (CP-PACS Collaboration),  Phys. Rev. D {\bf 65}, 054505 (2001).

\bibitem{Kaczmarek:1999mm}
  O. Kaczmarek, F. Karsch, E. Laermann and M. Lutgemeier,
  Phys.  Rev.  D {\bf 62}, 034021 (2000).

\bibitem{Nakamura1}
  A. Nakamura and T. Saito,
  Prog.  Theor.  Phys. {\bf 111} 733 (2004).

\bibitem{Nakamura2}
  A. Nakamura and T. Saito,
  Prog.  Theor.  Phys. {\bf 112} 183 (2004).

\bibitem{Kaczmarek:2005ui}
  O. Kaczmarek and F. Zantow,
  Phys.  Rev.  D {\bf 71}, 114510 (2005).

\bibitem{Doring}
  M. D\"{o}ring, S. Ejiri, O. Kaczmarek, F. Karsch and E. Laermann,
  Eur.  Phys.  J.  C {\bf 46}, 179 (2006).

\bibitem{Bornyakov:2004ii}
  V. G. Bornyakov {\it et al.}  (DIK Collaboration),
  Phys.  Rev.  D {\bf 71}, 114504 (2005).

\bibitem{Laine:2006ns}
  M. Laine, O. Philipsen, P. Romatschke and M. Tassler,
  preprint hep-ph/0611300.

\bibitem{Matsui-Satz}
  T. Matsui and H. Satz,
  Phys.  Lett.  B {\bf 178}, 416 (1986).

\bibitem{bali}
 G. S. Bali, J. Fingberg, U. M. Heller, F. Karsch and K. Schilling, 
 Phys. Rev. Lett. {\bf 71}, 3059 (1993).

\bibitem{karr}
  L. K\"{a}rkk\"{a}inen, P. Lacock, D. E. Miller, B. Petersson and T. Reisz,
 Phys. Lett. B {\bf 312}, 173 (1993).

\bibitem{karsch} 
  F. Karsch, E. Laermann and M. Lutgemeier,
  Phys.\ Lett.\ B {\bf 346}, 94 (1995).

\bibitem{boyd}
 G. Boyd, J. Engels, F. Karsch, E. Laermann, C. Legeland, M. Lutgemeier and B. Petersson,
  Nucl.\ Phys.\ B {\bf 469}, 419 (1996).

\bibitem{Laine:2005ai}
  M. Laine and Y. Schr\"{o}der,
  JHEP {\bf 0503}, 067 (2005).

\bibitem{rg}
 Y. Iwasaki, Nucl. Phys. B {\bf 258}, 141 (1985);
 University of Tsukuba Report No. UTHEP-118 (1983).

\bibitem{cl} 
 B. Sheikholeslami and R. Wohlert,
 Nucl. Phys. B {\bf 259}, 572 (1985).

\bibitem{Aoki:1983qi}
  S. Aoki,
  Phys.  Rev.  D {\bf 30}, 2653 (1984).

\bibitem{Aoki:1986xr}
  S. Aoki,
  Phys.  Rev.  Lett.   {\bf 57}, 3136 (1986).

\bibitem{Aoki:1987us}
  S. Aoki,
  Nucl.  Phys.  B {\bf 314}, 79 (1989).

\bibitem{Aoki:1995yf}
  S. Aoki, A. Ukawa and T. Umemura,
  Phys.  Rev.  Lett.   {\bf 76}, 873 (1996).

\bibitem{Aoki:1996pw}
  S. Aoki, T. Kaneda, A. Ukawa and T. Umemura,
  Nucl.  Phys.  Proc.  Suppl.   {\bf 53}, 438 (1997).

\bibitem{Nadkarni1}
  S. Nadkarni,
  Phys.  Rev.  D {\bf 33}, 3738 (1986).

\bibitem{Nadkarni2}
  S. Nadkarni,
  Phys.  Rev.  D {\bf 34}, 3904 (1986).

\bibitem{Aoki:1999ff}
  S. Aoki {\it et al.}  (CP-PACS Collaboration),
  Phys.  Rev.  D {\bf 60}, 114508 (1999).

\bibitem{Gockeler:2005rv}
  M. G\"{o}ckeler {\it et al.},
  Phys.  Rev.  D {\bf 73}, 014513 (2006).

\bibitem{Nakamura:2003pu}
  A. Nakamura, T. Saito and S. Sakai,
  Phys.  Rev.  D {\bf 69}, 014506 (2004).

\bibitem{Rebhan:1993az}
  A. K. Rebhan,
  Phys.  Rev.  D {\bf 48}, 3967 (1993).

\bibitem{Lepage:1992xa}
  G. P. Lepage and P. B. Mackenzie,
  Phys.  Rev.  D {\bf 48}, 2250 (1993).

\bibitem{Jahn:2004qr}
  O. Jahn and O. Philipsen,
  Phys.  Rev.  D {\bf 70}, 074504 (2004).

\bibitem{Umeda:2006xi}
  T. Umeda,
  PoS {\bf LAT2006}, 151.

\bibitem{ls}
 Y. Schr\"{o}der and M. Laine, 
 PoS {\bf LAT2005}, 180 (2006).

\bibitem{ejiri} 
 S. Ejiri {\it et al.}, 
 PoS {\bf LAT2006}, 132.

\bibitem{maezawa}
 Y. Maezawa {\it et al.}, in the proceedings of Quark Matter 2006, 
preprint TKYNT-0701, to be published in J. Phys. G. 

\end{thebibliography}
\end{document}